\documentclass[preprint]{elsarticle}
\usepackage{amssymb}
\usepackage{amsmath}
\usepackage{graphicx}
\usepackage{lipsum}
\usepackage{url}
\usepackage{mwe}
\usepackage{lineno}
\usepackage[greek,english]{babel}
\newcommand{\microm}{\textrm{\selectlanguage{greek}m\selectlanguage{english}m}}
\journal{Journal of Nuclear Instruments and Methods in Physics Research A}

\begin{document}
\begin{frontmatter}
\title{Modeling the Timing Characteristics of the PICOSEC Micromegas Detector}

\author[aff2]{J. Bortfeldt}
\author[aff2]{F. Brunbauer}
\author[aff2]{C. David}
\author[aff3]{D. Desforge}
\author[aff5]{G. Fanourakis}
\author[aff7]{M. Gallinaro}
\author[aff11]{F. Garc\'{i}a}
\author[aff3]{I. Giomataris}
\author[aff9]{T. Gustavsson}
\author[aff3]{F.J. Iguaz}
\author[aff3]{M. Kebbiri}
\author[aff1]{K. Kordas}
\author[aff1]{C. Lampoudis}
\author[aff3]{P. Legou}
\author[aff2]{M. Lisowska}
\author[aff4]{J. Liu}
\author[aff2,fn2]{\mbox{M. Lupberger}}
\author[aff3]{O. Maillard}
\author[aff1]{\mbox{I. Manthos}}
\author[aff2]{H. M\"{u}ller}
\author[aff1]{V. Niaouris}
\author[aff2]{\mbox{E. Oliveri}}
\author[aff3]{T. Papaevangelou}
\author[aff1]{K. Paraschou}
\author[aff10]{M. Pomorski}
\author[aff4]{B. Qi}
\author[aff2]{\mbox{F. Resnati}}
\author[aff2]{L. Ropelewski}
\author[aff1]{D. Sampsonidis}
\author[aff2]{L. Scharenberg}
\author[aff2]{T. Schneider}
\author[aff3]{L. Sohl}
\author[aff2]{M. van Stenis}
\author[aff6]{Y. Tsipolitis}
\author[aff1]{S.E. Tzamarias\corref{cor1}}
\ead{tzamarias@auth.gr}
\author[aff2]{A. Utrobicic}
\author[aff8,fn3]{\mbox{R. Veenhof}}
\author[aff4]{X. Wang}
\author[aff2]{\mbox{S. White}}
\author[aff4]{Z. Zhang}
\author[aff4]{Y. Zhou}

\address[aff3]{IRFU, CEA, Universit\'e Paris-Saclay, F-91191 Gif-sur-Yvette, France}
\address[aff2]{European Organization for Nuclear Research (CERN), CH-1211 Geneve 23, Switzerland}
\address[aff4]{State Key Laboratory of Particle Detection and Electronics, University of Science and Technology of China, Hefei CN-230026, China}
\address[aff1]{Department of Physics, Aristotle University of Thessaloniki, University Campus, GR-54124, Thessaloniki, Greece.}
\address[aff5]{Institute of Nuclear and Particle Physics, NCSR Demokritos, GR-15341 Agia Paraskevi, Attiki, Greece}
\address[aff6]{National Technical University of Athens, Athens, Greece}
\address[aff7]{Laborat\'{o}rio de Instrumentac\~{a}o e F\'{i}sica Experimental de Part\'{i}culas, Lisbon, Portugal}
\address[aff8]{RD51 collaboration, European Organization for Nuclear Research (CERN), CH-1211 Geneve 23, Switzerland}
\address[aff9]{LIDYL, CEA, CNRS, Universit Paris-Saclay, F-91191 Gif-sur-Yvette, France}
\address[aff10]{CEA-LIST, Diamond Sensors Laboratory, CEA Saclay, F-91191 Gif-sur-Yvette, France}
\address[aff11]{Helsinki Institute of Physics, University of Helsinki, FI-00014 Helsinki, Finland}

\cortext[cor1]{Corresponding author}
\fntext[fn2]{Now at University of Bonn, D-53115 Bonn, Germany.}
\fntext[fn3]{Also at National Research Nuclear University MEPhI, Kashirskoe Highway 31, Moscow, Russia; and Department of Physics, Uludağ University, 16059 Bursa,Turkey.}
\begin{abstract}

The PICOSEC Micromegas detector can time the arrival of Minimum Ionizing Particles with a sub-25 ps precision. A very good timing resolution in detecting single photons is also demonstrated in laser beams. The PICOSEC timing resolution is determined mainly by the drift field. The arrival time of the signal and the timing resolution vary with the size of the pulse amplitude.

Detailed simulations based on GARFIELD++ reproduce the experimental PICOSEC timing characteristics. This agreement is exploited to identify the microscopic physical variables, which determine the observed timing properties. In these studies, several counter-intuitive observations are made for the behavior of such microscopic variables.
In order to gain insight on the main physical mechanisms causing the observed behavior, a phenomenological model is constructed and presented. The model is based on a simple mechanism of ``time-gain per interaction" and it employs a statistical description of the avalanche evolution. It describes quantitatively the dynamical and statistical properties of the microscopic quantities, which determine the PICOSEC timing characteristics, in excellent agreement with the simulations. In parallel, it offers phenomenological explanations for the behavior of these microscopic variables.
The formulae expressing this model can be used as a tool for fast and reliable predictions, provided that the input parameter values (e.g. drift velocities) are known for the considered operating conditions. 
\end{abstract}

\begin{keyword}
gaseous detectors \sep Micromegas \sep modeling \sep timing resolution
\end{keyword}

\end{frontmatter}

\section{Introduction} \label{intro}
\begin{figure}[t]
    \centering
    \includegraphics[width=0.7\textwidth]{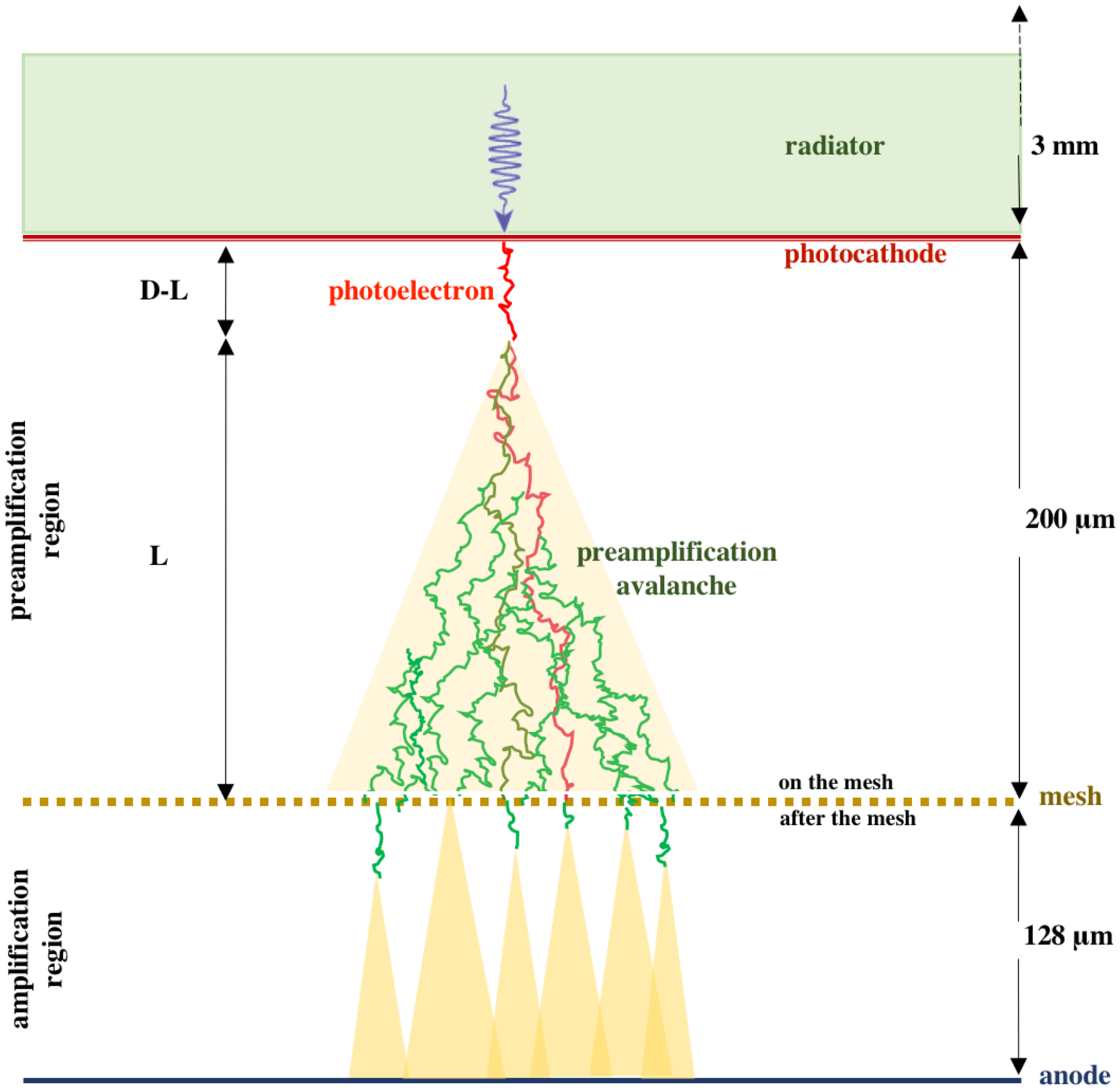}
    \caption{Illustration of the main PICOSEC detector components (dimensions are only indicative): the radiator of typical thickness $\approx 3~$mm, the photocathode, the pre-amplification (drift) region of depth  D ($200~\microm$), the mesh, the amplification region ($128~\microm$) and the anode. A photoelectron, after drifting a length D-L, produces a pre-amplification avalanche, of length L, ending on the upper surface of the mesh (on the mesh). A fraction of the avalanche electrons traverses the lower surface of the mesh (after the mesh) and produces avalanches in the amplification region.} 
    \label{fig:fig1}
\end{figure}

%\begin{linenumbers}
 The PICOSEC Micromegas detection concept is realized by a two-stage Migromegas detector \citep{micromegas} coupled to a front window that acts as Cherenkov radiator coated with a photocathode. The drift region is very thin ($\sim200~\microm$) minimizing the probability of direct gas ionization as well as diffusion effects on the signal timing. Due to the high electric field, photoelectrons undergo pre-amplification in the drift region. The readout is a bulk Micromegas \citep{bulk}, which consists of a woven mesh and an anode plane separated by a gap of  $\sim128~\microm$, mechanically defined by pillars. A relativistic charged particle traversing the radiator produces UV photons, which are simultaneously (RMS less than 10 ps) converted into primary photoelectrons at the photocathode. These primary photoelectrons produce  pre-amplification avalanches in the drift region (hereafter called pre-amplification region). A fraction of the pre-amplification electrons ($\sim25 \%$) traverse the mesh and are finally amplified in the amplification region. The main detector components along with a schematic representation of the  relevant microscopic processes producing the signal are shown in Fig. \ref{fig:fig1}.

 The arrival of the amplified electrons at the anode produces a fast signal component (with a rise-time of $\sim0.5$~ns) referred to as the electron-peak (``e-peak''), while the movement of the ions produced in the amplification gap generates a slower   ($\sim 100$~ns) ion-tail component. This type of detector operated with Neon or $CF_4$ based gas mixtures can reach high enough gains to detect single photoelectrons.
The PICOSEC Micromegas detector (hereafter PICOSEC) has the potential to time the arrival of Minimum Ionizing Particles (MIPs) with a sub-25 ps precision \citep{pico24}. Extensive tests with laser beams also demonstrated \citep{kostas} very good timing resolution in detecting single photons. These laser beam data are also used for detector calibration purposes, and are referred to as ``calibration data'' in the following.\\
 
It is not surprising that the PICOSEC approach to charged particle timing, results in a significant improvement over the time jitter obtained when using a gaseous detector sensitive to ionization produced by traversing charged particles in the gas volume. With multiple ionization and without pre-amplification in the drift region the timing resolution in a gaseous detector is of the order of a few nanoseconds  \citep{sauli}. With the above modifications to the design of a typical Micromegas, PICOSEC accomplishes a far better precision in timing for two reasons: i) the photoelectrons enter the drift region simultaneously and ii) the pre-amplification in the very thin drift region allows for time-averaging of the electrons arriving in the amplification region through the mesh structure.

The purpose of this paper is to give a full phenomenological description of the PICOSEC performance and to provide a detailed model to be used for further optimization of this device as a mature, robust detector. With this model in hand we are then able to address questions such as the following: a) What is the relative importance of the Drift stage and Amplification stage in the jitter of the PICOSEC Signal Arrival Time (SAT)? b) How does the SAT generated by a given photoelectron depend on the fluctuating distance to where it initiates the avalanche in the drift region? c) How the SAT jitter (i.e. the timing resolution) depends on the properties of the gas mixture that fills the detector, and  on the voltage settings? d) What is the effect of transmission through the mesh on time jitter? e) Which is the optimal structure?

As  will become obvious, a detailed microscopic description of the physics principles underlying the PICOSEC detector is a prerequisite to answering these questions.
\\
  \begin{figure}
\centering
\includegraphics[width=1.\textwidth]{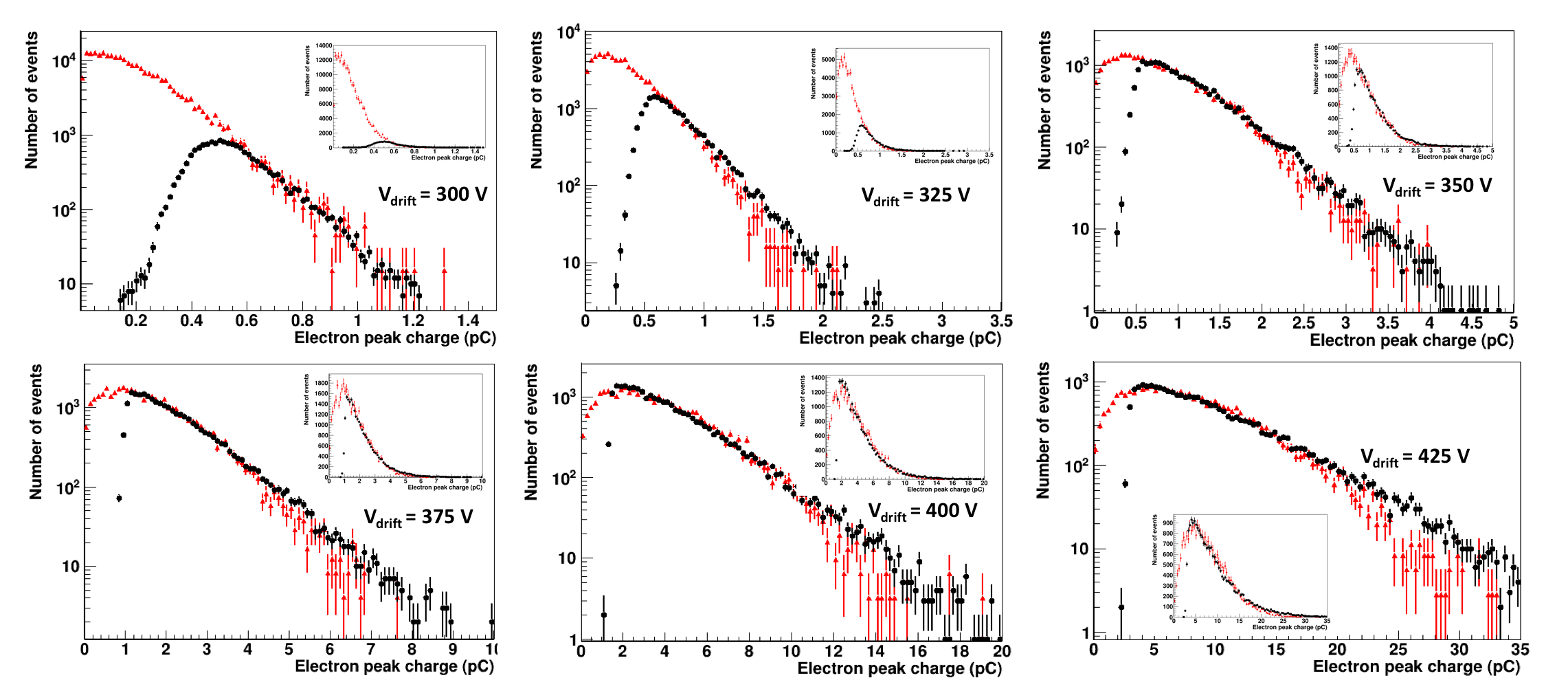}
\caption{Distributions of the e-peak charge induced by a single photoelectron, for several drift voltage settings (300 V, 325 V, 350 V, 375 V, 400 V and 425 V). The black points represent calibration data published in \citep{pico24} while the red triangles correspond to GARFIELD++ simulated PICOSEC e-peak waveforms treated the same way as the experimental data, as described in \citep{kostas}. The data distributions are affected, at low e-peak charge values, by the amplitude threshold applied for data collection.  
}
\label{fig:fig2a}
\end{figure}
 
 Naturally, the PICOSEC timing resolution depends on the drift and anode operating voltages. In the laser-beam tests, where the anode voltage was high ($>$400 V), it was found that the single-photoelectron timing resolution is determined mainly by the drift field. It was also observed that the PICOSEC signal arrival time (SAT) and the timing resolution vary as functions of the size of the e-peak, i.e. the e-peak voltage amplitude or the respective e-peak charge. These functional forms were found to be practically the same for drift voltages in the range of 300 V - 425 V. It should be emphasized that the above dependencies have been found \citep{pico24,kostas} not to be  systematic artifacts of the experimental timing technique but they stem from the physics determining the signal production.\\

\begin{figure}
\centering
\begin{minipage}{.48\textwidth}
\centering
\includegraphics[width=1.\textwidth]{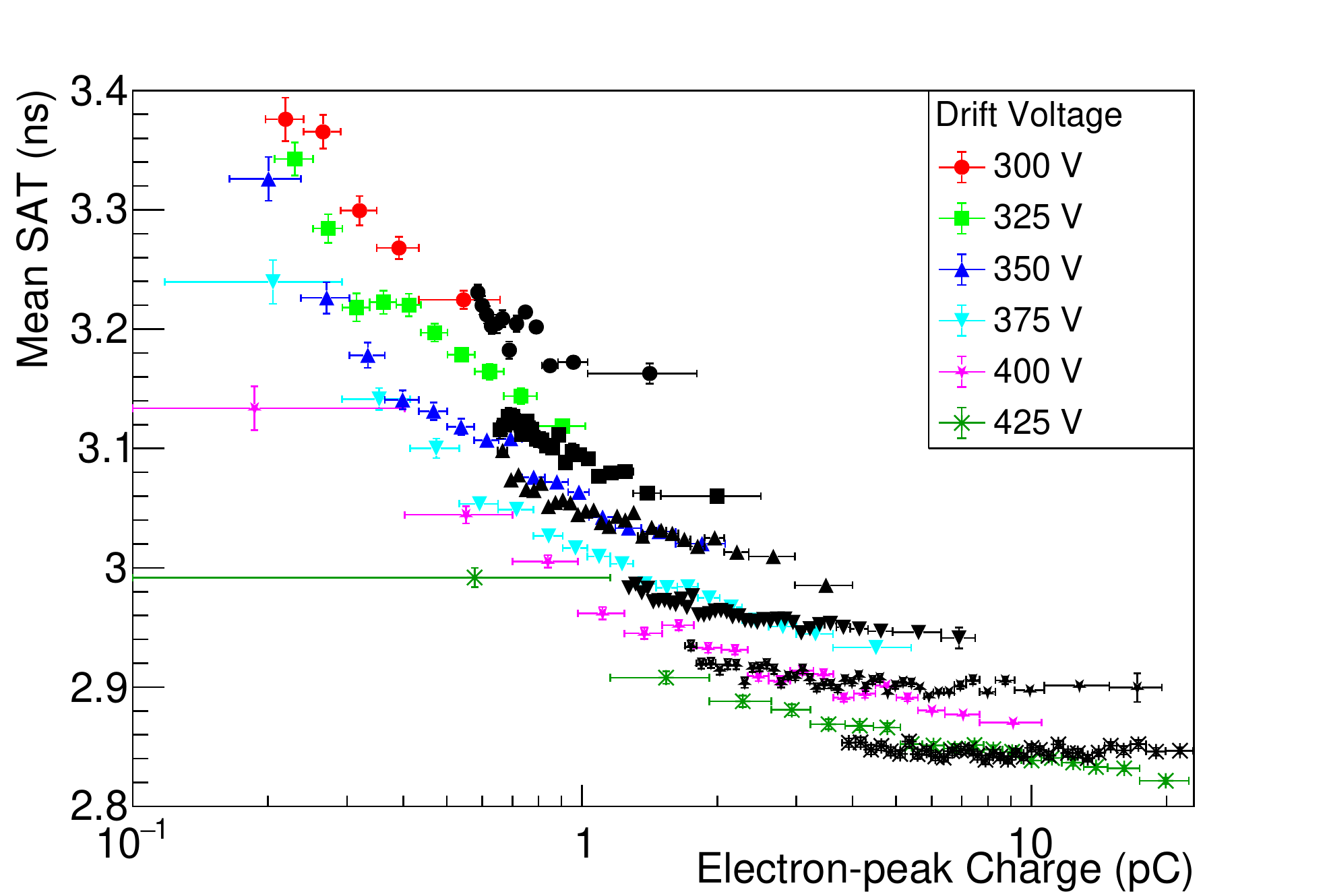}
\end{minipage}%
\begin{minipage}{.48\textwidth}
\centering
\includegraphics[width=1.\textwidth]{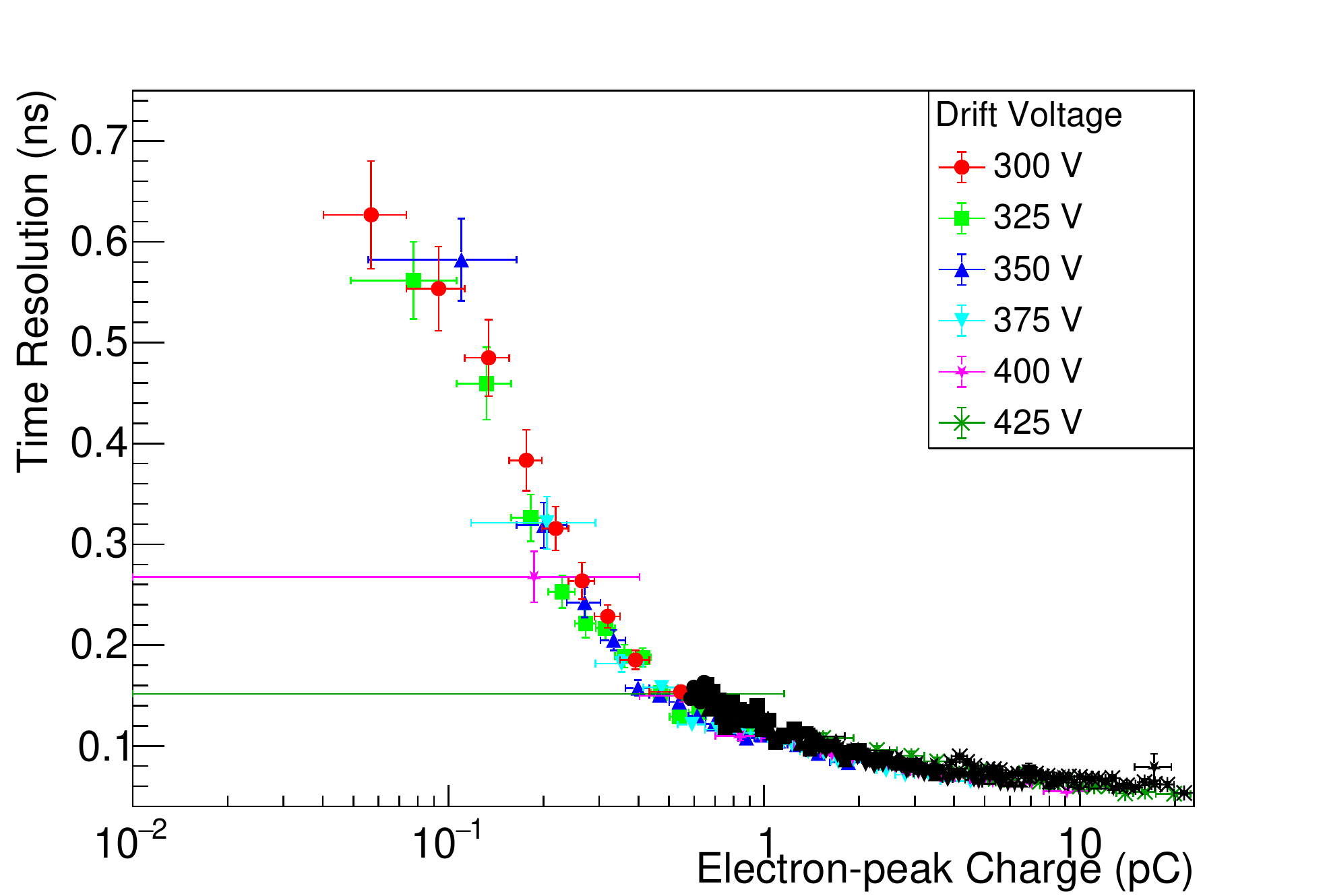}
\end{minipage}
\caption{(left) Mean SAT as a function of the electron peak charge. (right)
Time resolution as a function of the electron peak charge. In both figures
black points represent experimental measurements \citep{pico24} while colored symbols correspond
to simulations \citep{kostas}. The gas used is the COMPASS gas with an anode
voltage of 450 V and for drift voltages of (red) 300 V, (light green) 325 V, (blue)
350 V, (cyan) 375 V, (magenta) 400 V and (dark green) 425 V.}
\label{fig:fig2}
\end{figure}

Detailed simulations, based on the GARFIELD++ \citep{garfield} package, including the simulation of the electronic response of the detector and the noise contribution, were used to reproduce \citep{kostas} the observed PICOSEC performance  characteristics, when detecting single photons.
 Comparison of simulation predictions with the laser-beam calibration data resulted in estimating the Penning transfer rate (Ptr) \citep{penning}   of the used COMPASS gas\footnote{The term ``COMPASS gas'' refers to the mixture $80\%\, \textrm{Ne}, 10\%\, \textrm{C}_2 \textrm{H}_6, 10\%\, \textrm{CF}_4$, as used by the COMPASS Collaboration. The Ptr for this gas was estimated in \citep{kostas} to be $\sim 50\%$.}. The e-peak charge distribution of the simulated waveforms is in a good agreement with the calibration data, as it is illustrated in Fig. \ref{fig:fig2a}. Furthermore, as shown in Fig. \ref{fig:fig2}, the SAT and the timing resolution of the simulated waveforms depend on the e-peak size in exactly the same way as observed in the  data.
 
\begin{figure}
\centering
\begin{minipage}{.32\textwidth}
\centering
\includegraphics[width=1.\textwidth]{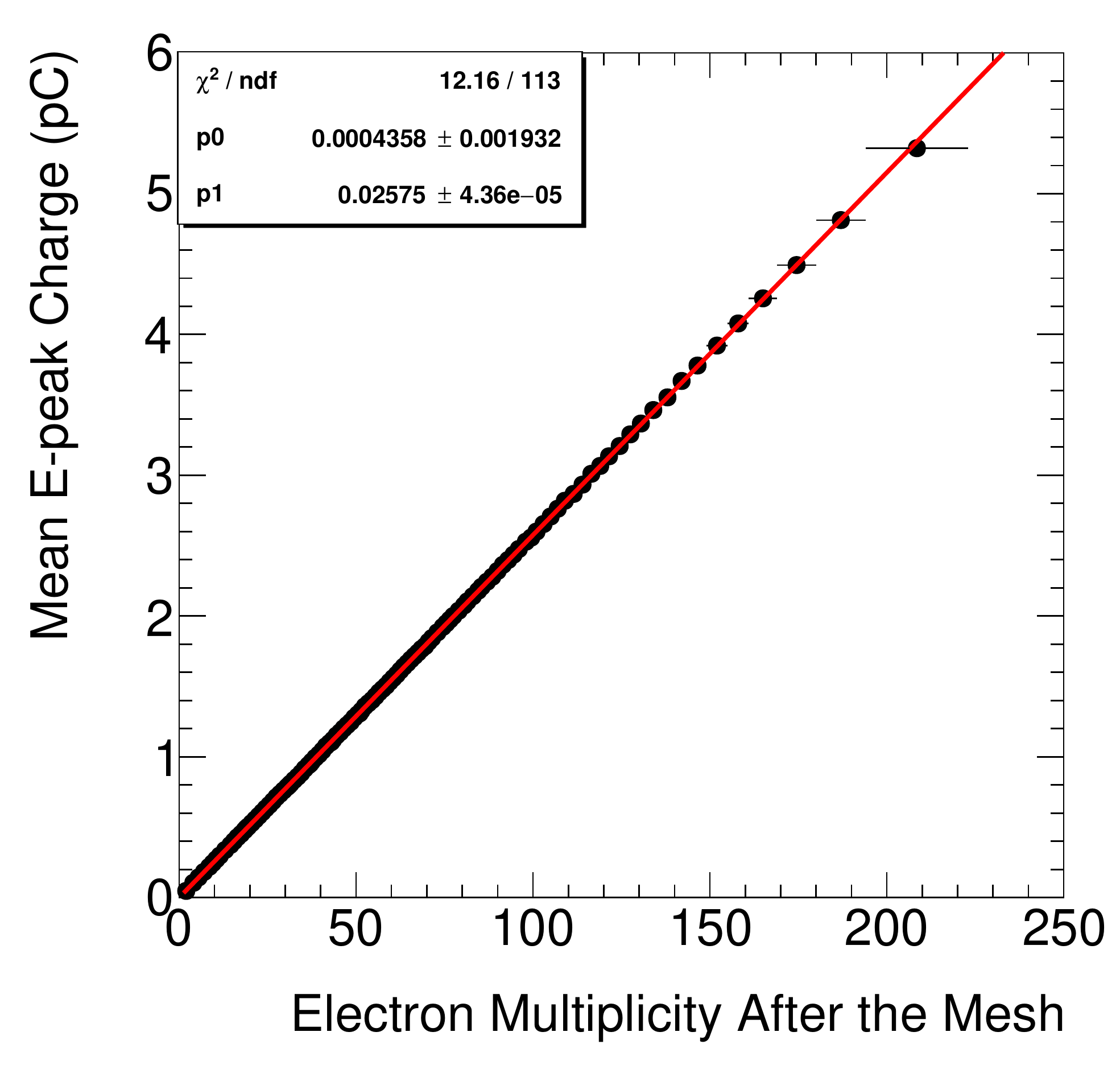}
\end{minipage}%
\begin{minipage}{.32\textwidth}
\centering
\includegraphics[width=1.\textwidth]{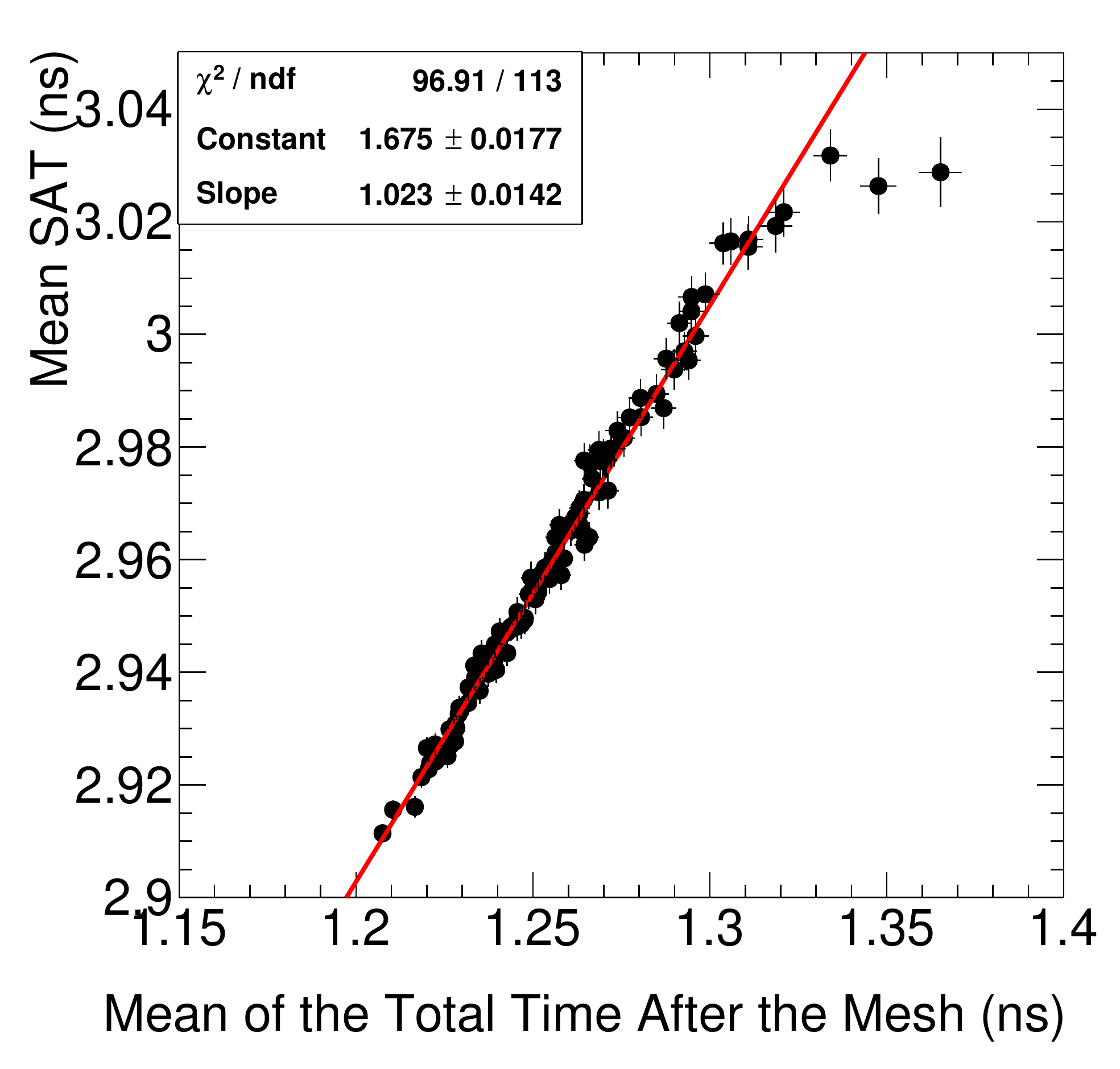}
\end{minipage}%
\begin{minipage}{.32\textwidth}
\centering
\includegraphics[width=1.\textwidth]{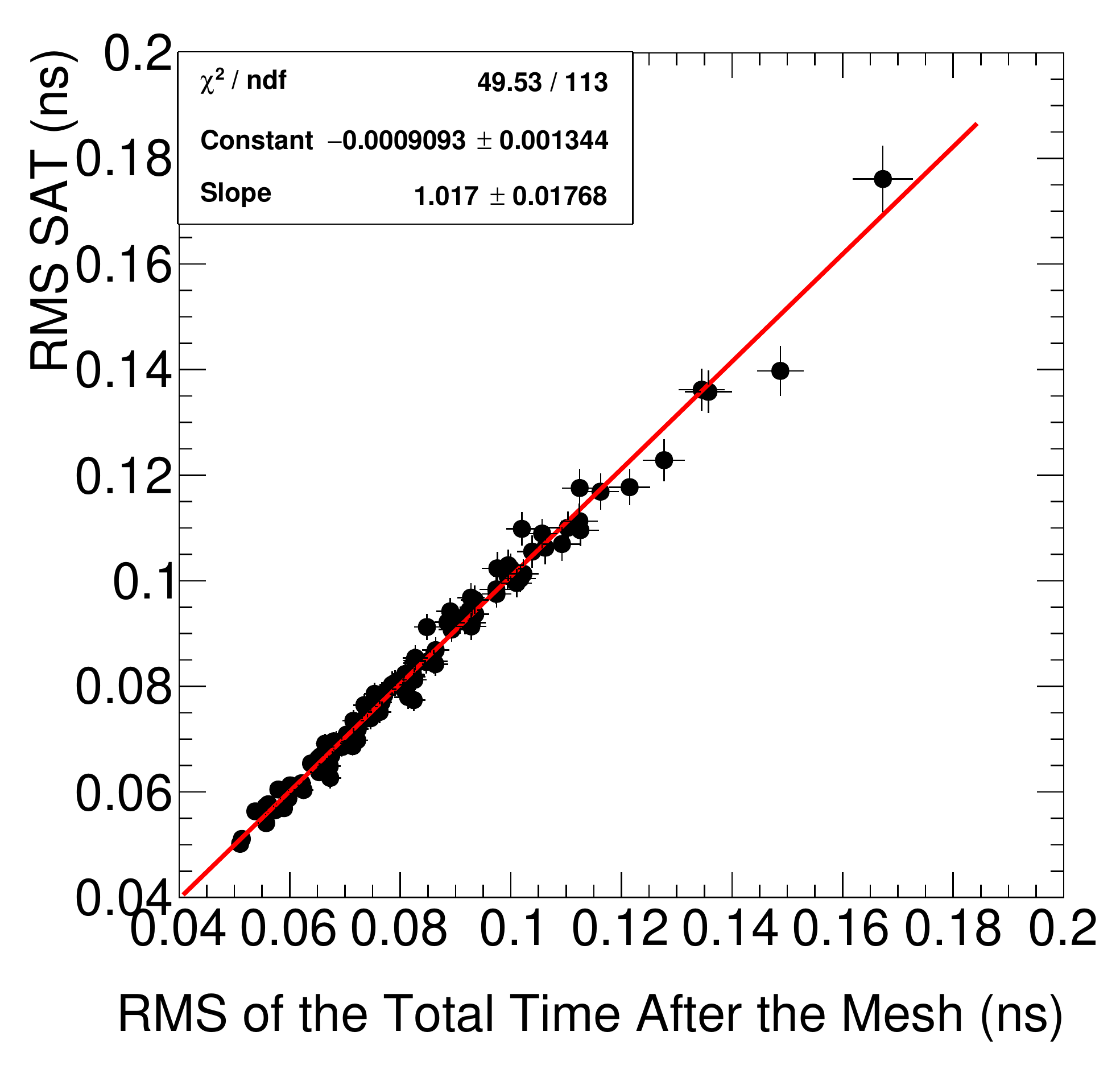}
\end{minipage}
\caption{(left) The mean e-peak charge of simulated PICOSEC signals versus the respective \textit{``electron multiplicity after the mesh"}. The middle and right plots demonstrate that  the macroscopically determined PICOSEC SAT has the same properties as the microscopic variable \textit{``total-time after the mesh"}, as it is described in the text. }
\label{fig:fig3}
\end{figure}

The agreement between simulation and experimental data is further exploited  in order to identify the microscopic physical variables that determine the observed timing characteristics. 
Specifically, GARFIELD++ simulations show that the number of pre-amplification electrons  traversing the mesh and initiating avalanches in the amplification region (a microscopic variable hereafter called \textit{``electron multiplicity after the mesh"}) determines the size of the PICOSEC e-peak (a macroscopic, observed quantity), as seen in the left plot of Fig. \ref{fig:fig3}. 

In the simulation, one has the ability, for each pre-amplification electron traversing the mesh, to determine the time it enters the anode region, measuring time from the instant of the photoelectron emission. The average of these times, for all pre-amplification electrons, defines the microscopic variable hereafter called \textit{``total-time after the mesh"}. This microscopic variable   has the same properties as the measured arrival time of the PICOSEC signal\footnote{The arrival time of the PICOSEC signal is defined at a constant fraction (20\%) of the e-peak amplitude, as described in \citep{pico24}.}.   
Indeed, as shown in the right plot of Fig. \ref{fig:fig3}, for simulated  single photoelectron events with the same e-peak size, the spread (RMS) of the microscopic ``total-time after the mesh" values is found to be equal to the spread of the corresponding signal arrival times, i.e. to the macroscopic PICOSEC timing resolution.
Furthermore, the mean values of the ``total-time after the mesh''   differ only by a constant time-offset from the respective mean values of the  PICOSEC signal arrival times, as demonstrated in the middle plot of Fig. \ref{fig:fig3}. This offset is independent of the e-peak size and it is due to the fact that the SAT also includes: a) the propagation time of the amplification avalanches and b) the rise-time of the signal up to the 20\% of the e-peak amplitude.

%Measuring time from the instant of the photoelectron emission, it was found that the average time the pre-amplification electrons take to enter the amplification region 
%  (a microscopic variable hereafter called \textit{``total-time after the mesh"}) has the same properties as the macroscopically determined timing of the PICOSEC signal\footnote{The timing of the PICOSEC signal is defined at a constant fraction ($\sim20 \%$) of the e-peak amplitude, as described in \citep{pico24}.}. 
%  The RM.S. (also called ``the spread" in the rest of the paper) of the microscopic ``total-time after the mesh", of synchronously produced photoelectrons and of the same e-peak size, is found to be equal to the spread of the SAT (i.e. the macroscopically observed timing resolution) determined in the same events.
%   Similarly, the mean value of the total-times after the mesh differ only by a constant delay from the respective, macroscopically observable mean value of the PICOSEC SAT.  
%  Based on these correspondences, the phenomena that determine the PICOSEC timing characteristics are studied in detail, on a microscopic level \citep{simu}, in the framework of GARFIELD++. 
  % Hereafter, any reference to GARFILED++ simulation results, should be considered as a citation to the work published in \citep{simu}. \\

Having identified the relevant microscopic variables that determine the PICOSEC timing characteristics,  the detailed GARFIELD++ simulation is further used in this work to study the dynamical evolution of the PICOSEC signal in terms of the electron multiplicities and other important variables, such as the primary photoelectron drift path and the length of the pre-amplification avalanches. Moreover, in order to gain insight on the  physical mechanisms causing: a) the dependence of the PICOSEC timing characteristics on the signal size and b) the weak influence of the mesh transparency on the timing resolution, a  stochastic model is constructed. The model is based on a simple concept of ``time-gain per interaction" and reproduces the PICOSEC timing characteristics equally well as  the detailed GARFIELD++ simulation. In addition, the model offers a phenomenological interpretation of a number of peculiar statistical properties found in the GARFIELD++ results. 

An overview of this article is given in Section \ref{overview} while  the remaining sections contain a detailed description of the stochastic modeling of all  relevant processes and  demonstrate the  model performance. The article finishes with concluding remarks in Section  \ref{concl}.

\section{An Overview} \label{overview}

In this work, the GARFIELD++ package\footnote{GARFIELD++ version: https://gitlab.cern.ch/garfield/garfieldpp, commit e018bcca (8 May 2017)} is used to describe microscopically  the PICOSEC  timing properties by simulating in detail all the relevant processes. Interpreting the simulation predictions statistically leads to several counter-intuitive observations, e.g. a) the primary photoelectron drift velocity  seems to depend on Ptr (Penning transfer rate), b) the  avalanche electrons  drift  faster than the primary photoelectron, c) the average speed of the avalanche as a whole is  larger than the drift velocity of its constituent electrons, d) the longitudinal diffusion of the avalanche is almost independent on its length and e) the 25\% transparency of the mesh has only a minor effect on the PICOSEC timing resolution. Furthermore, it is found that the PICOSEC timing resolution is mainly determined by the drift path of the primary photoelectron; however,  when expressing the timing resolution as a function of the number of electrons passing through the mesh (i.e. the e-peak size), the related   photoelectron and  avalanche contributions to the resolution were found to be heavily correlated.\\

In order to identify the main physical processes causing the observed behaviour, a simple phenomenological model is developed  and presented in this paper. The model is based on a simple mechanism of ``time-gain per interaction" and it employs a statistical description of the avalanche evolution. It describes well the above-mentioned phenomena in  excellent agreement with the GARFIELD++ simulation results, as demonstrated in the following sections.

	The input parameters of the model (i.e. drift velocities, ionization probabilities per unit length, multiplication and diffusion coefficients, mean value and variance of the ``time-gain per interaction", average mesh transparency and longitudinal diffusion around the mesh, etc.) are commonly used statistical variables with values that depend on the PICOSEC gas filling and the operating voltage settings. 
The values of these	parameters have been estimated from GARFIELD++ simulations, for the COMPASS gas mixture, assuming several values of Ptr (Penning  transfer rates: 0\%, 50\%, 100\%), anode voltage fixed to 450 V, and  various  drift voltages, i.e. 300 V, 325 V, 350 V, 375 V, 400 V and 425 V. A compilation of these input parameter values can be found in \ref{Appendix A}. The model predictions were compared with the GARFIELD++ results for all the above operating conditions (hereafter called ``considered operating conditions"). If the PICOSEC operating conditions are not specifically stated,  the following default values are implied: Ptr of $50\%$,   anode voltage of 450 V,  and drift voltage of 425 V. \\

	The model is based on the observation \citep{quench} that an electron, 
	drifting in  an homogeneous electric field and only undergoing elastic scatterings, drifts along the field with less average velocity than an electron suffering energy losses through its interactions. 
	In Section \ref{driftvel}, the above concept is quantified in terms of a ``time-gain per interaction". It is used to explain the  different   drift velocities between  a photoelectron prior to ionization and of an avalanche electron. It also explains the effect that  the Ptr seems to have on the drift velocities.

Section \ref{modelaval1} to Section \ref{modelaval3} describe the modelling of microscopic processes up to the mesh. At this stage, the important microscopic variables are: i) the number of pre-amplification electrons arriving on the mesh (hereafter called \textit{``electron multiplicity on the mesh"}), and ii) the average of the arrival times of the individual pre-amplification electrons on the mesh (hereafter called \textit{``total-time on the mesh"}). The transfer of the pre-amplification electrons through the mesh is modelled in Section \ref{modelaval5}.\\
Specifically, the average avalanche velocity is a statistical outcome of several dynamical effects, including those that determine the avalanche growth. Section \ref{modelaval1} examines the properties of GARFIELD++ simulated pre-amplification avalanches including the statistical distribution of the avalanche electron multiplicity before and after the mesh. The mean mesh transparency to pre-amplification  electrons is found to be constant and independent of the avalanche characteristics, for all the considered operating conditions. This implies that the signal size is determined effectively by ``the electron multiplicity on the mesh".  The simultaneous drift and growth of the pre-amplification avalanche is also modelled in  Section \ref{modelaval1} and   the ``avalanche transmission time''\footnote{The ``avalanche transmission time'' is defined as the average of the arrival times of the avalanche electrons on the mesh, starting from the instant of the first ionization which initiated the avalanche.} is expressed in terms of its length and its electron multiplicity. The model explains quantitatively the GARFIELD++ prediction that the avalanche, as a whole, runs faster than its constituent electrons. \\
In Section \ref{modelaval2}, by integrating properly the results of Section \ref{modelaval1}, the model predicts the dependence of the ``total time on the mesh" on the number of pre-amplification electrons.\\ 
The  arrival times of the avalanche electrons on a plane are mutually correlated, due to the sharing of common parent electrons. This correlation is quantified in Section \ref{modelaval3}. By evaluating  the avalanche contribution to the statistical spread of the ``total-time on the mesh", the model predicts that it is almost independent of the avalanche length. The  longitudinal diffusion of the primary photoelectron, along its drift path before the first ionization, is the major factor determining the PICOSEC timing resolution. However, due to the fact that the photoelectron drift path and the avalanche length sum up to the  pre-amplification region depth, the timing resolution indirectly depends on the avalanche length.\\
Although the length of the avalanche is an important physical parameter, it is not an experimental observable. In Section \ref{modelaval4},  the statistical spread of the ``total time on the mesh" is expressed as a function of the pre-amplification electron multiplicity, by modelling the dynamical growth of the avalanche.
%\  and it takes into account the statistical correlations arising from the dependence of both the photoelectron and the %\avalanche transmission times on their drift path.
The influence of the mesh on the PICOSEC timing properties is quantified in Section \ref{modelaval5}
in terms of the mesh transparency, the number of the pre-amplification electrons reaching the mesh and an extra time-spread term, due to the electron drift through the non-homogeneous electric field around the mesh. \\
Finally in Section \ref{discl}, limitations of the model to describe accurately the PICOSEC timing characteristics in the case of very small electron multiplicity on (and after) the mesh are discussed. In the same Section, the model extension to predict the complete probability density functions, which determine the timing properties of the PICOSEC signal, is also presented.
\\ 
The Section \ref{concl} comprises a summary of the model success to describe the PICOSEC timing characteristics   along with remarks on potential applications for studying related phenomena. 
% Finally, Appendix C comprise the MATHEMATICA Notebook, which can be used to produce model predictions, at $450\,V$ in the anode, by providing the drift voltage. The predictions are guaranteed to be accurate in a drift voltage region of $300$ to $450\,V$. Outside this region the user should provide values for each of the necessary parameters (i.e. those compiled in Table A-8) corresponding to the chosen drift voltage.

\section{Electron Drift Velocities and the Basic Model Assumptions} \label{driftvel}
Forward moving electrons lose more time when back-scattered elastically,  before the electric field or another collision sets them back to forward motion, compared to electrons losing energy to  interactions and also profit from longer mean-free paths at low energies due to the small scattering cross section (Ramsauer minimum).
	The fact that an electron gains in transmission time every time it loses energy is used  to explain the different drift velocities predicted by the detailed GARFIELD++ simulation. \\
\begin{figure}
\centering
\begin{minipage}{.52\linewidth}
\centering
\includegraphics[width=\linewidth]{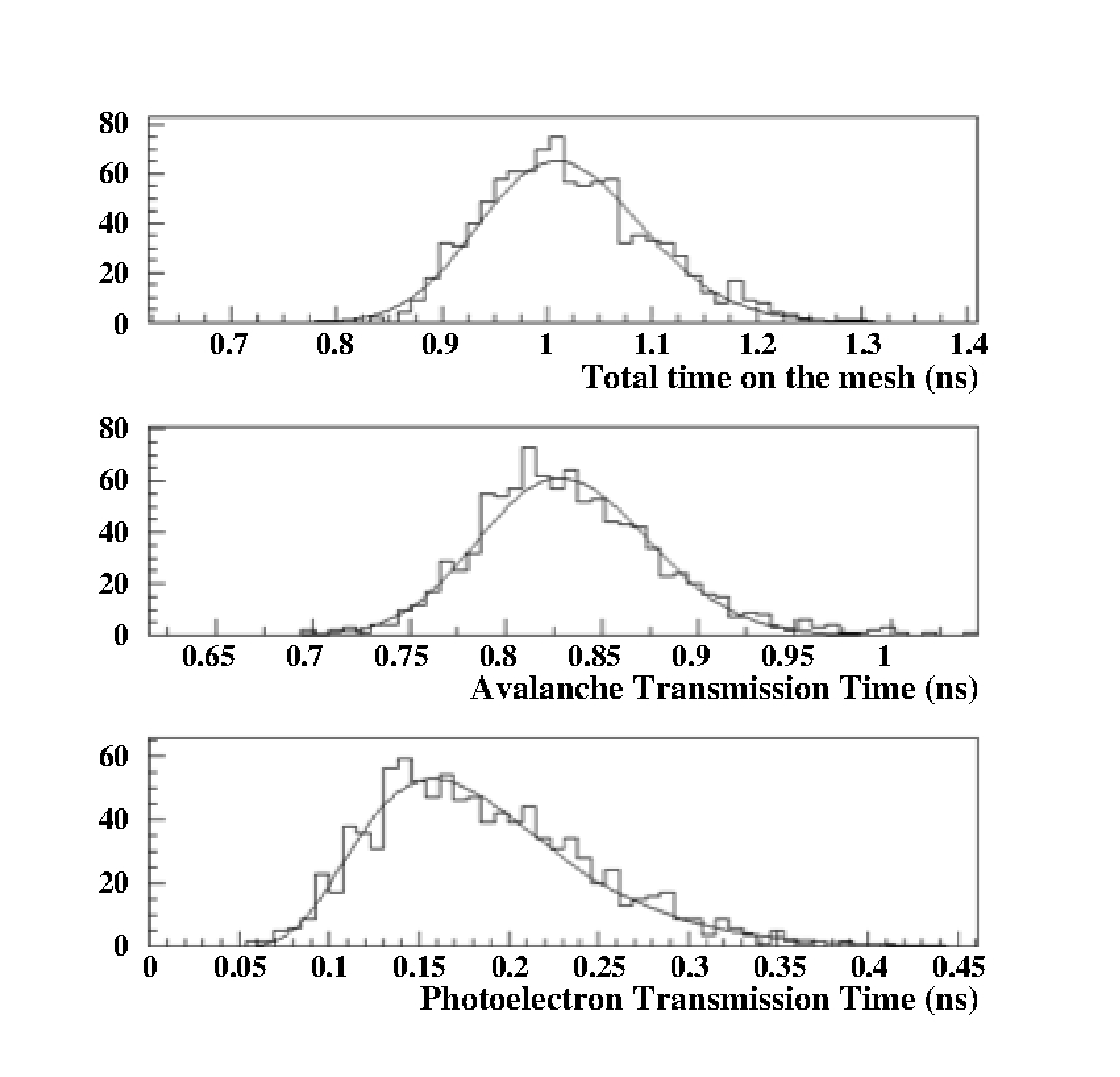}
\end{minipage}%
\begin{minipage}{.52\linewidth}
\centering
\includegraphics[width=\linewidth]{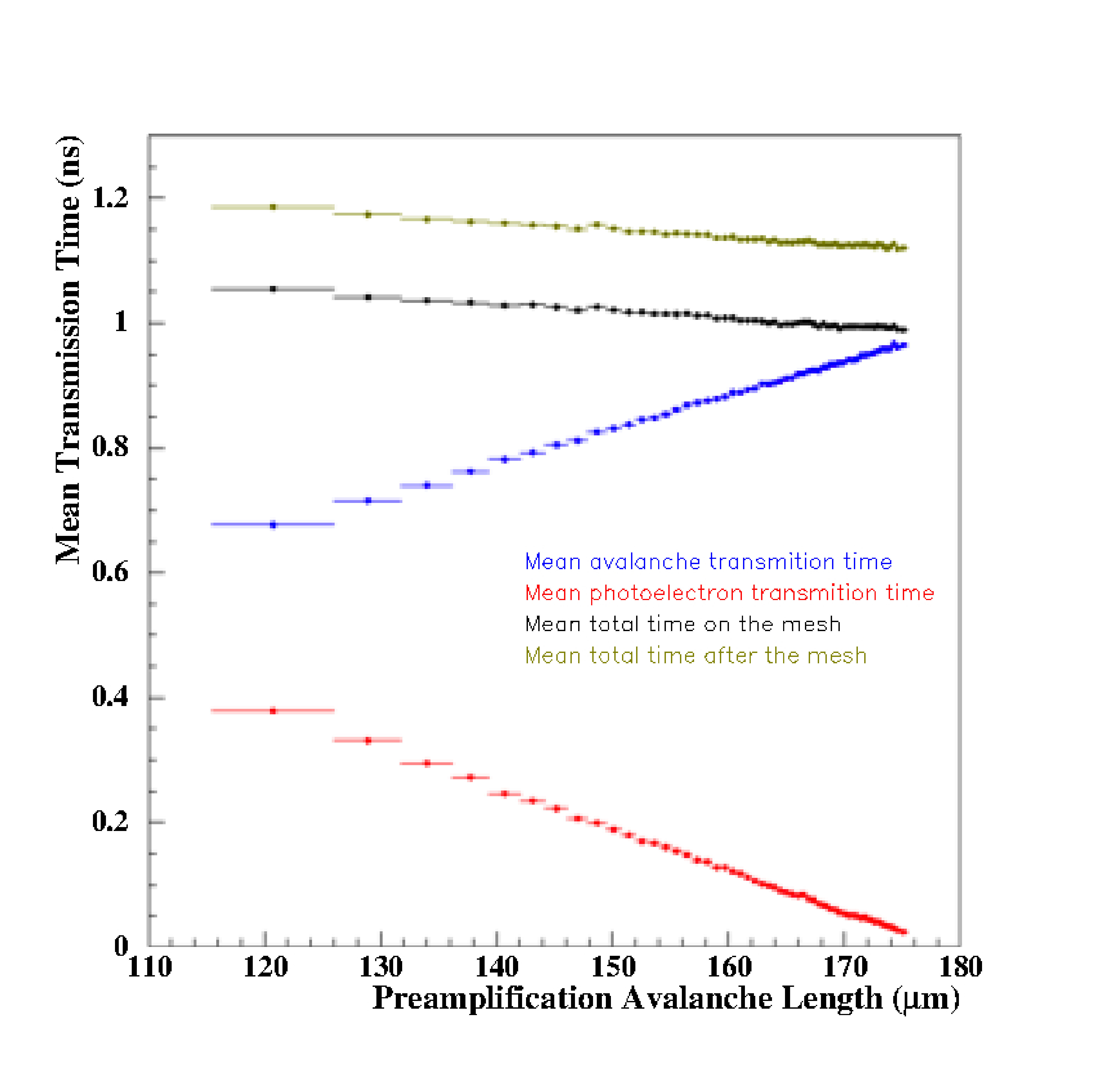}
\end{minipage}\quad
\caption{On the left, the plots show  the distributions of the ``total time on the mesh'' (top), the ``avalanche transmission time'' (middle) and the ``photoelectron transmission time'' (bottom), in the case that the length of the simulated  avalanche (L) is between 144.45 and 144.75 $\microm$.  The solid lines represent fits with the Wald distribution function. The right plot presents the mean values of the above times, as well as the mean of the ``total time after the mesh'', versus the length of the respective pre-amplification avalanche. It is worth noticing that the total time after the mesh differs only by a constant time-offset from the respective total time on the mesh, at all considered avalanche lengths.}
\label{fig:fig4}
\end{figure}
In a PICOSEC pre-amplification region of a certain depth $D$, let $L$ be the length of a pre-amplification avalanche and $D-L$ the corresponding drift length of the photoelectron before the first ionization initiating the avalanche. Let $T_{p}(L)$ be the time taken from the instant of the photoelectron emission to its first ionization (hereafter called \textit{``photoelectron transmission time"} or just \textit{``photoelectron time''}). Measuring time from the instant of the first ionization, let $T(L)$  be the average  time that the avalanche electrons take to reach the mesh (hereafter called \textit{ ``avalanche transmission time"} or just \textit{``avalanche time''}). Apparently the \textit{``total-time on the mesh"}, $T_{tot}(L)$ equals  the sum of the photoeletron and avalanche transmission times, i.e. $T_{tot}(L)=T_{p}(L)+T(L)$. All the above time-variables behave statistically as random variables following probability distributions that are well approximated by Inverse Gaussians (Wald) functions, as demonstrated in Fig. \ref{fig:fig4} (left) using GARFIEL++ simulations. The simulations also show that the mean values  of the above time distributions depend linearly on the avalanche length (see the right plot of Fig. \ref{fig:fig4}).  Similarly, the mean value of the time $T_{ea}\left( x\right) $, which is the time taken by an avalanche electron to cover a distance $x$ along the drift field, was found also to depend linearly on $x$.  The slopes of the aforementioned linear dependencies define the inverse of the respective drift velocities. 

Hereafter, $V_{p}$ stands for the \textit{``photoelectron drift velocity"}, $V_{a}$ denotes the \textit{``avalanche drift velocity"} and  $V_{ea}$ is the \textit{``drift velocity of an avalanche-electron"} assuming that every avalanche electron drifts with the same velocity. Estimated values of the  above drift velocities  are compiled in Table \ref{tab:tableA-1}, for three different Ptr (Penning transfer rate) values and default high voltage settings, and in Table \ref{tab:tableA-8} for 50\% Ptr, 450 V anode and several drift voltage settings.  The listed $V_{p}$, $V_{a}$ and $V_{ea}$ values  have been estimated by linear fits\footnote{The small, non-zero constant terms found in these linear fits were attributed to the fact that the stochastic description of the electron drift and the avalanche development starts to be valid after statistical equilibrium is reached.} to the $T_{p}(L)$ versus L, $T(L)$ versus L and  $T_{ea}\left( x\right) $ versus \textit{x} dependencies, observed in GARFIELD++ simulations, respectively.   Apparently, all the above drift velocities increase with the drift voltage;
%\\footnote{For 50\% Penning Transfer rate, the drift velocity values, in units of $\microm /ns$,  for drift voltages ($V_{drift}$) in the region 325-475 V, are well approximated by the following parametrizations: $V_p=36.12+0.27\cdot V_{drift}$, $V_a=21.916+0.38116\cdot V_{drift}$, $V_{ea}=27.75+0.3357\cdot V_{drift}.$} 
 however, the photoelectron drift velocity is smaller than the avalanche-electron drift velocity, which is in turn smaller than the drift velocity of the avalanche as a whole. Furthermore, as a function of Ptr, the photoelectron drift velocity decreases, the drift velocity of the avalanche as a whole increases, while the avalanche-electron drift velocity remains constant.\\

\begin{figure}[t]
    \centering
    \includegraphics[width=0.6\textwidth]{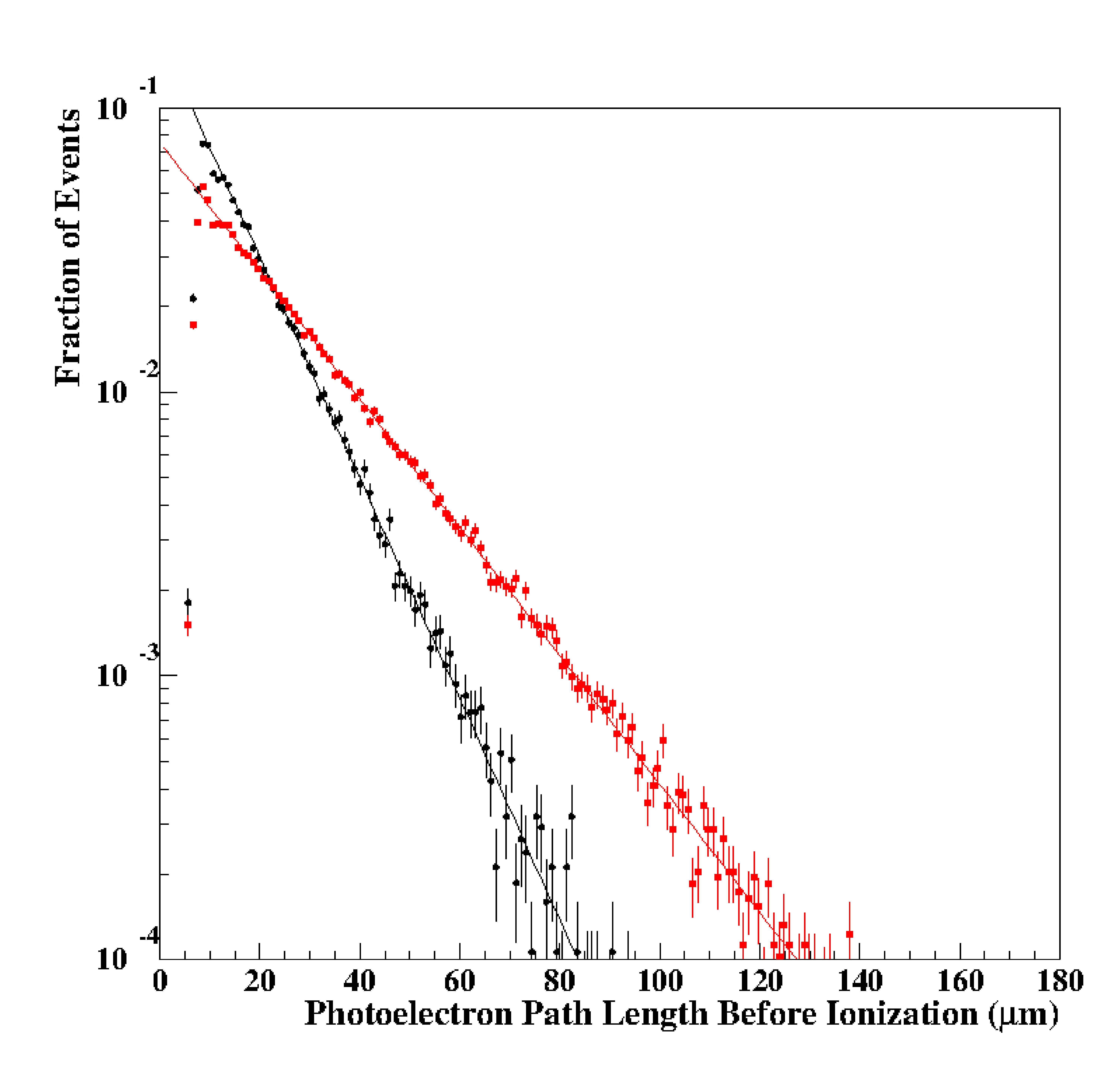}
    \caption{Distributions of the photoelectron drift path length, before the initiation of an avalanche, produced by GARFIELD++ simulations with 425 V drift voltage and Ptr equal to 100\% (black circles) and 0\% (red squares).The solid lines represent the results of exponential fits.} 
    \label{fig:fig5}
\end{figure}

Our model attributes the different values  of the above drift velocities to time-gains per inelastic interaction.  The frequency of such interactions is related to the probability per unit length that an existing electron provides enough energy for the production (by direct or indirect ionization) of a new, free electron in the gas. This probability per unit length (that is the first Townsend coefficient, hereafter denoted by $``\alpha"$), is estimated by an exponential fit to the distribution of the photoelectron (longitudinal) drift path length, up to the point of the  ionization  initiating the avalanche, as shown in Fig. \ref{fig:fig5}. Values of the parameter $\alpha$,  estimated with GARFIELD++ simulations, for different Ptr and drift voltage settings, are compiled in Tables \ref{tab:tableA-2} and \ref{tab:tableA-8}
%\footnote{The dependence of $\alpha$ on the drift voltage, for 50\% Penning Transfer rate, follows  the parametric %form: $\alpha=-0.7195\cdot 10^{-1}+0.3318\cdot 10^{-3}\cdot V_{drift}$,}. \\

The ionization probability per unit length depends on the  Ptr value, $r$, as: $\alpha\left( r\right) = \alpha\left( 0\right) + r\cdot \beta$; where $\beta = \alpha\left( 1\right) - \alpha\left( 0\right)$ is the increase of the ionization probability per unit length due to the Penning effect for $r=1$ ($100\%$) transfer rate. Indeed, the values of the first Townsend Coefficient in Table \ref{tab:tableA-2} exhibit such a linear dependence on $r$ and a linear fit results to $\alpha\left( 0\right) = 0.0519 \pm 0.0003 ~\microm^{-1}$ and $\beta =0.0366\pm 0.0007 ~\microm^{-1}$.  

An electron drifting in a noble gas mixture loses energy with probability $\beta$ per unit length, due to the excitation of the noble atoms, independently of the Ptr value. However,  when the first ionization occurs there is a probability $\dfrac{r\cdot \beta}{\alpha\left( 0\right) +r\cdot \beta}$ that the ionization was caused by the Penning effect.\\

Let us consider a photoelectron, before the first ionization, drifting by $\Delta x$ during a time interval $\Delta t$.  On average it undergoes $\left( 1-r\right) \cdot \beta\cdot\Delta x$ inelastic interactions, exciting noble atoms and providing enough energy for indirect ionization but without such an ionization to take place.  If the photoelectron does not lose any energy this way, it would  drift with a velocity, $V_{0}$. However, assuming that the photoelectron gains on average a time, $\tau$, after each of such energy loss, the following relation holds:
\begin{equation}\label{eq:eq1}
\Delta t=\dfrac{\Delta x}{V_{0}}-\left( 1-r\right) \cdot\beta\cdot\tau\cdot\Delta x \,\,\,\, \mbox{, or}\,\,\,\, \dfrac{1}{V_{eff}\left( r\right) }=\dfrac{\Delta t}{\Delta x}=\dfrac{1}{V_{0}}-\left( 1-r\right) \cdot\beta\cdot\tau
\end{equation}
where $V_{eff}(r)$ is the observed, effective drift velocity for Ptr equal to $r$.  Obviously, $V_{0}$ is the effective drift velocity for $r=1$, $V_{0}= V_{eff}(1)$. Eq. \ref{eq:eq1} indicates that by increasing the Ptr value, the effective drift velocity of the photoelectron decreases, in accordance with the GARFIELD++ results. Indeed, Eq. \ref{eq:eq1} fits well the drift velocity values of Table \ref{tab:tableA-1} resulting in an estimate of $V_{0}=142.6 \pm 0.6\,\microm/$ns and   a value for the mean time-gain per interaction of $\tau =17.9\cdot 10^{-3}\pm 1.2\cdot 10^{-3}\,$ns.\\

After the photoelectron has initiated an avalanche, its effective drift velocity is determined by the time-gains every time it looses energy, i.e. either due to excitation of  noble atoms or due to direct ionization. However, the energy loss effect on the drift velocity is independent of whether noble atom excitations result or not in subsequent ionizations via the Penning effect. Consequently, it is expected that the effective drift velocity of an avalanche electron is independent of the Ptr, in  agreement with the GARFIELD++ results, shown in Table \ref{tab:tableA-1}. \\

By definition, a photoelectron undergoes only non-new-electron-producing interactions before it initiates an avalanche. An avalanche electron undergoes the same number of such interactions per unit length but in addition ionizes directly atoms and molecules. Following the argument that more frequent energy losses  result in a larger drift velocity, it is expected that  the avalanche electrons drift faster than the photoelectron before the first ionization, for any Ptr value, in accordance with the GARFIELD++ results shown in Table \ref{tab:tableA-1}. 

The drift velocity of the avalanche as a whole is determined by the combination of the ``time-gain per interaction" and the electron multiplication processes during the avalanche evolution, as described in the following Section. 

\section{Drift and Development of the Pre-Amplification Avalanche} \label{modelaval1}
	Following the model assumption, every time an electron in the avalanche ionizes, it gains a time $\xi_{I}$ relative to an electron that undergoes elastic scattering only. 
	Any new electron produced by ionization starts with low energy.  At the start of its path, it suffers less delay due to elastic back-scattering compared to its parent. 
	Therefore, the model assumes that such a newly produced electron  will gain, relative to its parent, a time-gain $\rho_{I}$. 
	The parameters $\xi_{I}$ and $\rho_{I}$ in principle should follow a joint probability distribution determined by the physical process of ionization and the respective properties of interacting molecules. 
	As discussed in Section \ref{driftvel}, the collective effect of time-gains $\xi_{I}$ is a change in drift velocity from $V_{p}$, which is the photoelectron drift velocity before ionization, to an effective drift velocity $V_{ea}$, which is the drift velocity of an ionizing electron in the avalanche. By taking $V_{ea}$ to be the drift velocity of any electron in the avalanche, the energy-loss effect on the drift of the parent electron has been taken into account. On the other hand, the time gain $\rho_{I}$ of a newly produced electron is assumed to follow a distribution with mean value $\rho$ and variance $w^{2}$. From that moment onwards, this new electron propagates with drift velocity $V_{ea}$, as any other existing electron in the avalanche. Notice that this way, the model approximates the time gains of the parent and daughter electrons as uncorrelated variables. \\

\begin{figure}[h]
\centering\includegraphics[width=1.0\linewidth]{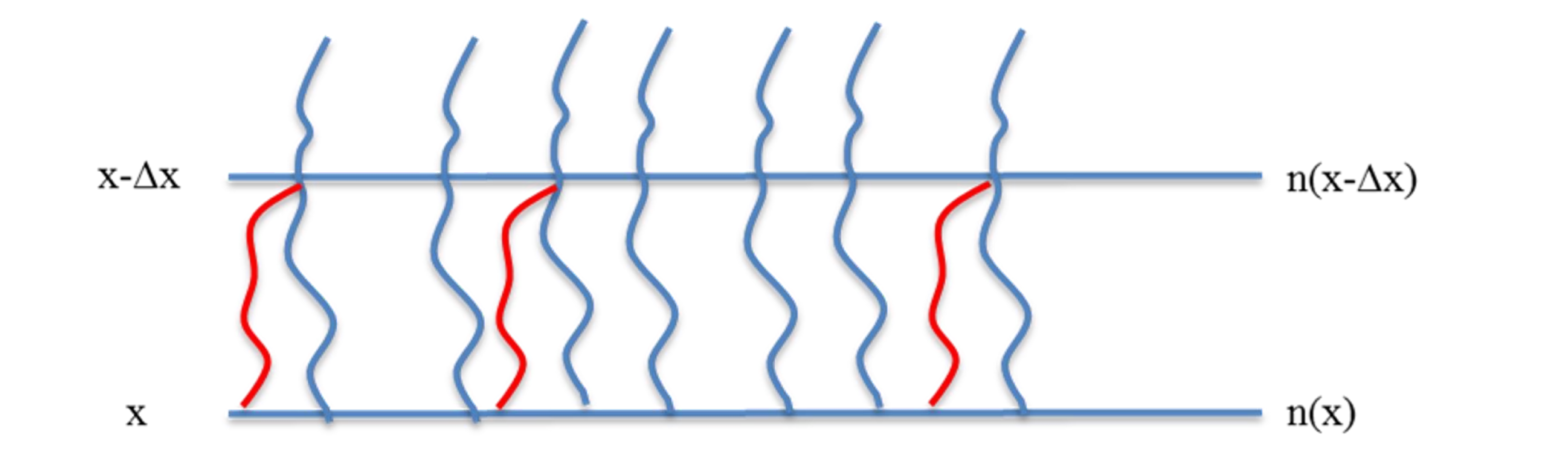}
\caption{Schematic representation of the change in the electron multiplicity in two stages of the avalanche evolution, depicted as a plane at $x - \Delta x$ and a plane  at $x$.}
\label{fig:fig6}
\end{figure}

	Let us consider an avalanche, which has been developed up to a length $x - \Delta x$ and let $n\left( x-\Delta x\right) $ be the number of electrons reaching the plane at $x - \Delta x$. Let $\Delta n$ be the number of electrons produced by ionization in the next development step, of length $\Delta x$. Without loss of generality, the production of the new electrons (shown in red in Fig. \ref{fig:fig6}) is assumed to take place on the plane at $x - \Delta x$.

The average arrival time of the $n\left( x\right) $ electrons at a plane on $x$ is expressed as:
\begin{equation}\label{eq:eq2}
\begin{array}{l}
T_{0}\left( x,n\left( x\right) \right) =\dfrac{1}{n\left( x\right) }\sum\limits_{k=1}^{n\left( x\right) }t_{k}\left( x\right) \\
=\dfrac{1}{n\left( x\right) }\left[ \sum\limits_{k=1}^{n\left( x-\Delta x\right) }\left( t_{k}\left( x-\Delta x\right) +\Delta t_{k}\right) +\sum\limits_{j=1}^{\Delta n} \left( t_{j}^{f}\left( x-\Delta x\right) +\Delta\tau_{j}\right) \right] \\
=\dfrac{1}{n\left( x \right) }\left[ \sum\limits_{k=1}^{n\left( x-\Delta x\right) }t_{k}\left( x-\Delta x\right) +\sum\limits_{j=1}^{\Delta n}t_{j}^{f}\left( x-\Delta x\right) +\sum\limits_{k=1}^{n\left( x-\Delta x\right) }\Delta t_{k} +\sum\limits_{j=1}^{\Delta n}\Delta\tau_{j}\right] 
\end{array}
\end{equation}
where all the times are measured from the instant of the first ionization that initiated the avalanche;  
$t_{k}\left( x\right)$ and $t_{k}\left( x-\Delta x\right)  $  are the times when the $k_{th}$ electron reaches the planes on $x$ and $x-\Delta x$ respectively; 
$t_{j}^{f}\left( x - \Delta x\right) $  is the time that the ``father" of the $j_{th}$ newly produced electron reaches the plane on $x -\Delta x$ (obviously $t_{j}^{f}\left( x - \Delta x\right) $  is one of the $t_{k}\left( x - \Delta x\right) $, ($k=1,2,3,...,n\left( x-\Delta x\right) $); 
$\Delta t_{k}$ is the time spent by the $k_{th}$ electron that reached the plane on $x-\Delta x$ to arrive at the plane on $x$;
$\Delta \tau_{j}$ is the time spent by the $j_{th}$ electron produced at $x-\Delta x$ to arrive at the plane on $x$.\\

 Due to the fact that a newly produced electron gains a certain time, $\rho_{i}$,($i=1,\Delta n$) relative to the parent electron, each $\Delta\tau_{j}$  can be expressed as $\Delta t_{j}^{f}-\rho_{j}$.  
Notice that:
a) since the set $\left\lbrace t_{1}^{f}\left( x - \Delta x\right),  t_{2}^{f}\left( x - \Delta x\right),  t_{3}^{f}\left( x - \Delta x\right),...,  t_{\Delta n}^{f}\left( x - \Delta x\right)\right\rbrace $  can be any size-$\Delta n$ subset of $\left\lbrace t_{1} \left( x - \Delta x\right),  t_{2} \left( x - \Delta x\right),  t_{3} \left( x - \Delta x\right),...,  t_{n\left( x - \Delta x\right) }\left( x - \Delta x\right)\right\rbrace $, any of the $n\left( x -\Delta x\right) $ pre-existing electrons has the same probability, $\Delta n/n\left( x-\Delta x\right) $, to produce a new electron,
 and 
 b) any one of the  $\Delta t_{j}^{f}$, $j=1,2,3,...,\Delta n$ coincides with one of the  $\Delta t_{k}$, $k=1,2,3,\ldots,n\left( x-\Delta x\right)$. 
 Therefore, by averaging Eq. \ref{eq:eq2} for all the possible configurations of $\Delta n$ newly produced electrons, one gets  $T_{1}\left( x,n\left( x\right) \right)  \equiv \langle T_{0}\left( x,n\left( x\right) \right) \rangle_{\Delta n}$, which is:
\begin{equation}\label{eq:eq4}
\begin{array}{l}
  T_{1}\left( x,n\left( x\right) \right)  
=\dfrac{1}{n\left( x - \Delta x\right) } \sum\limits_{k=1}^{n\left( x-\Delta x\right)}t_{k}\left( x-\Delta x\right) +\dfrac{1}{n\left( x-\Delta x\right) }\sum\limits_{k=1}^{n\left( x-\Delta x\right) }\Delta t_{k}-\dfrac{1}{n\left( x\right) }\sum\limits_{j=1}^{\Delta n} \rho_{j}
\end{array}
\end{equation}

Furthermore, averaging Eq. \ref{eq:eq4} over the possible values of $\Delta t_{k}$, the mean time that an avalanche drifts in order to reach a plane on $x$, 
$T\left( x,n\left( x\right) \right) \equiv \langle T_{1}\left( x,n\left( x\right) \right) \rangle_{\Delta t}$ follows the differential relation:
\begin{equation}\label{eq:eq5}
T\left( x,n\left( x\right) \right) =  T\left( x-\Delta x ,n\left( x- \Delta x \right) \right)+\langle \Delta t_{k}\rangle-\dfrac{\Delta n}{n\left( x \right) }\rho
\end{equation}
where $T\left( x-\Delta x ,n\left( x- \Delta x \right) \right)=\dfrac{1}{n\left( x-\Delta x\right) }\sum\limits_{k=1}^{n\left( x-\Delta x\right) } t_{k}\left( x -\Delta x\right) $ and $\rho=\langle\rho\rangle$  is the mean value of the time-gains.

\begin{figure}[h]
\centering\includegraphics[width=0.6\linewidth]{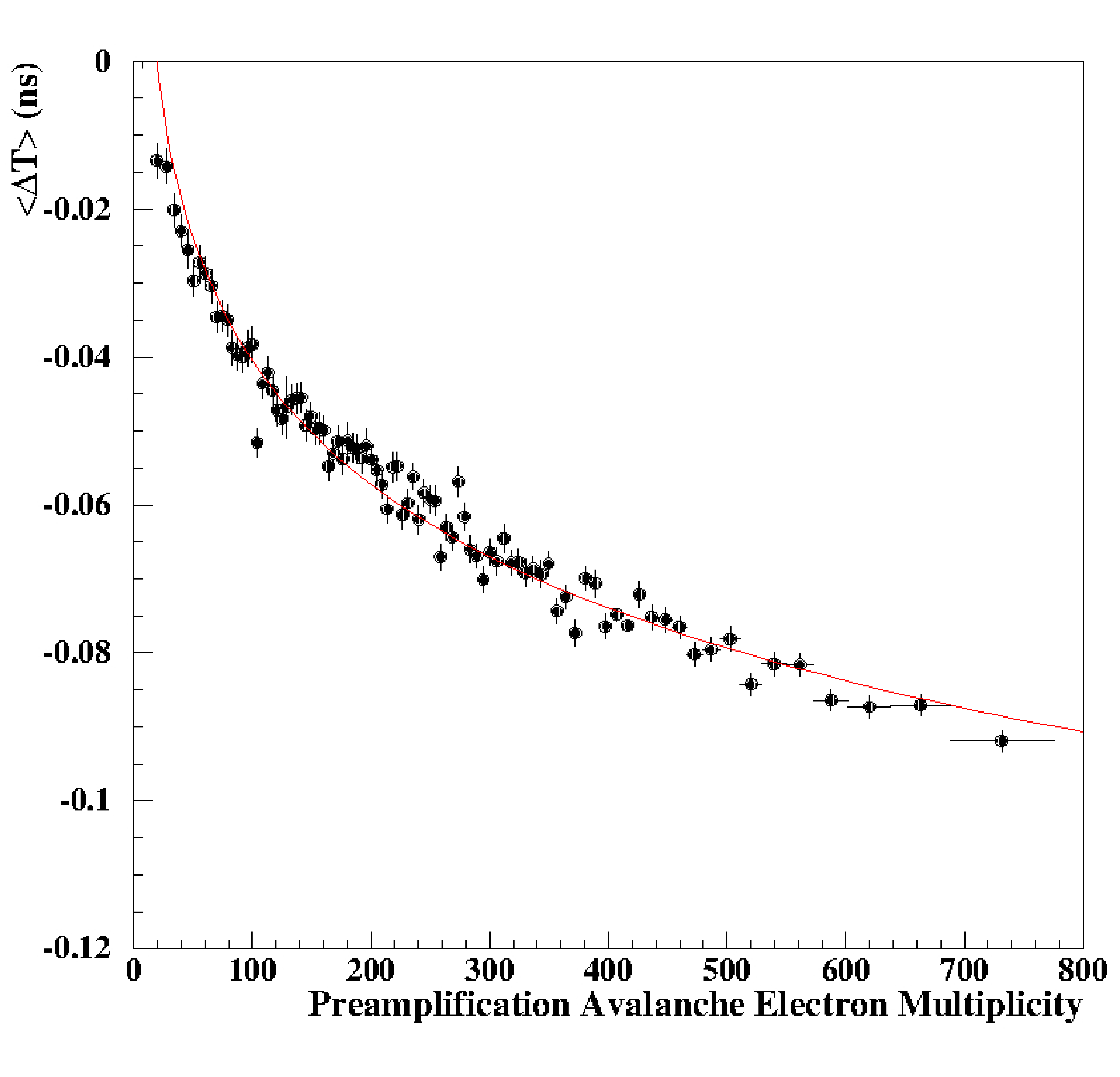}
\caption{Mean deviation ($\langle\Delta T\rangle$) of the  avalanche transmission time from the naively expected time (see text) versus the respective avalanche electron-multiplicity. The points represent results of  GARFIELD++ simulations, assuming 50\% Ptr, anode voltage 450 V and drift voltage 375 V. The line represents a fit using Eq. \ref{eq:eq7}.}
\label{fig:fig7}
\end{figure}

 Finally, using the definition  $V_{ea}= \langle \Delta x /\Delta t_{k}\rangle$, taking the limit for infinitesimal $\Delta x$ and integrating up to an avalanche length L, the following result is obtained:
\begin{equation}\label{eq:eq6}
d T \left( x,n\left( x \right) \right) =\dfrac{dx}{V_{ea}}-\dfrac{dn}{n\left( x\right) }\rho
\Rightarrow
T\left( L,N_{L}\right) =\dfrac{L}{V_{ea}}-\rho\cdot ln\left( N_{L}\right) +C
\end{equation}
where $N_{L}$ is the number of the avalanche electrons reaching a plane on L, and C is an integration constant, which is approximated as independent of L for reasons that will be discussed later in this Section.
Eq. \ref{eq:eq6} predicts that the avalanche transmission time depends linearly on the drift length, L, like it is the case for any individual avalanche electron, but it also depends logarithmically on the electron multiplicity of the avalanche. 
However, the quantity $\Delta T\left( N_{L}\right) =T\left( L,N_{L}\right) - L/V_{ea}$ does not depend explicitly on the avalanche length.  
Consequently, the average residual time $\langle \Delta T \left( N_{L}\right)  \rangle_{L}$, for all avalanches with $N_L$ electrons arriving on the mesh,  depends only on the electron multiplicity, $N_{L}$.
Indeed, symbolizing by $G\left( L\vert N_{L}\right) dL$ the conditional probability of an avalanche with $N_{L}$ electrons reaching the mesh to have a length in the region $\left[ L,L+dL\right] $, the average residual time is:
\begin{equation}\label{eq:eq7}
\begin{array}{l}
\langle \Delta T\left( N_{L}\right) \rangle_{L}=\int\limits_{0}^{\infty}\left[ -\rho \ln \left( N_{L}\right) +C\right] \cdot G\left( L\vert N_{L}\right) dL
=-\rho\ln \left( N_{L}\right) +C
\end{array}
\end{equation}
Eq. \ref{eq:eq7} expresses the mean deviation of the avalanche time from the time expected in case the avalanche speed  is equal to the drift velocity of its constituent electrons.  GARFIELD++ simulations show that this mean time-deviation is described  well, for all considered operating parameters, by the logarithmic expression given in Eq. \ref{eq:eq7}, as shown in Fig. \ref{fig:fig7}.
The mean value of the time-gain $\rho$ and the constant term C, were estimated  by fitting such 
GARFIELD++ simulation results with Eq. \ref{eq:eq7}.
The estimated values of the above parameters are compiled in Tables \ref{tab:tableA-3} and \ref{tab:tableA-8} for a variety of Ptr values and drift voltages, respectively. 

%It is worth noticing that the estimated values of the mean time-gain $\rho$ are, within statistical errors, independent of the Penning Transfer Rate. This is in accordance to the basic ideas of the model. Indeed, a new electron in the avalanche should gain on average the same mean-time relatively to the existing electrons, independently of the production mechanism (direct ionization or ionization through the Penning effect). 
The newly produced electrons  would gain in average the same time, at the beginning of their path, independently of their production mechanism, i.e. via direct ionization or Penning transfer. Consequently the estimated values of the parameter  $\rho$ should be independent of the Ptr value, as indeed it was found by fitting GARFIELD++ simulation results (see Table \ref{tab:tableA-3}).  
 Moreover, as the newly produced electrons accelerate and reach equilibrium faster at higher rather than at lower drift fields, it is expected that the value of the time-gain parameter, $\rho$, should decrease as the drift voltage increases, in agreement with the estimated values shown in Table \ref{tab:tableA-8}.
%\footnote{ The dependence of the $\rho$ estimated values of the parameter $\rho$  on the drift voltage is described \%well, by the following  empirical parametrization: $\rho(V_{drift})=-0.37038+23.317\cdot V^{-1}_{drift}$, where $\rho$ %is given in ns and the drift voltage is measured in Volts.} . 

Eq. \ref{eq:eq6} has been derived by treating  the  simultaneous drift and growth of the avalanche  differentially. 
%Such an approach approximates successfully the avalanche development only after it has reached an electron multiplicity that allows a differential description.  
Thus, the integration constant, C, depends on a minimum avalanche length, after which the growth of the mean avalanche electron multiplicity  allows for a differential treatment. 
Such a minimum avalanche length depends on the avalanche electron multiplication that, in turn, depends on the Ptr and the drift voltage, as it can be seen in Tables \ref{tab:tableA-3} and \ref{tab:tableA-8}.\\

\begin{figure}[h]
\centering
\begin{minipage}{.52\textwidth}
\centering
\includegraphics[width=.85\textwidth]{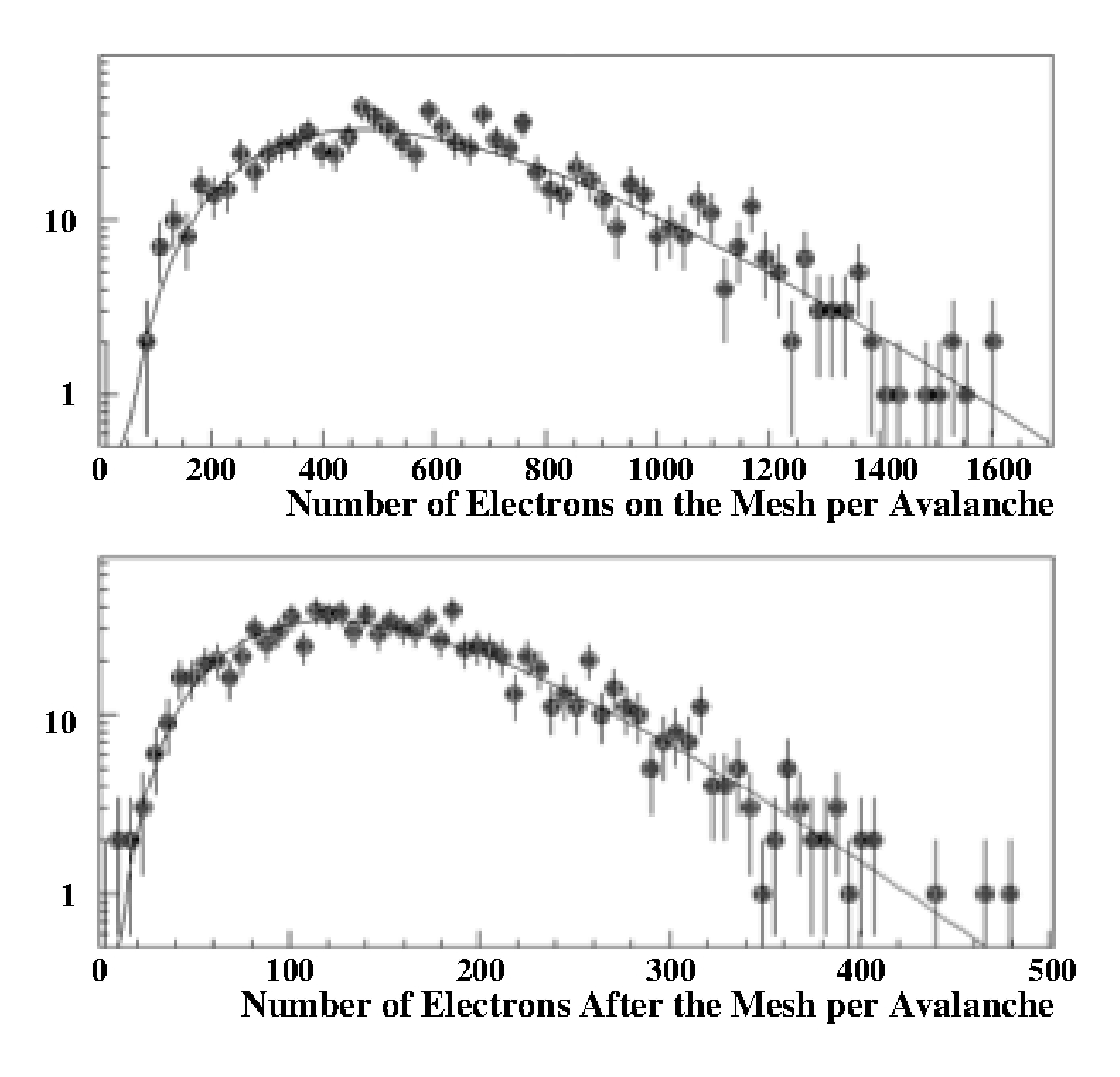}
\end{minipage}%
\begin{minipage}{.52\textwidth}
\centering
\includegraphics[width=.85\textwidth]{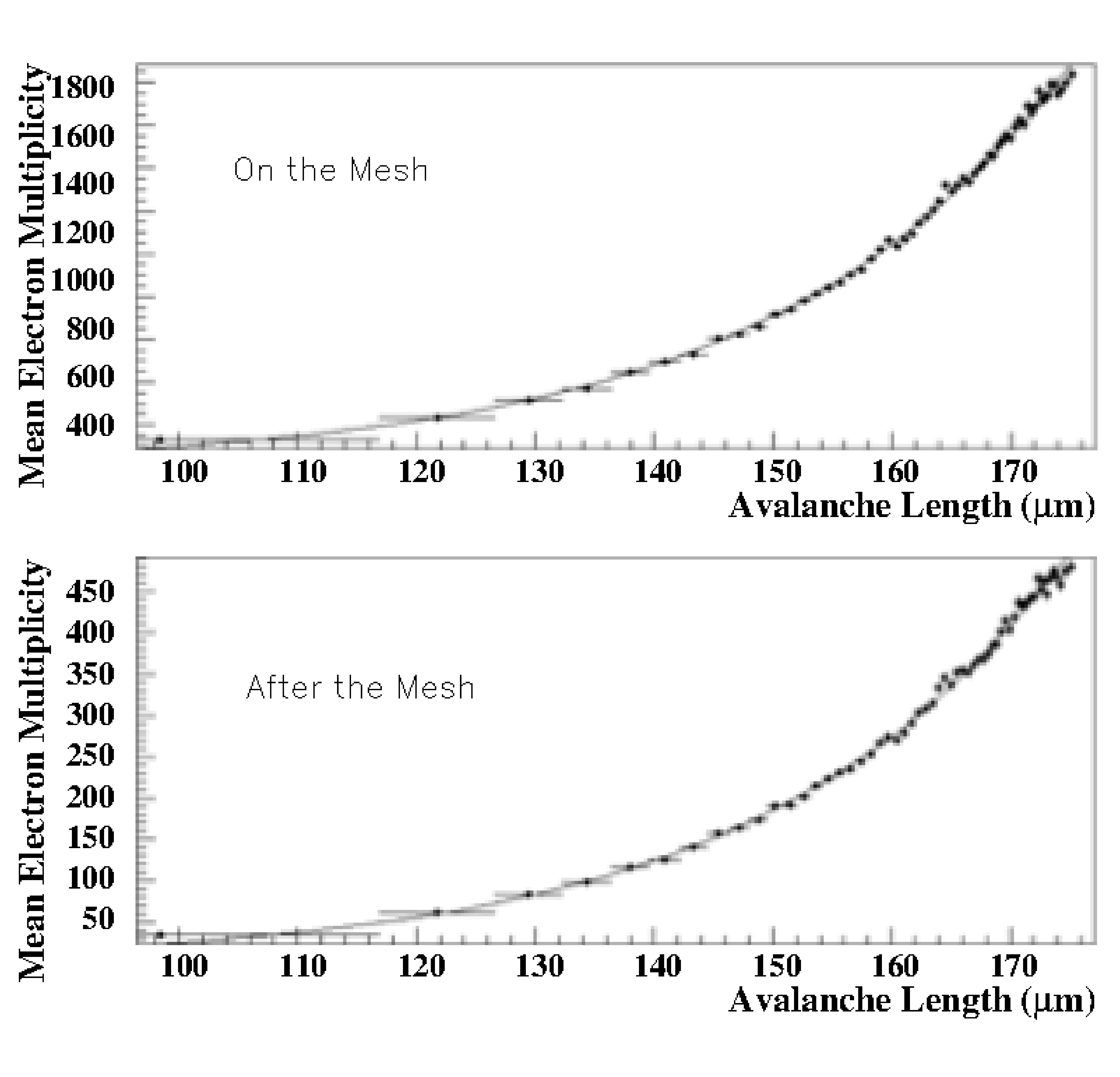}
\end{minipage}
\caption { The points represent GARFIELD++ simulation results. (top-left) Distribution of the number of electrons arriving on the mesh, produced in avalanches with length between 144.45 and 144.75 $\microm$. The solid line represents  a Gamma distribution function fitted to the simulation results. (top-right) The mean value of the avalanche electron multiplicity on the mesh versus the length of the respective avalanche. The solid line represents exponential fit to the simulation results, as described in the text. For completeness, GARFIELD++ simulation results, related to the electron multiplicity after the mesh, are also presented in the bottom-row plots. 
    }  
\label{fig:fig8}
\end{figure}

The avalanche drift velocity is determined by expressing the mean avalanche transmission time, $\langle T\left( L\right) \rangle$,  as a function of the avalanche length, L, i.e. by averaging  Eq. \ref{eq:eq6} over all possible values of the avalanche electron multiplicity,
\begin{equation}\label{eq:eq6_a}
\begin{array}{l}
\langle T\left( L\right) \rangle=$ $\int\limits_{0}^{L}T\left( L,N_{L}\right)\cdot \Pi\left( N_{L}\vert L\right) dN_{L}
\end{array}
\end{equation}

where $\Pi\left( N_{L}\vert L\right)$ denotes the conditional probability density function (p.d.f.)  of the number of electrons, $N_{L}$,  produced in an avalanche, given the length of the avalanche, L.\\

  As deduced from GARFIELD++, $\Pi\left( N_{L}\vert L\right)$ is well approximated by the Gamma distribution function, $P\left( N_{L}; q(L),\theta\right)$, with $q(L)$ being the mean value, and $\theta$  the shape parameter. This successful approximation is demonstrated in the top-left plot of Fig. \ref{fig:fig8}. \\
  While the shape parameter is found to be independent of the avalanche length, the mean value depends exponentially on the the length of the avalanche, i.e. $q\left( L;a_{eff}\right) =2\cdot e^{a_{eff}L}$, as shown in the right plot of Fig. \ref{fig:fig8}.  The exponential slope $a_{eff}$ (hereafter called ``multiplication factor") is the probability per unit length for the net production of a new electron. Estimated values of  $a_{eff}$ and $\theta$, using GARFIELD++ simulations with different values of  Ptr and  drift voltage,  are compiled in Tables \ref{tab:tableA-4}, \ref{tab:tableA-5} and \ref{tab:tableA-8}.
%\footnote{As expected, both the $a_{eff}$ and $\theta$ take larger values as the drift voltage increases. However, the relative spread, i.e. the RMS over the mean value of the Gamma distribution ($=(1+\theta)^{-1/2}$), decreases as a function of the drift field.}.
\\ 

It should also be  noticed that, as the left-bottom plot of Fig. \ref{fig:fig8} demonstrates,  the electron multiplicity after the mesh  also follows a Gamma distribution function with the same $\theta$ value as the corresponding distribution of the electron multiplicity on the mesh (see also Table \ref{tab:tableA-5}). The mean electron multiplicity after the mesh depends exponentially on the avalanche length, as it is shown in the bottom-right plot of Fig. \ref{fig:fig8}. Moreover,  the exponential slope is found to be equal to the multiplication factor, $a_{eff}$, which implies that the mesh transparency is independent of the avalanche length. Furthermore, as it is deduced from GARFIELD++ simulations for all considered operating conditions,  the mean electron multipliplicity after the mesh is consistently 25\% of the number of the avalanche electrons arriving on the mesh, (see Tables \ref{tab:tableA-4} and \ref{tab:tableA-8}). Taking into account that the PICOSEC e-peak signal size was found (see Fig. \ref{fig:fig3}) to depend linearly on the electron multiplicity after the mesh, the constant mesh transparency also implies that the observed signal size is determined by the electron multiplicity on the mesh. \\

Having expressed the term $\Pi\left( N_{L}\vert L\right)$ of Eq. \ref{eq:eq6_a} as a Gamma distribution function, $P\left( N_{L};q(L)=2e^{a_{eff}L},\theta\right)$, and substituting $T\left( L,N_{L}\right)$ from Eq. \ref{eq:eq6}, the average time taken by an avalanche to drift along a length L, for any avalanche electron multiplicity, $N_{L}$, is written as:
\begin{equation}\label{eq:eq11}
\langle T\left( L\right) \rangle = 
\dfrac{L}{V_{ea}}-\rho\cdot\int\limits_{0}^{L}\ln\left( N_{L}\right) P\left( N_{L};q(L)=2e^{a_{eff}L},\theta\right) dN_{L}+C
\end{equation}
Using the properties of the Gamma distribution function,  Eq. \ref{eq:eq11} becomes:
\begin{equation}\label{eq:eq13}
\begin{array}{l}
\langle T\left( L\right) \rangle = 
L\left[ \dfrac{1}{V_{ea}} -\rho \cdot a_{eff}\right] +\left[ -\rho\ln 2+C+\rho\ln\left( \theta+1\right) -\rho\psi\left( \theta+1\right) \right] 
\end{array}
\end{equation}
where $\psi\left(x\right)$ denotes the digamma function. \\

\begin{figure}[h]
\centering\includegraphics[width=0.6\linewidth]{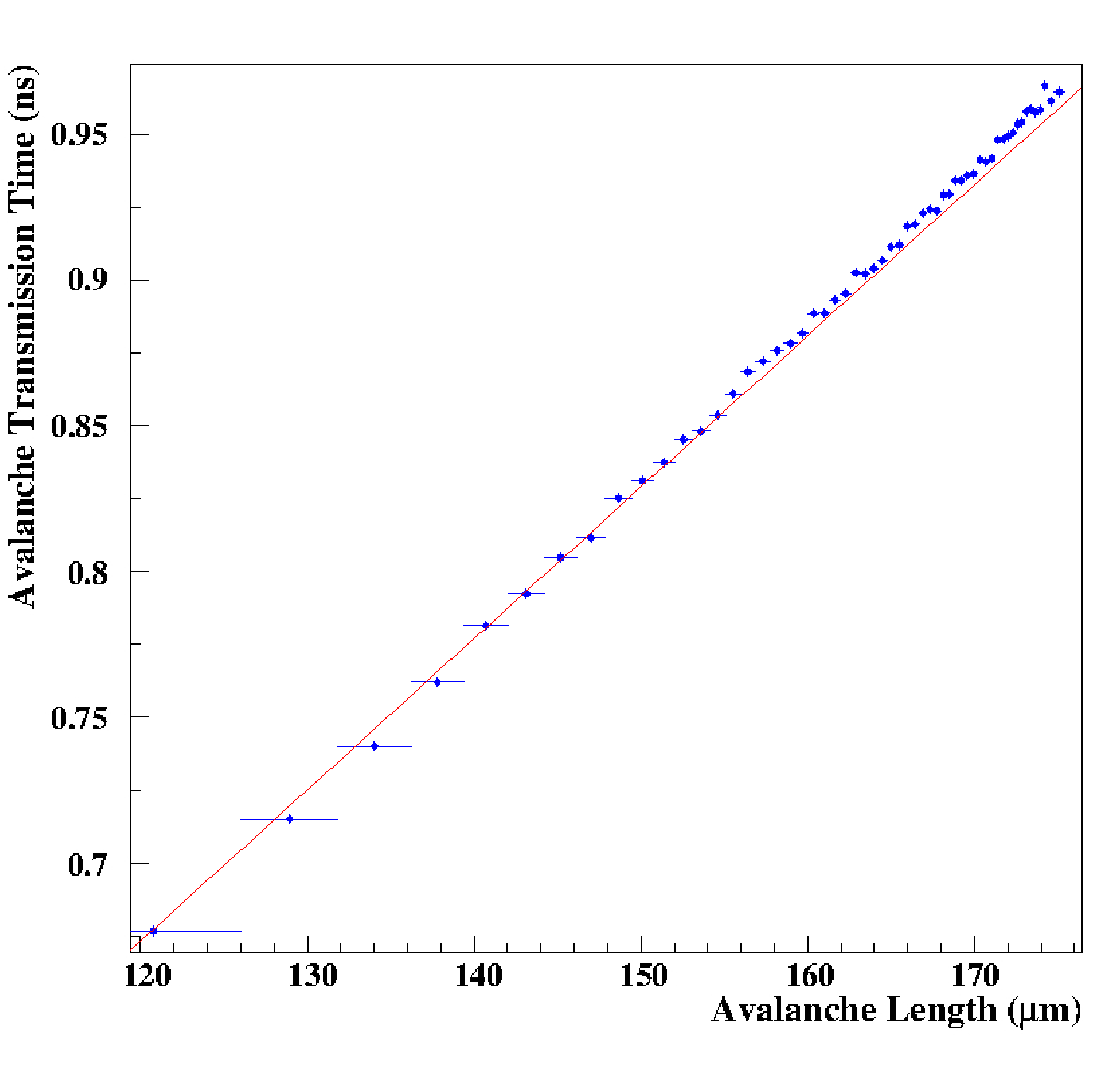}
\caption{ The average time needed by an avalanche, of a certain length, to arrive on the mesh (the avalanche transmission time) as a function of the length of the avalanche.  The points are GARFIELD++ simulation results  for 50\% Ptr and  a drift voltage of 425 V. 
 The solid line represents the model prediction, expressed by  Eq. \ref{eq:eq13}. }
\label{fig:fig9}
\end{figure}

Eq. \ref{eq:eq13} relates linearly the mean value of the avalanche transmission time to the  avalanche length. As it is easily verified by using numerical values for the model parameters ($\rho$, $\theta$, $a_{eff}$, $V_{ea}$ and $C$) from \ref{Appendix A}, the constant term, $\left[ -\rho\ln 2+C+\rho\ln\left( \theta+1\right) -\rho\psi\left( \theta+1\right) \right]$, takes very small values for all considered drift voltages and Ptr values. Therefore, the effective avalanche drift velocity is determined by 
the inverse of the term $\left[ \dfrac{1}{V_{ea}} -\rho \cdot a_{eff}\right]$. 
Since both $\rho$ and $a_{eff}$ are positive-value parameters, the model predicts that the avalanche, as a whole, drifts with higher velocity than the velocity $V_{ea}$ of its constituent electrons, as it was also found using GARFIELD++ simulations.
Furthermore, the GARFIELD++ simulation results are found to be in a  good quantitative agreement with the model predictions expressed by Eq. \ref{eq:eq13}, as demonstrated in Fig. \ref{fig:fig9}. The same agreement holds for all  considered operating conditions.

%Since the drift velocity $V_{ea}$ of the individual avalanche electrons is independent of the Penning Transfer Rate, Eq. \ref{eq:eq13} predicts that the effective avalanche drift velocity should increase with the Penning Transfer Rate. Indeed, the model requires that the time-gain, $\rho$, should be independent of the Penning Transfer Rate. On the other hand, the multiplication coefficient, $a_{eff}$, increases with the Penning Transfer Rate, as shown in Table \ref{tab:tableA-4}, causing the increase of the product $\rho\cdot a_{eff}$. Consequently, the slope term in Eq. \ref{eq:eq13} decreases and the effective avalanche drift velocity increases with the Penning Transfer Rate, in agreement with the GARFIELD++ results. 

\section{Transmission Times vs the Avalanche Electron Multiplicity} \label{modelaval2}

In Section \ref{intro} it was shown that the total time after the mesh determines the PICOSEC signal arrival time (SAT). Nevertheless, as it will be  discussed in detail in Section \ref{modelaval5}, the total time after the mesh differs from the respective total time on the mesh  by a constant  interval, which is independent of  electron multiplicities and drift lengths.  Moreover, in Section \ref{modelaval1} it was shown that the mean electron multiplicity after the mesh, which determines the signal size, is a constant fraction (25\%) of the electron multiplicity on the mesh. Thus, the expression of the mean  total time  as a function of the electron multiplicity on the mesh, by  properly integrating Eq. \ref{eq:eq6}, will provide the microscopic description of the   PICOSEC SAT dependence on the signal size (shown in Fig. \ref{fig:fig2}).

By employing Bayes' theorem, the conditional p.d.f., $G\left( L\vert N\right)$, that an avalanche with $N$ electrons reaching the mesh has a length in the region $\left[ L,L+dL\right] $, is expressed as: 
\begin{equation}\label{eq:eq14}
G\left( L\vert N\right) =\dfrac{p\left( N\vert L\right) R\left( L\right) }{p\left( N\right) }
\end{equation}
Here $R\left( L\right)$  is the p.d.f. of any avalanche to have a length L; $p\left( N\vert L\right)$  is the conditional p.d.f. that an avalanche produced $N$ electrons reaching the mesh, given that its length equals L. 
The normalizing term $p\left( N\right) $, defined as $p\left( N\right) = \int\limits_{x_{1}}^{x_{2}} p\left( N\vert L\right) R\left( L\right) dL$, expresses the p.d.f. that an avalanche has $N$ electrons reaching the mesh and any length within the region $x_{1}\leq L \leq x_{2}$ \footnote{The lower integration limit is $x_{1}=0$. However, as the GARFIELD++ simulations indicate, the maximum avalanche length, $x_{2}$, does not reach  the full depth of the pre-amplification region, D, because the initial photoelectron takes a minimum distance before it gains enough energy to start an avalanche. Naturally, this limit depends on the drift voltage, as shown in Table \ref{tab:tableA-8}.}.

In this model, $p\left( N\vert L\right)$  is approximated by the the Gamma distribution function $P\left( N;q=2e^{a_{eff}L},\theta\right)$, as discussed in Section \ref{modelaval1}.
% which has been found to describe very well the PICOSEC calibration data and GARFIELD++ simulations. 
$R\left( L\right) $  is expressed in terms of the first Townsend coefficient, $a$, as:
\begin{equation}\label{eq:eq15}
R\left( L\right) =R\left( L; a\right) =a\cdot\dfrac{\exp\left[ a\cdot L\right] }{\exp\left[ a\cdot x_{2}\right] -\exp\left[ a\cdot x_{1}\right]} 
\end{equation}
Then,  the conditional p.d.f. $G\left( L\vert N\right) $  takes the form: 
\begin{equation}\label{eq:eq16}
G\left( L\vert N\right) =\dfrac{P\left( N;q=2e^{a_{eff}L},\theta\right) R\left( L; a\right) }{\int\limits_{x_{1}}^{x_{2}}P\left( N;q=2e^{a_{eff}L},\theta\right) R\left( L; a\right)dL}
\end{equation}
Using Eq. \ref{eq:eq6}, the average transmission time, $\langle T\left( N\right) \rangle=\int\limits_{x_{1}}^{x_{2}} T\left( N,L\right)  G\left( L\vert N\right) dL$
%$\langle T\left( N\right) \rangle$, of an avalanche of any length but with N electrons arriving on the mesh, 
is written as follows:
\begin{equation}\label{eq:eq17}
\begin{array}{l}
\langle T\left( N\right) \rangle=\dfrac{\langle L\left( N\right) \rangle}{V_{ea}}-\rho\ln N+C
\end{array}
\end{equation}
where $\langle L\left( N\right) \rangle=\int\limits_{x_{1}}^{x_{2}}L\cdot G\left( L\vert N\right) dL$  is the average length of all avalanches resulting to N electrons on the mesh.\\ 

As discussed in Section \ref{driftvel}, the mean transmission time of the photoelectron before it ionizes, depends linearly on its  drift path, D-L, as:
\begin{equation}\label{eq:eq18}
T_{p}\left( L\right) =\dfrac{D-L}{V_{p}}+d_{off}
\end{equation}
where   the constant term, $d_{off}$, is attributed to the fact that the drift velocity is a statistical variable, which characterizes the drift of an electron after it has undergone enough scatterings\footnote{That is after an initial stage of 3.6 - 4.7 $\microm$ along the drift field, as indicated by the GARFIELD++ simulations for the drift voltages considered in this work.} in order to be described statistically. 
The mean transmission time, from the emission up to  the first ionization, of a photoelectron that initiates an avalanche with N electrons on the mesh, is given as:
\begin{equation}\label{eq:eq19}
\langle T_{p}\left( N\right) \rangle=\int\limits_{x_{1}}^{x_{2}} T_{\rho}\left( L\right) G\left( L\vert N\right)dL=\dfrac{D-\langle L\left(N\right) \rangle }{V_{p}}+d_{off} 
\end{equation}
%Due to the averaging in Eq. \ref{eq:eq19}, the mean transmission time of photoelectrons, which initiate avalanches with a certain number of electrons on the mesh, depends on this electron multiplicity. 
		
	The total time on the mesh, $\langle T_{tot}\left( N\right) \rangle$, is the sum of the two terms given by Eq. \ref{eq:eq17} and \ref{eq:eq19}:
\begin{equation}\label{eq:eq20}
\langle T_{tot}\left( N\right) \rangle=\langle T_{p}\left( N\right) \rangle+\langle T\left( N\right) \rangle=\langle L\left( N\right) \rangle\left[ \dfrac{1}{V_{ea}}-\dfrac{1}{V_{p}}\right] -\rho\ln N+\left[ \dfrac{D}{V_{p}}+C+d_{off}\right] 
\end{equation}

%All the parameters in Eqs. \ref{eq:eq17}, \ref{eq:eq19} and \ref{eq:eq20} have been estimated from GARFIELD++ simulations  (see Appendix A) and the integral $\int\limits_{x_{1}}^{x_{2}}L\cdot G\left( L\vert N\right) dL$, which defines the mean value $\langle L\left( N\right) \rangle$ is easily evaluated numerically for any value of N.

\begin{figure}[h]
\centering\includegraphics[width=0.6\linewidth]{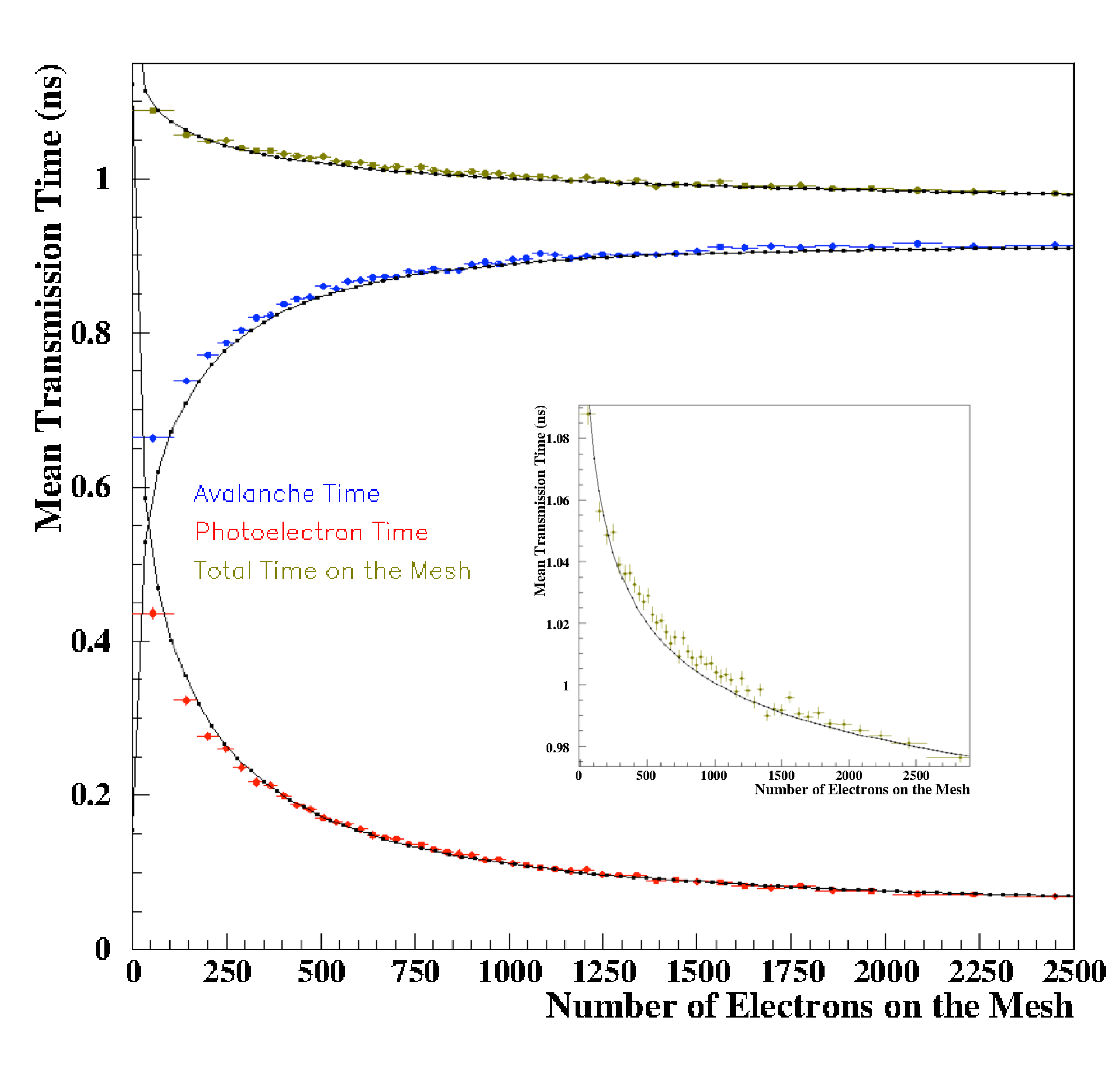}
\caption{The points represent GARFIELD++ simulation results related to the  mean transmission times versus the respective multiplicity of the avalanche electrons arriving on the mesh, for 50\% Ptr; 425 V and 450 V drift and anode voltages respectively: (red) the transmission time of the photoelectron before the first ionization,  (blue) the transmission time of the avalanche from its beginning until the mesh and (golden) the transmission time of the whole process,  from the photoelectron emission until the avalanche reaches the mesh. The solid lines represent the predictions of Eq. \ref{eq:eq17}, \ref{eq:eq19}, \ref{eq:eq20} respectively. The inset plot details the dependence of the total time on the mesh on the number of electrons arriving on the mesh.}
\label{fig:fig10}
\end{figure}
The third term in the right hand side of Eq. \ref{eq:eq20} represents the total time on the mesh in absence of any ``time gain'' caused by interactions. In such a case the SAT should be constant ($\simeq~D/V_{p}$), determined only by the photoelectron drift velocity ($V_{p}$) and independent of the signal size. However, due to  time gains because of  inelastic interactions, the avalanche electrons drift faster than the photoelectron. Thus, the first term represents the total time gain by a collection of electrons drifting with $V_{ea}$ relative to a photoelectron drifting the same distance. Finally, the second term represents an extra time gain, due to the fact that each newly produced electron in the avalanche gains on average a time $\rho$ relative to its parent. Taking also into account that the average avalanche length is a positive, increasing function of N, both the above time gain terms increase in absolute value as N increases. Equivalently, Eq. \ref{eq:eq20} predicts that, due to the time gain concepts employed by our model, large size PICOSEC signals should arrive earlier than smaller pulses in accordance with the experimental observations and the GARFIELD++ simulation results.

Furthermore, the model  predicts, as demonstrated in Fig. \ref{fig:fig10}, that the photoelectron (Eq. \ref{eq:eq19}), the avalanche (Eq. \ref{eq:eq17}), and the total (Eq. \ref{eq:eq20}) transmission times and their dependence on the  electron multiplicity on the mesh are in good agreement with the GARFIELD++ simulation results. Moreover, setting appropriate values to the model-parameters, e.g. from Table \ref{tab:tableA-8}, the model successfully reproduces  the respective GARFIELD++ results for all the considered PICOSEC operating conditions.

\section{Timing Resolution as a function of the Avalanche Length.} \label{modelaval3}

	As it was shown in Fig. \ref{fig:fig3}, the PICOSEC timing resolution is determined by the spread of the total-time after the mesh. However, the processes occurring  in the pre-amplification region influence the statistical SAT fluctuations in a much stronger way than the passage of the pre-amplification electrons through the mesh, as shown in Section \ref{modelaval5}. This Section focuses on describing stochastically  the  spread of the  total-time on the mesh as a function of the avalanche length. The longitudinal diffusion of the primary photoelectron and  the spread of the avalanche transmission time are the sources of this spread.
	The latter  emerges as the combination of: a) the individual avalanche electrons diffusion, b) the electron multiplicity increase as the avalanche grows, and  c) the statistical correlation between the drift times of the individual electrons. Also notice that the avalanche length (L or its residual D-L) is the natural parameter to express the photoelectron diffusion, as well as  the avalanche growth and the correlation between its electrons.\\
	
\begin{figure}[h]
\centering
\begin{minipage}{.49\textwidth}
\centering
\includegraphics[width=.85\textwidth]{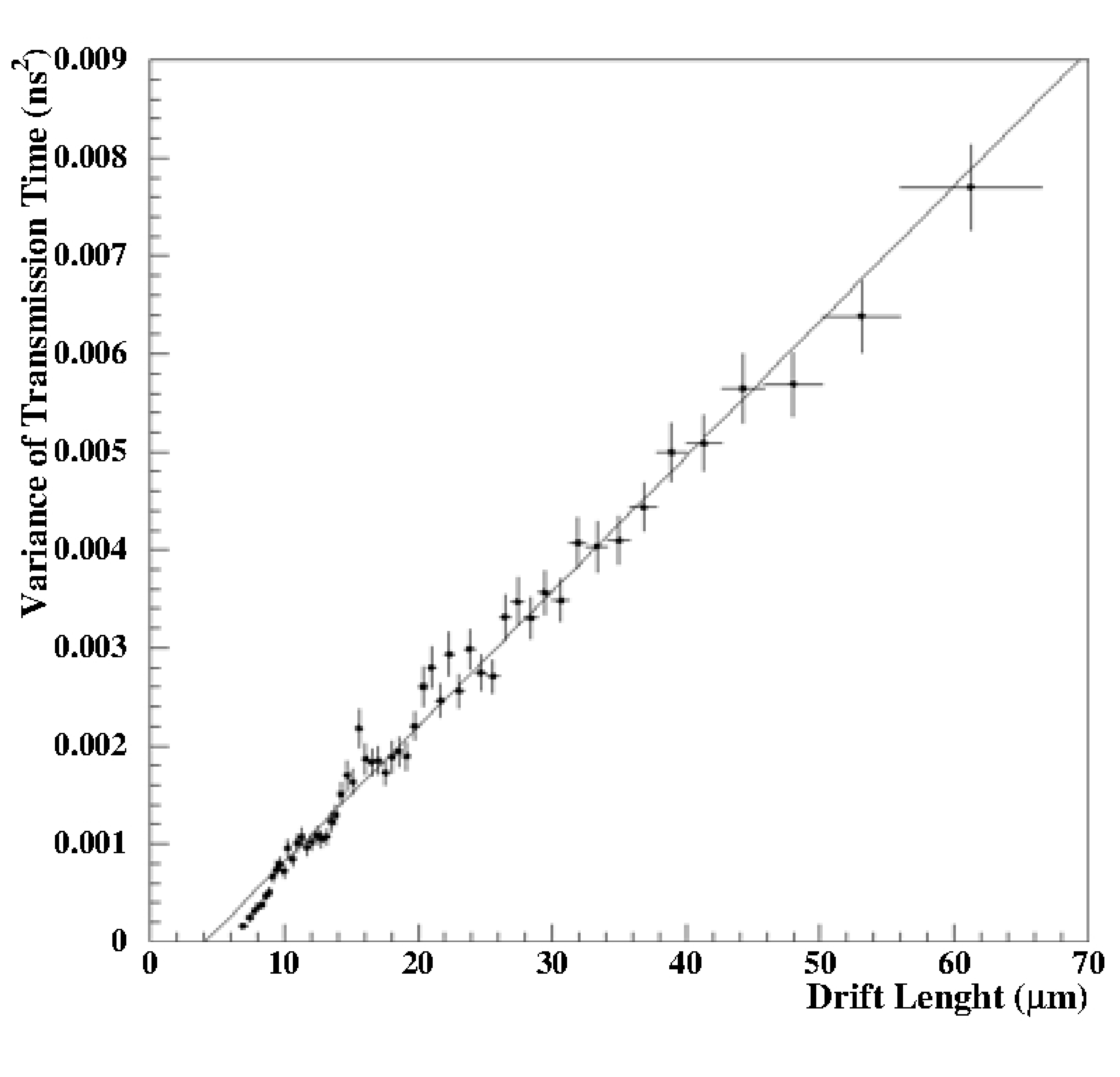}
\end{minipage}
\begin{minipage}{.49\textwidth}
\centering
\includegraphics[width=.85\textwidth]{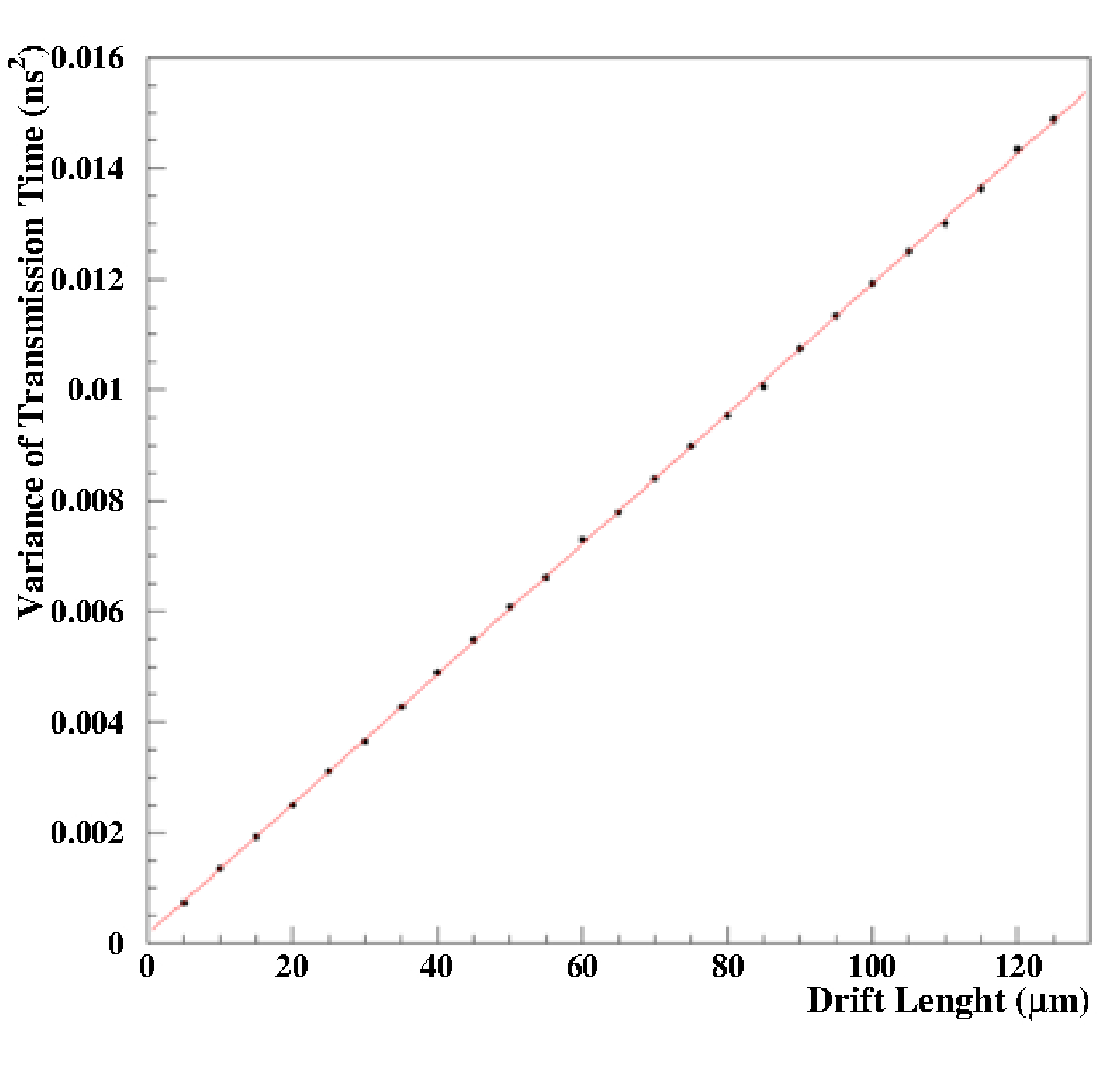}
\end{minipage}
\caption { The points represent GARFIELD++ simulation results. (left) The  variance of the photoelectron transmission time at the point of the first ionization versus the respective drift length. (right) The variance of the time taken by an  avalanche electron to drift a certain length versus the respective length. The solid curves  represent  linear fits to the points.
    }  
\label{fig:fig11}
\end{figure}
In GARFIELD++ simulations the variance of the photoelectron transmission time $V\left[ T_{p}\left( L\right)\right] $, and the variance of the drift time of an avalanche electron $V\left[ T_{ea}\left( x\right)\right] $ , depend linearly on the respective  drift lengths:  
\begin{equation}\label{eq:eq21}
V\left[ T_{p}\left( L\right) \right] =\left( D-L\right) \cdot \sigma_{\rho}^{2}+\Phi
\end{equation}
\begin{equation}\label{eq:eq22}
V\left[ T_{ea}\left( x\right) \right] =\sigma_{0}^{2}\cdot x+\phi
\end{equation}
The slopes  ($\sigma_{p}^{2}$, $\sigma_{0}^{2}$)  and  the constant terms ($\Phi$, $\phi$) in the above relations are  evaluated by linear fits  to   GARFIELD++ simulation results\footnote{The ``simulation results'' are the variances of the respective time distributions, estimated by fits with a Wald function, as described in Section \ref{driftvel}, Fig. \ref{fig:fig4}}. 
	Estimated values of these parameters, for all considered PICOSEC operating conditions,  are compiled in Tables \ref{tab:tableA-6}, \ref{tab:tableA-7} and \ref{tab:tableA-8}. \\
	In all above estimations, the  variable $\Phi$ acquires negative values. This is due to the fact that the photoelectron motion at its initial part has not yet reached statistical equilibrium, as it is apparent in the left plot of Fig. \ref{fig:fig11}.  
	On the other hand, only positive values were found for $\phi$, as it is demonstrated with the right plot of Fig. \ref{fig:fig11}.  A positive $\phi$ value  implies that an avalanche electron inherits time spread before it starts drifting which is, however, consistent with the phenomenological model advocated in this study. 
	Indeed, all the terms expressing time-gains  in this model are random variables, with 
variances  contributing to the variance of the respective drift times. 
	Thus, the constant term $\phi$ corresponds to the variance of the time gained by the first avalanche electron when it initiates the avalanche.
	 Nevertheless, the contribution of the constant term, $\phi$, in Eq. \ref{eq:eq22} is much smaller than the  part which is proportional to the drift length\footnote{According to GARFIELD++ simulations, at all  voltage settings considered in this study,  the vast  majority of the avalanches have lengths greater than $100\,\microm$, even in the case of 0\% Ptr. For a $100\,\microm$ long avalanche, the time variance of an avalanche electron that arrives on the mesh, is more than 70 times larger than the contribution of the constant term $\phi$.} and it will be ignored in the following.

	For an avalanche of length L, initiated by a photoelectron after  drifting  a length $D-L$, the  avalanche time $T\left( L\right)$ and the photoelectron time $T_{p}\left( L\right) $ are statistically, mutually  uncorrelated. 
	Therefore, the total time on the mesh, $T_{tot}\left( L\right) $, and its variance, $V\left[ T_{tot}\left( L\right) \right] $, are expressed as:
\begin{equation}\label{eq:eq23}
\begin{array}{l}
T_{tot}\left( L\right) =T_{p}\left( L\right) +T\left(L\right) \\
V\left[ T_{tot}\left( L\right) \right] =V\left[ T_{p}\left( L\right) \right] +V\left[ T\left( L\right) \right]   
\end{array}
\end{equation}
where  $V\left[ T_{p}\left( L\right) \right] $ is given by Eq. \ref{eq:eq21}.\\ 

	The term  $V\left[ T\left( L\right) \right] $ will be evaluated by considering the evolution of the avalanche between two planes, on $x-\Delta x$ and on $x$, as presented in Section 3 and depicted in Fig. \ref{fig:fig6}. 
	The average of the electron arrival times at a plane on $x$, expressed by Eq. \ref{eq:eq2}, is factorized as the sum of five terms (A, B, C, D and E), as follows: 
\begin{equation}\label{eq:eq24}
\begin{array}{l}
T_{0}\left( x,n\left( x\right) \right) = \\
\dfrac{1}{n\left( x\right) }\left[  \underbrace{\sum\limits_{k=1}^{n\left( x-\Delta x\right)}t_{k}\left( x-\Delta x\right)  }_{A} +\underbrace{\sum\limits_{j=1}^{\Delta n}t_{j}^{f}\left( x-\Delta x\right)  }_{B}+\underbrace{\sum\limits_{k=1}^{n\left( x-\Delta x\right)}\Delta t_{k}}_{C}+\underbrace{\sum\limits_{j=1}^{\Delta n}\Delta t_{j}^{f}}_{D}+\underbrace{\sum\limits_{j=1}^{\Delta n}\rho_{j}}_{E} \right]
\end{array}
\end{equation}
	As in Section 3, the model treats the times $\Delta t_{k}\,\left( k=1,2,3,...,n\left( x-\Delta x\right) \right)$ as mutually uncorrelated and independent of the  history of pre-existing electrons. 
%	No distinction is made between ionizing and non-ionizing pre-existing electrons. 
Recall that the times $\Delta \tau$, taken by the newly produced electrons to drift between the planes on $x-\Delta x$ and $x$, is the difference of two random variables: $\Delta \tau_{j} = \Delta t_{j}^{f} - \rho_{j} \left( j=1,2,...,\Delta n\right)$. 
The first variable $\Delta t_{j}^{f}$ has the same statistical properties as the times $\Delta t_{k}$ %\left( k=1,2,3,...,n\left( x-\Delta x\right) \right)$ 
of the pre-existing electrons. 
	The time-gains acquired by the new electrons $\rho_{j} \left( j=1,...,\Delta n\right)$ are mutually uncorrelated, and they are also uncorrelated with anyone of the $\Delta t_{k}$’s. 
	
As in Section 3, the model assigns a probability $\Delta n/n\left( x-\Delta x\right)$ to each of the pre-existing electrons at the plane on $x-\Delta x$ to ionize and produce a new electron. Under these assumption, the terms B and D in Eq. \ref{eq:eq24}, when averaged for all possible configurations of $\Delta n$ newly produced electrons, are transformed to:
\begin{equation}\label{eq:eq25}
\begin{array}{l}
B_{1}=\langle \sum\limits_{j=1}^{\Delta n}t_{j}^{f}\left( x-\Delta x\right)\rangle_{\Delta n} = \dfrac{\Delta n}{n\left( x-\Delta x\right) }\sum\limits_{k=1}^{n\left( x-\Delta x\right) }t_{k}\left( x-\Delta x\right) \\
D_{1}= \langle \sum\limits_{j=1}^{\Delta n}\Delta t_{j}^{f} \rangle_{\Delta n} = \dfrac{\Delta n}{n\left( x-\Delta x\right) }\sum\limits_{k=1}^{n\left( x-\Delta x\right) }\Delta t_{k}
\end{array}
\end{equation}

Considering the aforementioned correlation relations between the individual drift times and time gains, the covariances $\text{cov}[A,B_1]$ and  $\text{cov}[C,D_1]$ are non-zero, while all the other term combinations have zero covariances. 
Consequently, the variance  of $T_{1} \left( x,n\left( x\right) \right)=\langle T_{0} \left( x,n\left( x\right) \right) \rangle_{\Delta n} $ is expressed as:\\

\begin{equation}\label{eq:eq26}
\begin{array}{l}
V\left[ T_{1}\left( x,n\left( x\right) \right) \right] = \\
\dfrac{1}{n^{2}\left( x\right) }\left( V\left[ A\right] +V\left[ B_{1}\right] +V\left[ C\right] +V\left[ D_{1}\right] +V\left[ E\right] +2\text{cov}\left[ A,B_{1}\right] +2\text{cov}\left[ C,D_{1}\right] \right) 
\end{array}
\end{equation}
where
\begin{equation}\label{eq:eq27}
\begin{array}{l}
V\left[ A\right] =\sum\limits_{k=1}^{n\left( x-\Delta x\right) }\underbrace{\left( E\left[ t_{k}^{2}\left( x-\Delta x\right) \right] -E^{2}\left[ t_{k}\left( x-\Delta x\right) \right] \right) }_{\sigma_{k}^{2}\left( x-\Delta x\right)}\\
+\sum\limits_{k=1}^{n\left( x-\Delta x\right)} \sum\limits_{l=1,k\neq l}^{n\left( x-\Delta x\right)} \underbrace{\left( E\left[ t_{k}\left( x-\Delta x\right) t_{l}\left( x-\Delta x\right) \right] -E\left[ t_{k}\left( x-\Delta x\right) \right] E\left[ t_{l}\left( x-\Delta x\right) \right] \right) }_{c_{kl}}\\
=\sum\limits_{k=1}^{n\left( x-\Delta x\right)} \sigma_{k}^{2}\left( x-\Delta x\right) +\sum\limits_{k=1}^{n\left( x-\Delta x\right)} \sum\limits_{l=1,k\neq l}^{n\left( x-\Delta x\right)}c_{kl}
\end{array}
\end{equation}

\begin{equation}\label{eq:eq28}
\begin{array}{l}
V\left[ B_{1}\right] = 
\left( \dfrac{\Delta n}{n\left( x-\Delta x\right) }\right)^{2}\cdot\left( \sum\limits_{k=1}^{n\left( x-\Delta x\right) }\sigma_{k}^{2}\left( x-\Delta x\right) +\sum\limits_{k=1}^{n\left( x-\Delta x\right)} \sum\limits_{l=1,k\neq l}^{n\left( x-\Delta x\right)}c_{kl}\right) = \\ 
\left( \dfrac{\Delta n}{n\left( x-\Delta x\right) }\right)^{2}\cdot V\left[ A\right] 
\end{array}
\end{equation}

\begin{equation}\label{eq:eq29}
V\left[ C\right] =\sum\limits_{k=1}^{n\left( x-\Delta x\right)} \underbrace{\left( E\left[ \left( \Delta t_{k}\right)^{2}\right] -E^{2}\left[ \Delta t_{k}\right] \right)}_{\delta_{k}^{2}}=\sum\limits_{k=1}^{n\left( x-\Delta x\right)}\delta_{k}^{2}
\end{equation}

\begin{equation}\label{eq:eq30}
V\left[ D_{1}\right] =\left( \dfrac{\Delta n}{n\left( x- \Delta x\right) }\right)^{2}\,\sum\limits_{k=1}^{n\left( x-\Delta x\right) }\delta_{k}^{2}=\left( \dfrac{\Delta n}{n\left( x-\Delta x\right) }\right)^{2}V\left[ C\right] 
\end{equation}

\begin{equation}\label{eq:eq31}
V\left[ E\right] =\sum\limits_{j=1}^{\Delta n}\underbrace{\left( E\left[ \left( \rho_{j}\right) ^{2}\right] -E^{2}\left[ \rho_{j}\right] \right) }_{d_{j}^{2}}=\sum\limits_{j=1}^{\Delta n}d_{j}^{2}
\end{equation}

Similarly, the covariance terms are expressed as:
\begin{equation}\label{eq:eq32}
\begin{array}{l}
\text{cov}\left[ A,B_{1}\right] =
%\dfrac{\Delta n}{n\left( x-\Delta x\right) }\sum\limits_{k=1}^{n\left( x-\Delta \right) }\sum\limits_{l=1}^{n\left( x-\Delta x\right) }\left( E\left[ t_{k}\left( x-\Delta x\right) t_{l}\left( x-\Delta x\right) \right] -E\left[ t_{k}\left( x-\Delta x\right) \right] E\left[ t_{l}\left( x-\Delta x\right) \right] \right) \\
%=\dfrac{\Delta n}{n\left( x-\Delta x\right) }\left( \sum\limits_{k=1}^{n\left( x-\Delta x\right) }\sigma_{k}^{2}\left( x-\Delta x\right) +\sum\limits_{k=1}^{n\left( x-\Delta x\right) }\sum\limits_{l=1,k\neq l}^{n\left( x-\Delta x\right) }c_{kl}\right) \\
\dfrac{\Delta n}{n\left( x-\Delta x\right) }V\left[ A\right] 
\end{array}
\end{equation}
\begin{equation}\label{eq:eq33}
\begin{array}{l}
\text{cov}\left[ C,D_{1}\right] =
%\dfrac{\Delta n}{n\left( x-\Delta x\right) }\sum\limits_{k=1}^{n\left( x-\Delta x\right) }\sum\limits_{l=1}^{n\left( x-\Delta x\right) }\left( E\left[ \Delta t_{k}\Delta t_{l}\right] -E\left[ \Delta t_{k}\right] E\left[ \Delta t_{l}\right] \right) \\
%=\dfrac{\Delta n}{n\left( x-\Delta x\right) }\left( \sum\limits_{k=1}^{n\left( x-\Delta x\right) }\delta_{k}^{2}\right) \\
\dfrac{\Delta n}{n\left( x-\Delta x\right) }V\left[ C\right] 
\end{array}
\end{equation}
Substituting Eq. \ref{eq:eq27} $-$ \ref{eq:eq33} into Eq. \ref{eq:eq26}, the variance becomes:
\begin{equation}\label{eq:eq34}
\begin{array}{l}
V\left[ T_{1}\left( x,n\left( x\right) \right) \right]
=\dfrac{1}{n^{2}\left( x-\Delta x\right) }\left( \sum\limits_{k=1}^{n\left( x-\Delta x\right) }\sigma_{k}^{2}\left( x-\Delta x\right) +\sum\limits_{k=1}^{n\left( x-\Delta x\right) }\sum\limits_{l=1,k\neq l}^{n\left( x-\Delta x\right) }c_{kl}\right) \\ 
+  \dfrac{1}{n^{2}\left( x-\Delta x\right) }\sum\limits_{k=1}^{n\left( x-\Delta x\right) }\delta_{k}^{2}+\dfrac{1}{n^{2}\left( x\right) }\sum\limits_{j=1}^{\Delta n}d_{j}^{2}
%=\dfrac{1}{n^{2}\left( x\right)}\left( V\left[ A\right] +\left( \dfrac{\Delta n}{n\left( x-\Delta x\right) }\right)^{2}\cdot V\left[ A\right] +V\left[ C\right] +\left( \dfrac{\Delta n}{n\left( x-\Delta x\right) }\right) ^{2}\cdot V\left[ C\right] +V\left[ E\right] +2\dfrac{\Delta n}{n\left( x-\Delta x\right) }V\left[ A\right] +2\dfrac{\Delta n}{n\left( x-\Delta x\right) }V\left[ C\right]\right) \\
%=\dfrac{1}{n^{2}\left( x\right) }\left( \left( \dfrac{n\left( x\right) }{n\left( x-\Delta x\right) }\right)^{2}\cdot V\left[ A\right] +\left( \dfrac{n\left( x\right) }{n\left( x-\Delta x\right)} \right) ^{2} \cdot V\left[ C\right] +V\left[ E\right]\right)  \\
%=\dfrac{1}{n^{2}\left( x-\Delta x\right) }\left( \sum\limits_{k=1}^{n\left( x-\Delta x\right) }\sigma_{k}^{2}\left( x-\Delta x\right) +\sum\limits_{k=1}^{n\left( x-\Delta x\right) }\sum\limits_{l=1,k\neq l}^{n\left( x-\Delta x\right) }c_{kl}\right) +\dfrac{1}{n^{2}\left( x-\Delta x\right) }\sum\limits_{k=1}^{n\left( x-\Delta x\right) }\delta_{k}^{2}+\dfrac{1}{n^{2}\left( x\right) }\sum\limits_{j=1}^{\Delta n}d_{j}^{2}
\end{array}
\end{equation}
\\

Taking into account that all $\Delta t_{k}$ % \left( k=1,2,3,...,n\left(  x-\Delta x\right) \right) $ 
follow the same distribution with a variance ($\delta^{2}$)  proportional to the
corresponding drift distance ($\Delta x$),  i.e. $\delta^{2}=\sigma_{0}^{2}\cdot \Delta x$ , and 
that the time-gains $\rho_{j} \left( j=1,2,3,...,\Delta n \right)$ follow a distribution with a variance $w^{2}$, 
the two last terms in Eq. \ref{eq:eq34} are written as:
\begin{equation}\label{eq:eq35}
\dfrac{1}{n^{2}\left( x-\Delta x\right)}\sum\limits_{k=1}^{n\left( x-\Delta x\right) }\delta_{k}^{2}=\dfrac{\sigma_{0}^{2}\cdot\Delta x}{n\left( x-\Delta x\right) }  \text{  \; and \;   } 
\dfrac{1}{n^{2}\left( x\right) }\sum\limits_{j=1}^{\Delta n}d_{j}^{2}=\dfrac{\Delta n}{n^{2}\left( x\right) }w^{2}
\end{equation}

In addition, the total avalanche time variance at the plane on $x-\Delta x$ is:
\begin{equation}\label{eq:eq37}
V\left[ T_{1}\left( x-\Delta x,n\left( x-\Delta x\right) \right) \right] =\dfrac{1}{n^{2}\left( x-\Delta x\right) }\left( \sum\limits_{k=1}^{n\left( x-\Delta x\right) }\sigma_{k}^{2}\left( x-\Delta x\right) +\sum\limits_{k=1}^{n\left( x-\Delta x\right) }\sum\limits_{l=1,k\neq l}^{n\left( x-\Delta x\right) }c_{kl}\right) 
\end{equation}
Then, substituting Eqs. \ref{eq:eq35}, Eq. \ref{eq:eq37},  
and the approximation  $n^2(x) \simeq n(x) \cdot n(x-\Delta x)$  into Eq. \ref{eq:eq34},  one gets:
\begin{equation}\label{eq:eq38}
\begin{array}{l}
V\left[ T_{1}\left( x,n\left( x\right) \right) \right] -V\left[ T_{1}\left( x-\Delta x,n\left( x-\Delta x\right) \right) \right]  \\
%\dfrac{\sigma_{0}^{2}\cdot\Delta x}{n\left( x-\Delta x\right) }+\dfrac{\Delta n}{n^{2}\left( x\right) }w^{2}\\
%\simeq \dfrac{\sigma_{0}^{2}\cdot\Delta x}{n\left( x-\Delta x\right) }+\dfrac{\Delta n}{n\left( x\right)n\left( x-\Delta x\right)  }w^{2}\\
\simeq \dfrac{\sigma_{0}^{2}\cdot\Delta x}{n\left( x-\Delta x\right) }-w^{2}\left( \dfrac{1}{n\left( x\right) }-\dfrac{1}{n\left( x-\Delta x\right) }\right) 
\end{array}
\end{equation}
which expresses the increase of the avalanche-time variance as the avalanche grows between two planes, on $x-\Delta x$ and on $x$, given that $n\left( x-\Delta x\right) $ electrons reach the first plane and $\Delta n$ more electrons reach the second plane. 

For all the avalanches evolving up to a length $x$, the variance of the avalanche-time can be obtained by averaging Eq. \ref{eq:eq38} for all possible values of $n\left( x-\Delta x\right) $ and $\Delta n$. Specifically:
\begin{equation}\label{eq:eq39}
\begin{array}{l}
\dfrac{\langle V\left[ T_{1}\left( x,n\left( x\right) \right) \right] -V\left[ T_{1}\left( x-\Delta x,n\left( x-\Delta x\right) \right) \right] \rangle_{n,\Delta n}}{\Delta x}\\
=\sigma_{0}^{2}\langle\dfrac{1}{n\left( x-\Delta x\right)} \rangle_{n,\Delta n}-\dfrac{w^{2}}{\Delta x}\langle\dfrac{1}{n\left( x\right) }-\dfrac{1}{n\left( x-\Delta x\right) } \rangle_{n,\Delta n}
\end{array}
\end{equation}

\begin{figure}[h]
\centering\includegraphics[width=0.6\linewidth]{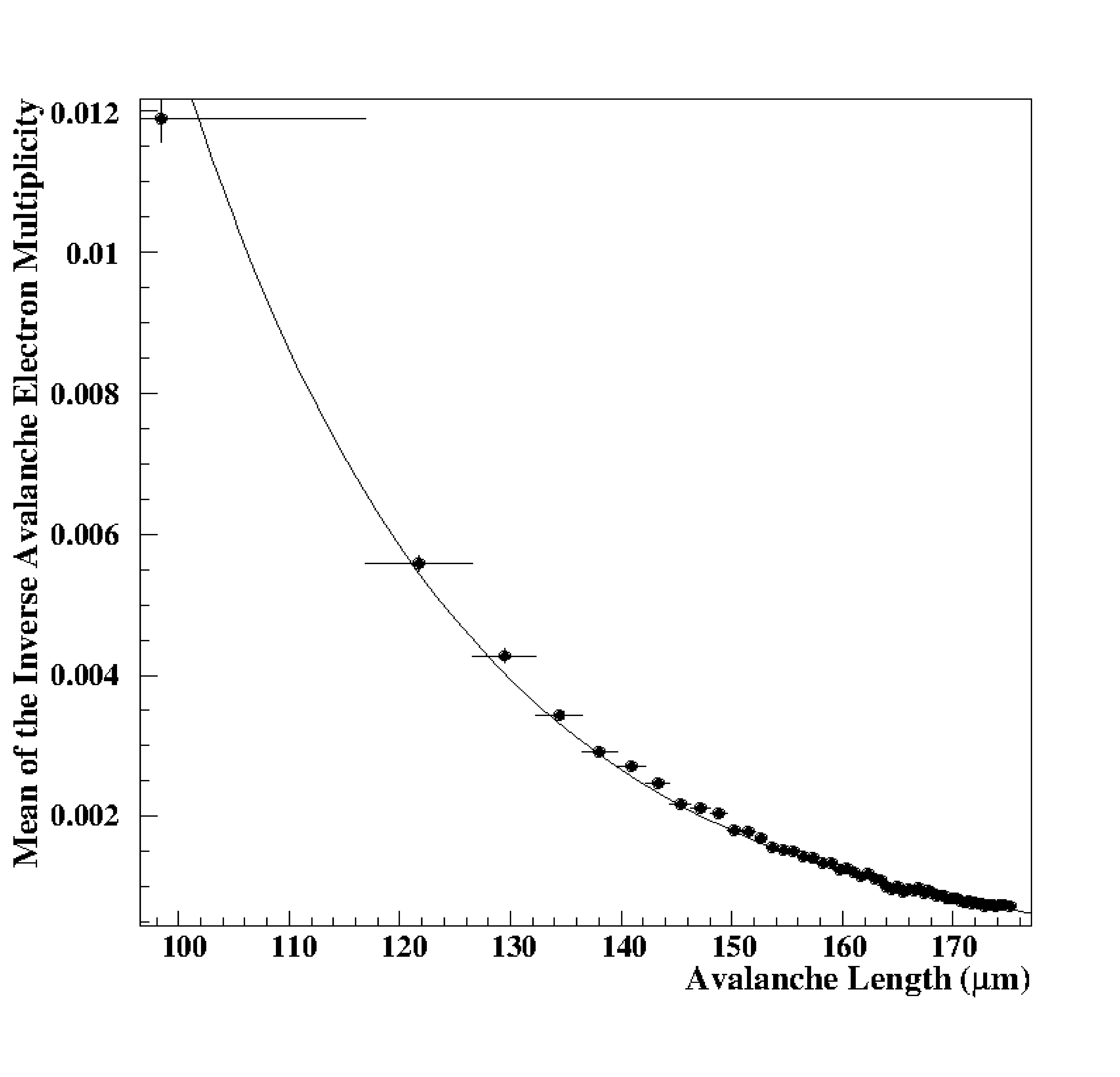}
\caption{Each point represents the mean value of the inverse avalanche-electron multiplicity for simulated avalanches of a certain length.  The  GARFIELD++ simulation package has been used, assuming  50\% Ptr, a drift voltage of 425 V and anode voltage of 450 V. The solid curve represents graphically Eq. \ref{eq:eq40} with the proper values for the physical parameters, from Table \ref{tab:tableA-8}.}
\label{fig:fig12}
\end{figure}

Assuming that $n\left( x\right) $ follows the Gamma distribution function,
%of Eq. \ref{eq:eq8}, 
the mean value of the inverse multiplicity, $1/n\left( x\right) $, is given by the formula:
\begin{equation}\label{eq:eq40}
\langle\dfrac{1}{n\left( x\right) }\rangle_{n}=\dfrac{\left( \theta+1\right) }{2\theta}\exp\left( -a_{eff}\cdot x\right) 
\end{equation}
which describes well the GARFIELD++ simulation results, as it is shown in Fig. \ref{fig:fig12}.

Substituting Eq. \ref{eq:eq40} in Eq. \ref{eq:eq39}, the differential increase of the variance is expressed as:
\begin{equation}\label{eq:eq41}
\begin{array}{l}
\dfrac{\langle V\left[ T_{1}\left( x\right) \right] -V\left[ T_{1}\left( x-\Delta x\right) \right] \rangle_{n,\Delta n}}{\Delta x}= 
\sigma_{0}^{2}\dfrac{\left( \theta+1\right) }{2\theta}\exp\left( -a_{eff}\cdot x\right) \cdot\exp\left(  a_{eff}\cdot\Delta x\right)  \\
- \dfrac{w^{2}}{\Delta x}\dfrac{\left( \theta+1\right) }{2\theta}\exp\left( -a_{eff}\cdot x\right) \cdot\left(1-\exp\left(a_{eff}\cdot\Delta x\right)\right) 
\end{array}
\end{equation}

Expanding the right hand side of Eq. \ref{eq:eq41} with respect to $\Delta x$, keeping up to first order terms, and letting $\Delta x$ going to zero,
the differential equation that expresses the evolution of the avalanche-time variance is deduced to:
\begin{equation}\label{eq:eq42}
\dfrac{d V\left[ T\left( x\right) \right] }{dx}=\dfrac{\left( \theta+1\right) }{2\theta}\text{exp}(-a_{\text{eff}}\cdot x)\left[ \sigma_{0}^{2}+w^{2}a_{eff}\right] 
\end{equation}

Then, by integrating up to an avalanche-length L, the variance of the  avalanche-time at avalanche-length L is:
\begin{equation}\label{eq:eq43}
V\big[T(L)\big] = \frac{(\theta+1)}{2\theta}[\sigma_0^2 + w^2a_{\text{eff}}]\frac{1 - \text{exp}(-a_{\text{eff}}\cdot L)}{a_{\text{eff}}}
\end{equation}

Therefore,  the variance of the total time on the mesh, according to Eq. \ref{eq:eq23}, is:
\begin{equation}\label{eq:eq44}
\begin{array}{l}
V\big[T_{tot}(L)\big] = V\big[T(L)\big] + V\big[T_P(L)\big] \\  \\
=\frac{(\theta + 1)}{2\theta}[\sigma_0^2 + w^2a_{\text{eff}}]\frac{1 - \text{exp}(-a_{\text{eff}}\cdot L)}{a_{\text{eff}}}+(D-L) \cdot \sigma_P^2 +\Phi
\end{array}
\end{equation}
which is expected to describe the GARFIELD++ simulations for photoelectron drift lengths long enough to guarantee statistical equilibrium (typically $\left( D-L\right) >10\,\microm$.\\

\begin{figure}[h]
\centering\includegraphics[width=0.6\linewidth]{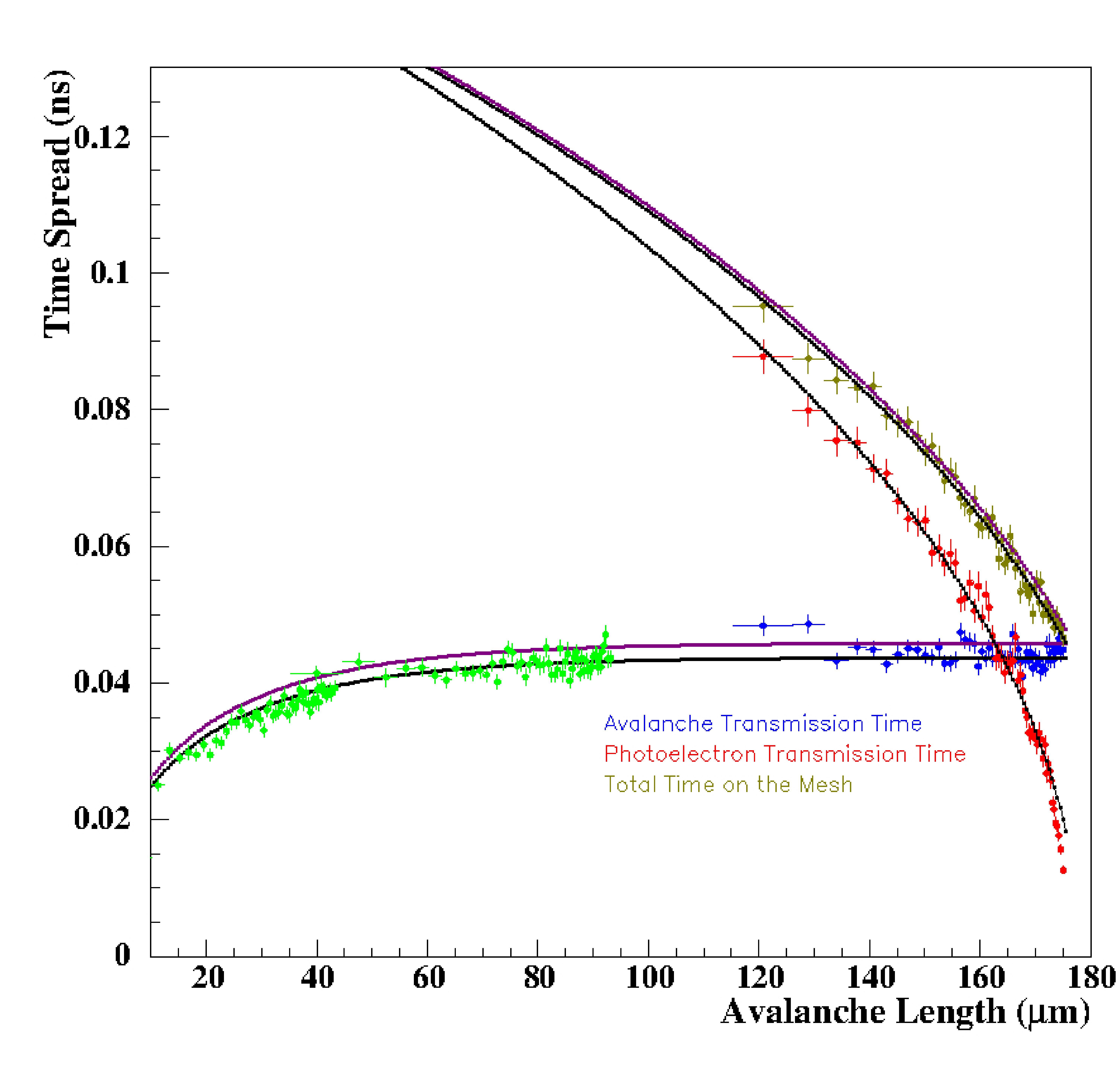}
\caption{The points show results of GARFIELD++ simulations assuming 50\% Ptr, 425 V drift and 450 V anode voltages, versus the respective length of the avalanche. The golden points depict the spread of the total time on the mesh. The red and blue (plus bright green) points represent   spreads of the primary photoelectron time and of the avalanche time, respectively. The  corresponding model predictions, for the two $w$ values discussed in the text, are presented  as solid lines.}
\label{fig:fig13}
\end{figure}

  Model predictions for the time spreads, expressed by Eqs. \ref{eq:eq21}, \ref{eq:eq43} and \ref{eq:eq44}, are shown in Fig. \ref{fig:fig13} to be in good agreement with the GARFIELD++ simulation results. The same  good agreement is found for all Ptr values and drift voltages considered in this work.
   
    While the mean value of the time-gain parameter $\rho$ has been evaluated from GARFIELD++ simulations (see Fig. \ref{fig:fig7}), there is no similar, straightforward way to estimate the value of its variance ($w^{2}=V\left[ \rho\right]$). 
    As an alternative, the double lines in Fig. \ref{fig:fig13} represent the predictions of Eqs. \ref{eq:eq43} and \ref{eq:eq44} for $w=0$ and $w=\rho$, i.e. either assuming that the time-gain per newly produced electron is a constant or that it follows a very broad physical distribution with an RMS equal to  100\% of its mean value. Apparently, by imposing a 100\% spread on $\rho$, only a small change is induced to the model predictions.\\

	As Fig. \ref{fig:fig13} indicates,  signals produced by  long avalanches achieve good resolution because they are associated with short drifting photoelectrons, which suffer small longitudinal diffusion. The model predicts that the contribution of short avalanches  to the timing resolution depends on their length. However, as the avalanche length grows, the variance of the avalanche time reaches a plateau.  
	At the operational parameter settings considered in this study, the vast majority of the GARFIELD++ simulated avalanches in the PICOSEC pre-amplification region are too long to reveal the increase of the avalanche time spread. 
	In order to check the model prediction in detail, special GARFIELD++ simulations of shorter pre-amplification avalanches were performed. 
	Two groups of such simulation results are also shown, as bright green points in the same Figure, demonstrating the success of the model in predicting the avalanche time spread at all avalanche lengths.
	Nevertheless, the predicted spread of the photoelectron time seems to deviate from the GARFIELD++ points at very large avalanche lengths (short photoelectron drift paths), due to the inadequacy of Eq. \ref{eq:eq21} to describe the photoelectron longitudinal dispersion at the beginning of its drift path, i.e. before it reaches statistical equilibrium through multiple scatterings. 
	However, this small deviation appears in the region of very large avalanche lengths, where the timing resolution is practically determined by the avalanche time spread. 

\section{Timing Resolution versus Electron Multiplicity on the Mesh} \label{modelaval4}

GARFIELD++ simulations showed that the electron multiplicity on the mesh determines the PICOSEC signal size (see Sections \ref{intro} and \ref{modelaval1}). To get insight on the dependence of the timing resolution on the signal amplitude, the effects of the photoelectron drift and the pre-amplification avalanche development  are expressed as  functions  of the electron multiplicity on the mesh.
% by properly integrating equations \ref{eq:eq21} and \ref{eq:eq38}. 
The other, weaker effect on the timing resolution, i.e.  the passage of the avalanche electrons through the mesh, is discussed in Section \ref{modelaval5}.\\

	The variance of the avalanche  time can be evaluated as a function of the  electron multiplicity on the mesh, $N_{L}$, by averaging Eq. \ref{eq:eq38} over  $n\left( x\right) $, under the condition that at the end of the avalanche development, i.e. at $x=L$, the observed electron multiplicity, $n\left( L\right) $, should  equal $N_{L}$. 
	Measuring from the point of the first ionization, the conditional p.d.f., $\Pi\left( n\left( x\right) \vert n\left( L\right) = N_{L} \right) $, that an avalanche has $n\left( x\right) $ electrons at a plane on $x$ given that it has $N_{L}$ electrons at a plane on L $\left( L>x\right) $, can be expressed as:
\begin{equation}\label{eq:eq45}
\Pi \big(n(x)|n(L) = N_L\big) = \frac{\Pi \big(n(L) = N_L|n(x)\big) \cdot \Pi \big( n(x)\big)}{\Pi \big(n(L) = N_L \big)}
\end{equation}
The term $\Pi \left( n\left( x\right) \right) $ denotes the p.d.f. that an avalanche has $n\left( x\right) $ electrons at a plane on $x$. It is approximated (see Fig. \ref{fig:fig8}) by the  Gamma distribution function, i.e. 
 $\Pi   \left( n\left( x\right) \right) = P(n(x); q=2 e^{a_{eff}x}, \theta)$.
 
The other term in the numerator of Eq. \ref{eq:eq45}, $\Pi \left( n\left( L\right) = N_{L} \vert n\left( x\right) \right) $, is the conditional p.d.f. that an avalanche has $N_{L}$ electrons at a plane on L, given that it has $n\left( x\right) $ electrons at a plane on $x$. 
Assuming that each of the $n\left( x\right)$ electrons will initiate an independent avalanche and each avalanche will evolve until it reaches the plane on L, there will be $n\left( x\right) $ statistically identical and independent avalanches, each of length $L-x$. Then, $ \Pi \left( n\left( L\right) = N_{L} \vert n\left( x\right) \right) $ can be approximated by the convolution of $n\left( x\right)$ Gamma distributions, resulting in the expression:
\begin{equation}\label{eq:eq46}
\begin{array}{l}
\Pi\big(n(L) = N_L|n(x) \big) = \overbrace{P_1(n)\otimes P_1(n)\otimes \dots \otimes P_1(n)}^\text{n(x)times} \\
\\
= \frac{1}{q\big(L-x\big)} \frac{(\theta + 1)^{n(x)(\theta + 1)}}{\Gamma \big( n(x)\cdot (\theta + 1)\big)}\cdot\Bigg( \frac{N_L}{q\big(L - x \big)} \Bigg)^{n(x)(\theta + 1)-1}\cdot exp\Bigg[ -(\theta+1)\frac{N_L}{q\big( L-x\big)} \Bigg]
\end{array}
\end{equation}
where, $q\left( L-x\right) $ is the mean multiplicity of a single avalanche of length $L-x$. 
The mean value and the variance of the above p.d.f. are $n\left( x\right)\cdot q\left( L - x\right) $ and $n\left( x\right)\cdot\dfrac{q^{2}\left( L-x\right) }{\theta+1}$, respectively. 
A drawback in expressing $\Pi \left(n\left( L\right)= N_{L}\vert n\left( x\right) \right) $ as in Eq. \ref{eq:eq46} is that $n\left( x\right) $ should be treated as an integer while $N_{L}$ as a real number. Alternatively, by invoking the Central Limit Theorem, a Gaussian distribution can be used, in case that $n\left( x\right)$ is a large number, i.e.
\begin{equation}\label{eq:eq47}
\Pi \big(n(L) = N_L|n(x) \big) = \frac{1}{\sqrt{2\pi\cdot n(x)\cdot\sigma^2(L-x)}}exp\Bigg[-\frac{\big(n(x)\cdot q\big(L-x\big) -N_L\big)^2}{2\cdot n(x)\cdot \sigma^2\big( L - x \big)}  \Bigg]
\end{equation}
where $\sigma^{2}\left( L-x\right)$ is the variance of a single avalanche of length $L-x$. The p.d.f. expressed by Eq. \ref{eq:eq47} has the same mean value and variance as the p.d.f. of Eq. \ref{eq:eq46}. It should be emphasized that Eq. \ref{eq:eq47} is strictly valid only in case that $n\left( x\right)$ is an integer parameter. However, in order to simplify numerical calculations, $n\left( x\right)$ is treated as a continuous variable.

The normalizing term, $\Pi \big( n(L) = N_L \big)$, in the denominator of Eq. \ref{eq:eq45} is defined as:
\begin{equation}\label{eq:eq48}
\begin{array}{l}
\Pi \big( n(L) = N_L \big) = \sum\limits_{n=0}^{\infty} \Pi \big( n(L)= N_L|n\big) \cdot \Pi(n) \simeq \displaystyle \int\limits_{0} ^{\infty} \Pi \big( n(L) = N_L |n(x)\big) \cdot \Pi \big(n(x)\big) dn(x)
\end{array}
\end{equation}
%ensures that the p.d.f. $\Pi \left( n\left( x\right) \vert n\left( L\right)  = N_{L}\right)$ is normalized to unity.
%\footnote{The denominator in Eq. \ref{eq:eq45}, $\Pi\left( n\left( L\right) \right) $, could be approximated by a Gamma distribution function with $q\left( L\right) =2\exp\left[a_{eff}L\right] $. However, due to the fact that the description of the electron multiplicity distribution by a Gamma distribution function is an approximation, the integral of Eq. \ref{eq:eq45} will be only approximately equal to one, bringing extra inaccuracies in the model predictions. It is thus preferable to define the normalizing factor by Eq. \ref{eq:eq48}.}. 

Then, having determined the functional form of $\Pi\left( n\left( x\right) \vert n\left( L\right) = N_{L} \right) $, it is straightforward to  properly average Eq. \ref{eq:eq38} by imposing the condition that the electron multiplicity at an avalanche length L  equals  $N_{L}$. 

Using Eq. \ref{eq:eq38} and the following definitions:
\begin{equation}\label{eq:eq49}
\begin{array}{l}
\big< V(x)\big>_{n(L)=N_L} \equiv \displaystyle \int\limits_{0}^{\infty} V\big[ T_{1}\big( x,n(x)\big)\big] \cdot P\big( n(x)|n(L)=N_L\big)dn(x)\\
\big< V(x-\Delta x)\big>_{n(L)=N_L} \equiv \\ 
\displaystyle  \int\limits_{0}^{\infty}V\big[T_{1}(x-\Delta x,n(x-\Delta x)\big)\big]\cdot P\big(n(x-\Delta x)|n(L)=N_L\big)dn(x-\Delta x)\\
\Big< \frac{1}{n(x)}\Big>_{n(L)=N_L} \equiv \displaystyle \int\limits_{0}^{\infty}\frac{1}{n(x)}\cdot P\big(n(x)|n(L)=N_L\big)dn(x)
\end{array}
\end{equation}
the average increase of the avalanche time variance, between the planes on $x-\Delta x$ and $x$, under the condition that at $x=L$ the electron multiplicity equals $N_{L}$, is written as:
\begin{equation}\label{eq:eq50}
\begin{array}{l}
\langle V\left( x\right) \rangle_{n\left( L\right) =N_{L}} - \langle V\left( x-\Delta x\right) \rangle_{n\left( L\right) =N_{L}} \\
=\sigma_{0}^{2}\cdot \Delta x\langle\dfrac{1}{n\left( x-\Delta x\right) }\rangle_{n\left( L\right) =N_{L}}-w^{2}\left( \langle\dfrac{1}{n\left( x\right) }\rangle_{n\left( L\right) =N_{L}}-\langle\dfrac{1}{n\left( x-\Delta x\right) }\rangle_{n\left( L\right) =N_{L}}\right) 
\end{array}
\end{equation}
\\

Notice that the imposed condition, $n\left( L\right) =N_{L}$, has forced the averages,  $\langle 1/n\left( x\right) \rangle_{n\left( L\right)=N_{L}}$ and $\langle V\left( x\right) \rangle_{ n\left( L\right) =N_{L}}$ , to be also function of $N_{L}$. Hereafter, terms symbolized as $\langle\bullet\left(  x\right) \rangle_{n\left( L\right) =N_{L}}$ must be considered as functions of both $x$ and $N_{L}$.

 A recursive summation of Eq. \ref{eq:eq50}, starting at $x=L$ and stopping at $x=0$, in steps of size $\Delta x$, results to:
\begin{equation}\label{eq:eq51}
\begin{array}{l}
\big<V (L) \big>_{n(L)=N_L} - \big< V(0)\big>_{n(L)=N_L} \\
= \sigma^2_0 \cdot \Delta x \sum\limits_{i-1}^{L/\Delta x} \Bigg< \frac{1}{n(L - i \cdot \Delta x)} \Bigg>_{n(L) = N_L} - w^2 \Bigg(\Bigg< \frac{1}{n(L)} \Bigg>_{n(L)=N_L} - \Bigg< \frac{1}{n(0)}\Bigg>_{n(L)=N_L}\Bigg)
\end{array}
\end{equation}

At the limit of $\Delta x$ going to zero and using that
\begin{equation*}
\Bigg<V(0)\Bigg>_{n(L)=N_L}=0 , \,\,\,\,\,\,\,\,\,\,\,\,\,\,\Bigg<\dfrac{1}{n(0)} 
\Bigg>_{n(L)=N_L}=\frac{1}{2} , \,\,\,\,\,\,\,\,\,\,\,\,\,\,\Bigg<\dfrac{1}{n(L)}\Bigg>_{n(L)=N_L}=\dfrac{1}{N_L}
\end{equation*}
Eq. \ref{eq:eq51} becomes
\begin{equation}\label{eq:eq52}
\big< V(L) \big>_{n(L)=N_L} = \sigma^{2}_{0} \cdot \int\limits_{0}^{L}\Big<\frac{1}{n(x)}\Big>_{n(L)=N_L}dx-w^2\Big(\frac{1}{N_L}-\frac{1}{2}\Big)
\end{equation}
expressing the variance of the avalanche time, when the electron multiplicity on the mesh is $N_{L}$ and given that the  avalanche length equals L. 

The first term in the above equation is a double integral, which is easily evaluated by numerical integration, for any L and $N_L$ values, using Eq. \ref{eq:eq49} with the definition expressed by either Eq. \ref{eq:eq47} or Eq. \ref{eq:eq48},  as well as setting appropriate values to the relevant model parameters ($\sigma_0$, $\theta$, and $a_{eff}$) from Table \ref{tab:tableA-8}.\\

 In order to express the variance of the avalanche time  as a function of only the electron multiplicity on the mesh, N, Eq. \ref{eq:eq52} should be integrated  considering the contribution of any avalanche, of any length L, which produces N electrons arriving on the mesh ($N_{L}=N$). Naturally, each such contribution should be weighted by the likelihood that such an avalanche is produced, which is given by the p.d.f. $G(L\vert N)$ defined by Eq. \ref{eq:eq16}.\\

	Let us consider a sample of avalanches with N electrons on the mesh. Schematically, this sample comprises many (infinite) sets, each set consisting of avalanches with a certain length, L, having a  population proportional to $G(L\vert N)$. The mean avalanche time in a set is $T(N,L)$ and the respective variance is $\big<V(L)\big>_{n(L)=N}$. 
	In the hypothetical case that all the above subsets had the same mean avalanche time, the time variance of the whole sample will be given simply by the weighted sum of the respective variances of the subsets. 
	However, due to the fact that the mean avalanche time varies among the sets, the variance of the avalanche time considering all avalanches in the sample  should be evaluated according to Eq. \ref{eq:eqB5} (see \ref{Appendix B}). Thus, the variance of the avalanche time, $V[T(N)]$, when the electron multiplicity on the mesh is N, is given by the following expression: 
\begin{equation}\label{eq:eq53}
\begin{array}{l}
V[T(N)]=\displaystyle \int\limits_{x_1}^{x_2} \big<V(L)\big>_{n(L)=N} \cdot G(L|N)dL + \\
\displaystyle \int\limits_{x_1}^{x_2}T(N,L)^2\cdot G(L|N)dL- \Bigg[ \displaystyle \int\limits_{x_1}^{x_2}T(N,L)\cdot G(L,N)dL\Bigg]^2
\end{array}
\end{equation}
\\
\begin{figure}[h]
\centering\includegraphics[width=0.8\linewidth]{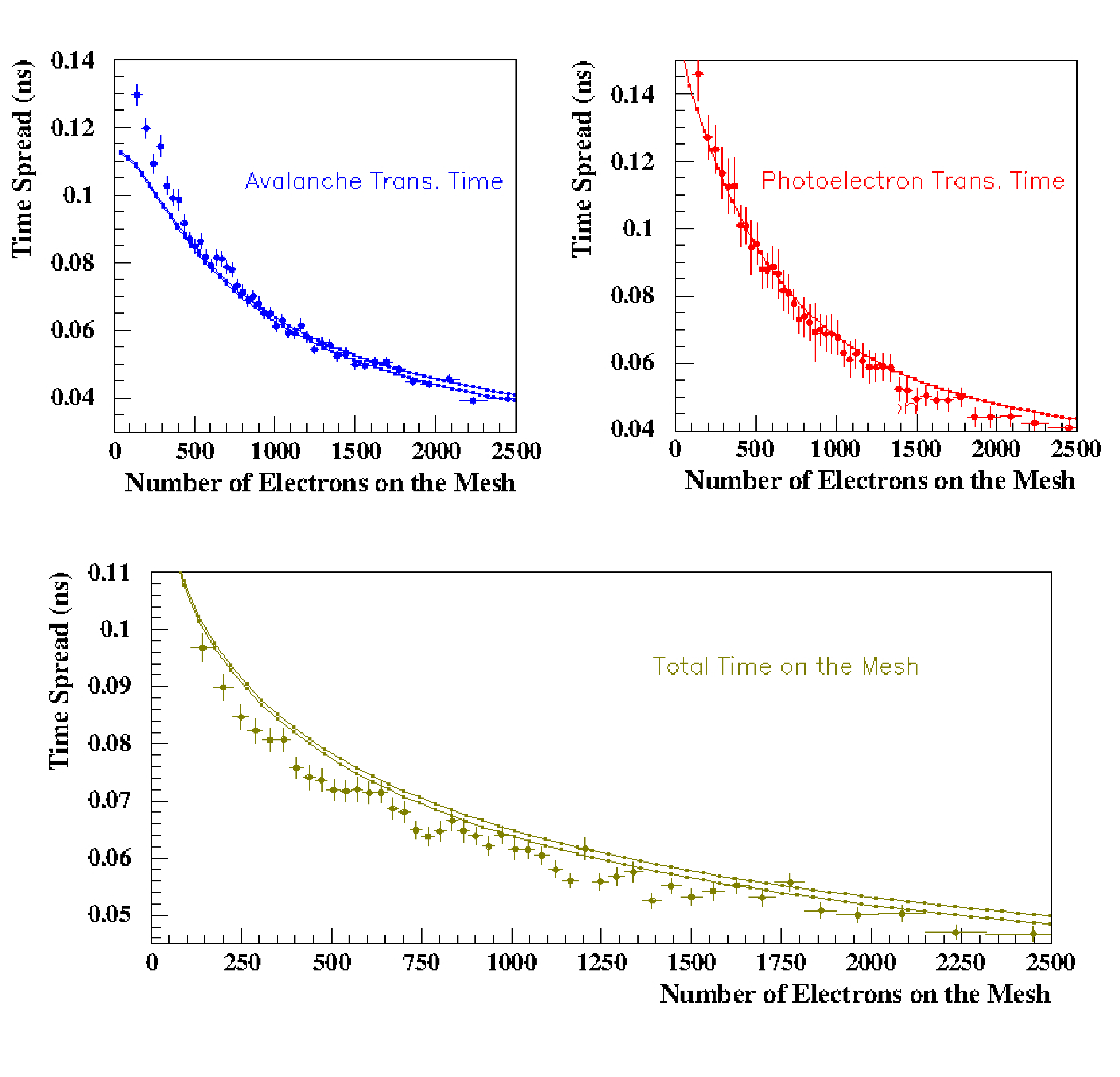}
\caption{The points represent the transmission time spread evaluated using GARFIELD++ simulations, with 50\% Penning Transfer Rate, 425 V drift and 450 V anode voltage. The double lines  present model predictions for $w=0$ and $w=\rho$ as discussed in Section \ref{modelaval3}. The top-left (blue), the top-right (red) and the bottom (golden) plots show the  avalanche time spread,  the photoelectron time spread and  the spread of the total time on the mesh, respectively, versus the number of pre-amplification electrons arriving on the mesh.}
\label{fig:fig14}
\end{figure}

Physically, the variance of the photoelectron time, $V[T_{p}(L)]$, depends only on its drift length, D-L, as expressed in Eq. \ref{eq:eq21}. 
Since the photoelectron drift length D-L is the residual of the respective avalanche length L, which in turn determines the mean multiplicity of the avalanche electrons, 
the variance of the photoelectron time is indirectly related to the electron multiplicity on the mesh, N.\\  

The variance of the photoelectron time, $V[T_p(N)]$, is expressed in Eq. \ref{eq:eq54} as a function of N, by weighting Eqs. \ref{eq:eq18} and \ref{eq:eq21} with  $G(L|N)$, integrating over the avalanche length and applying Eq. \ref{eq:eqB5} as before. 

\begin{equation}\label{eq:eq54}
V[T_p(N)]  =\displaystyle  \int\limits_{x_1}^{x_2}V[T_p(L)]\cdot G(L|N)dL +\displaystyle  \int\limits_{x_1}^{x_2} T_p^2(L) \cdot G(L|N)dL - \Bigg[ \displaystyle \int\limits_{x_1}^{x_2}T_p(L) \cdot G(L|N) dL\Bigg]^2
\end{equation} 
\\
Finally, the variance of the total time on the mesh is expressed in accordance to Eq. \ref{eq:eqB5} as:
\begin{equation}\label{eq:eq55}
\begin{array}{l}
 V[T_{tot}(N)]  = \displaystyle \int\limits_{x_1}^{x_2}\big[ V[T_p (L)] + \big< V(L)\big>_{n(L)=N} \big] \cdot G(L|N)dL\\
 + \displaystyle \int\limits_{x_1}^{x_2}\Big[ T(N,L) +T_p(L)\Big]^2 \cdot G(L|N)dL - \Bigg[\displaystyle  \int\limits_{x_1}^{x_2}\Big[T(N,L)+T_p(L)\Big] \cdot G(L|N)dL\Bigg]^2
\end{array}
\end{equation}

Notice that  Eq. \ref{eq:eq55} is not simply the sum of Eq. \ref{eq:eq53} and Eq. \ref{eq:eq54}, as it would be the case if the photoelectron and avalanche contributions to the total-time,  expressed as functions of the electron multiplicity on the mesh, were uncorrelated. 
This correlation is also apparent in the GARFIELD++ simulations shown in Fig. \ref{fig:fig14} and it is caused by the fact that the same number of pre-amplification electrons arriving on the mesh can be produced by avalanches of different length, while  the  mean avalanche  time depends on the avalanche length. 

The predictions of Eqs. \ref{eq:eq53} - \ref{eq:eq55} are  in good agreement with the corresponding GARFIELD++ simulation results, as shown in Fig. \ref{fig:fig14}.  Moreover, the model reproduces successfully  the related GARFIELD++ simulation results at all  operational conditions considered in this study.

However, for small values of electron multiplicity on the mesh, the  time-spreads predicted by our model  are systematically smaller than the related  GARFIELD++ simulation results. 
This underestimation stems from the inadequacy of the  p.d.f.'s employed to approximate  the avalanche statistical properties at its very beginning (i.e. at small avalanche length and low electron multiplicity)  and it is discussed in Section \ref{discl}.

\section{Effects related to  electrons traversing the mesh} \label{modelaval5}

\begin{figure}[h]
\centering
\begin{minipage}{.52\textwidth}
\centering
\includegraphics[width=.88\textwidth]{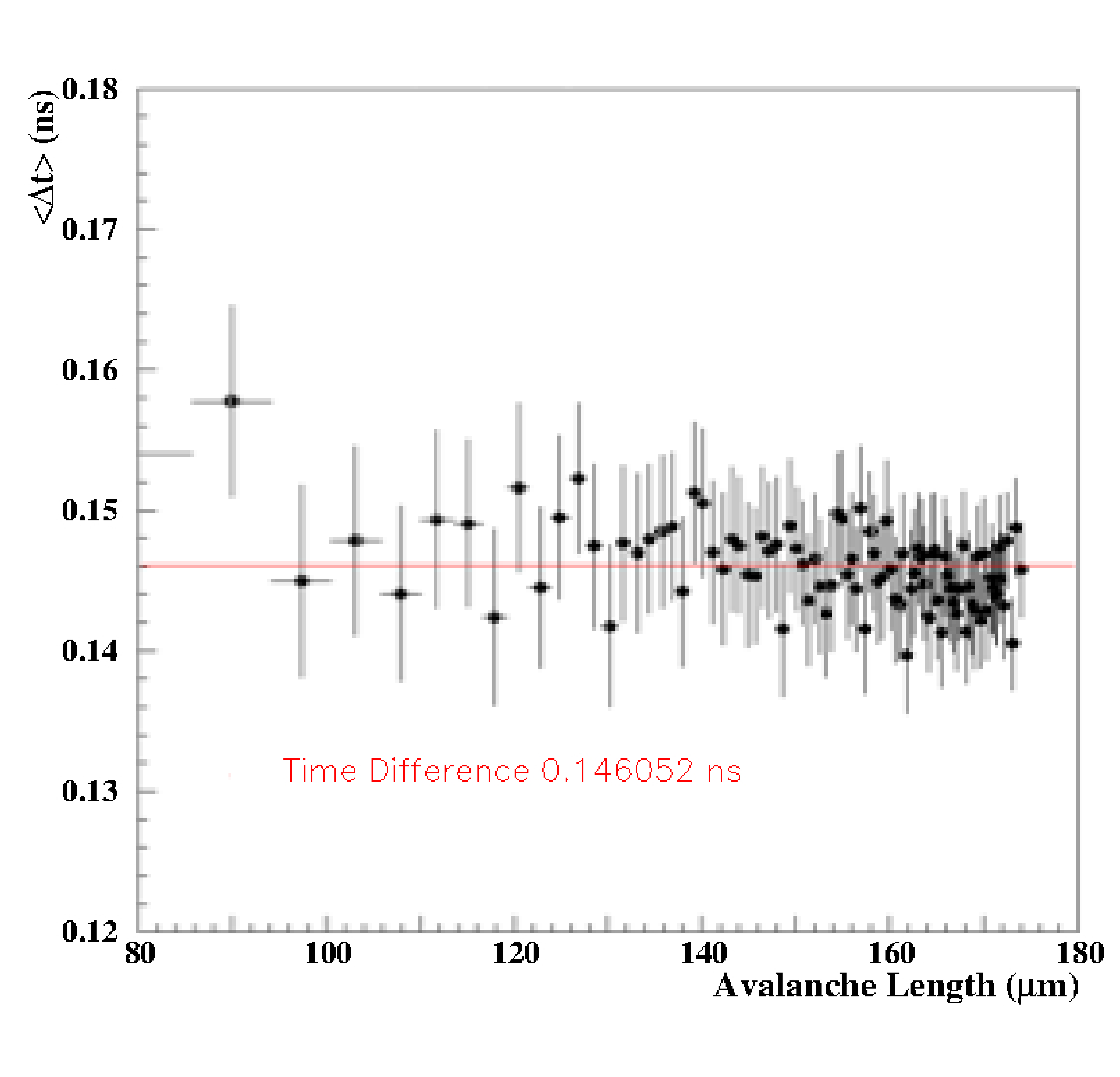}
\end{minipage}%
\begin{minipage}{.52\textwidth}
\centering
\includegraphics[width=.85\textwidth]{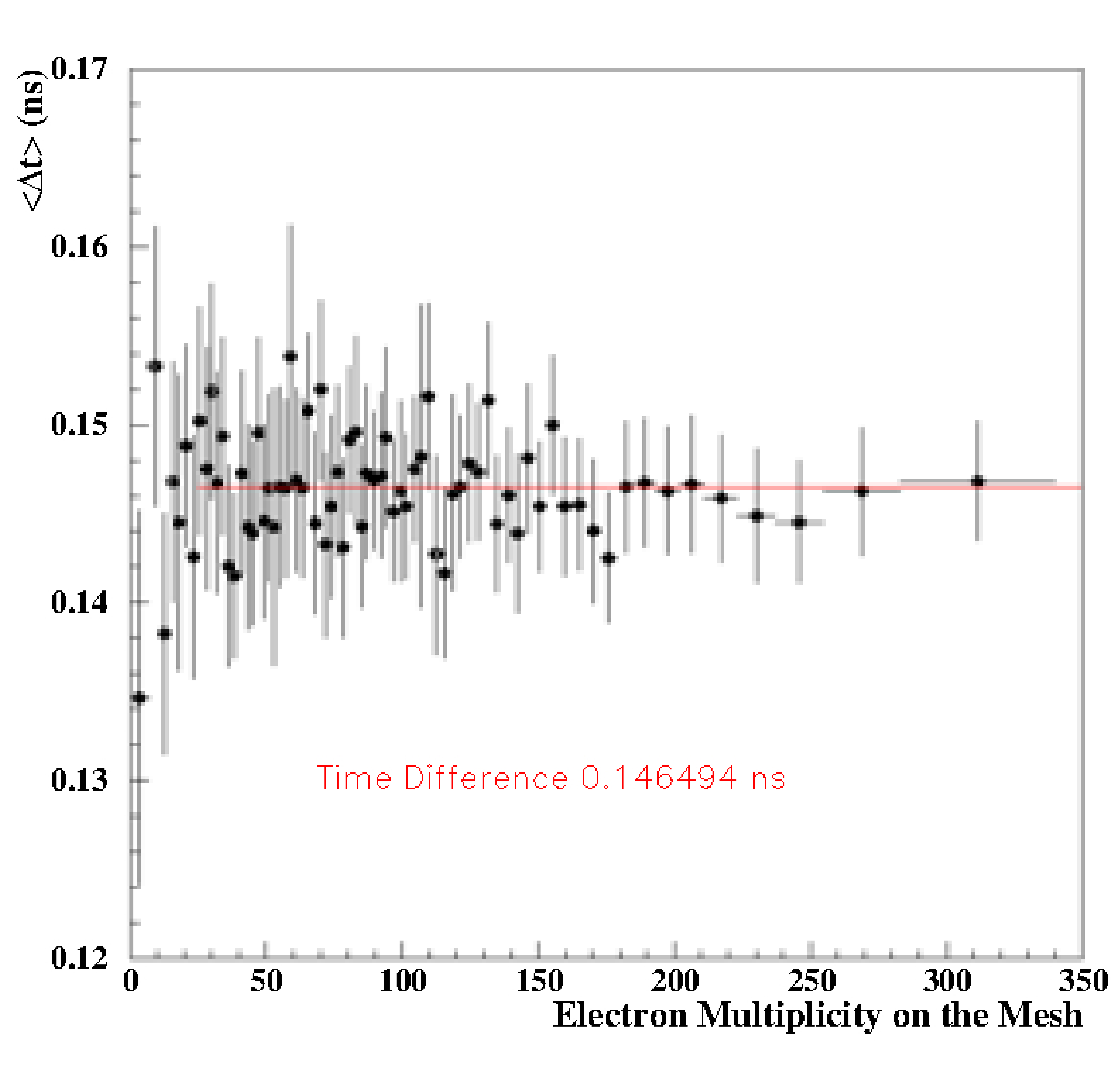}
\end{minipage}
\caption { The points represent GARFIELD++ simulation results, assuming 50\% Ptr, 450 V and 350 V drift and anode voltage, respectively.  The time to pass through the mesh (i.e. $\langle \Delta t \rangle$ is the difference between the total time after the mesh and the total time on the mesh) is shown versus the respective avalanche length (left plot) and the electron multiplicity on the mesh (right plot). The solid curves represent fits by a constant function.
    }  
\label{fig:fig15}
\end{figure}

\begin{figure}
\centering
\begin{minipage}{.33\textwidth}
\centering
\includegraphics[width=1.\textwidth]{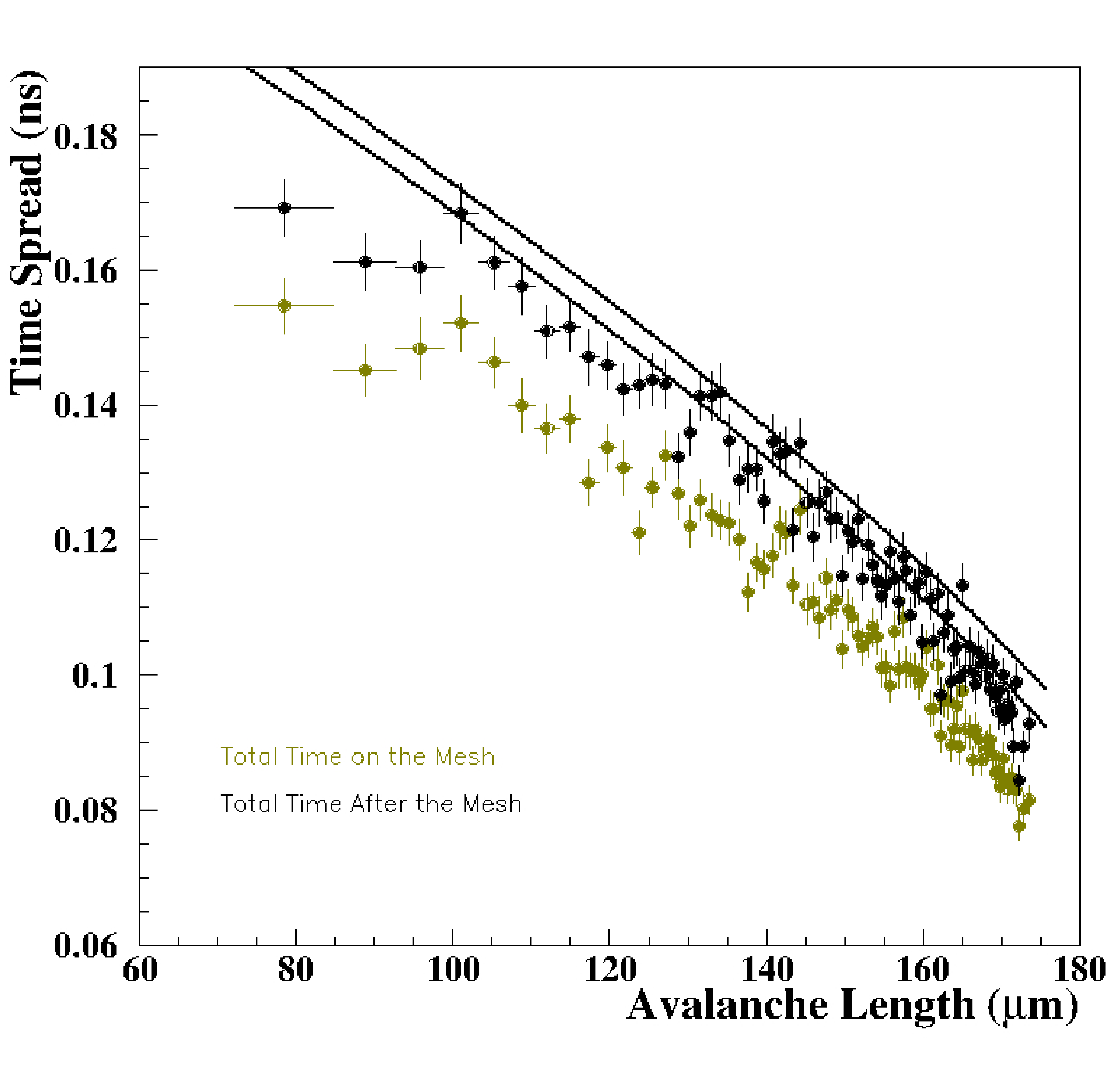}
\end{minipage}%
\begin{minipage}{.33\textwidth}
\centering
\includegraphics[width=1.\textwidth]{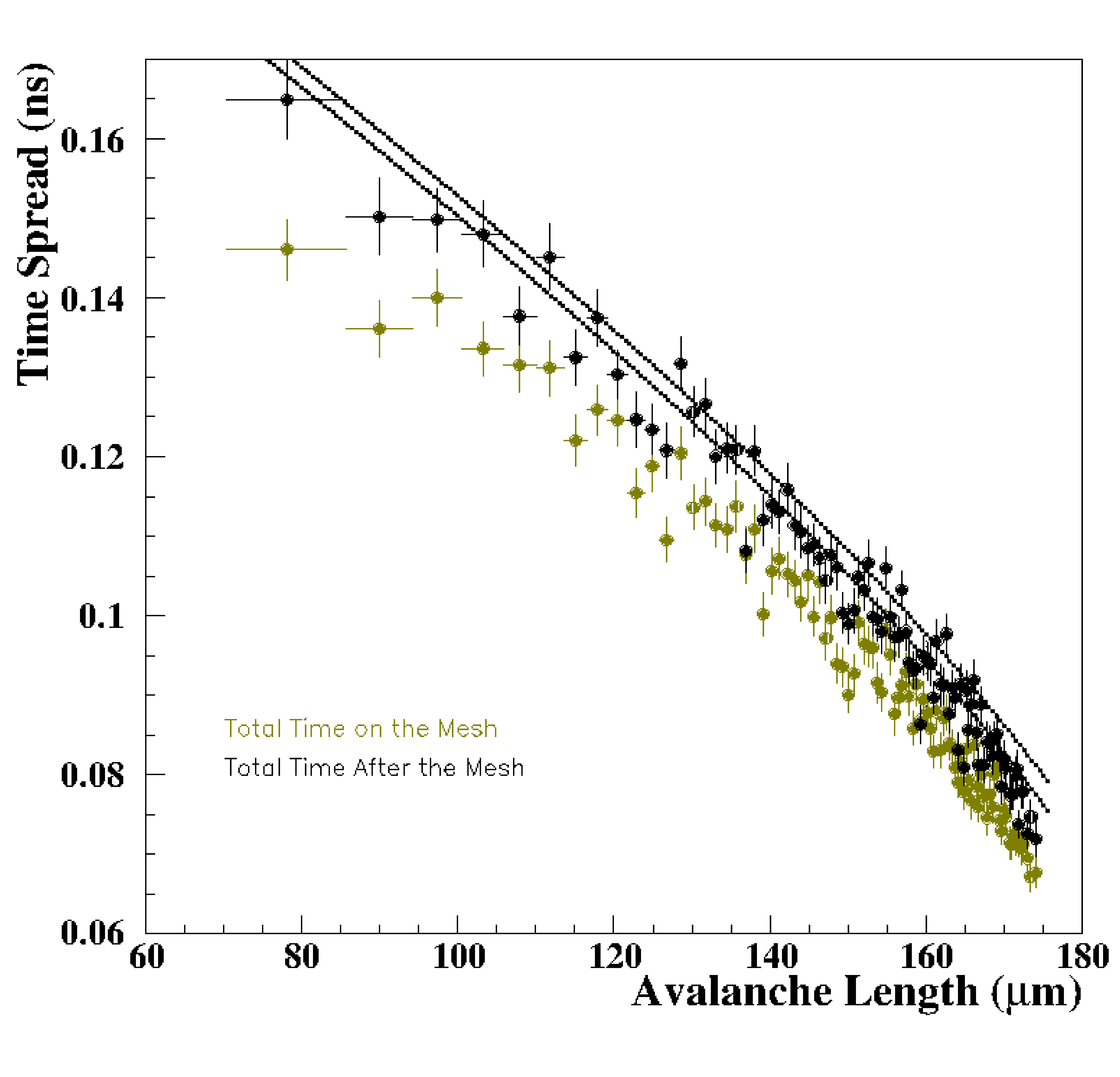}
\end{minipage}%
\begin{minipage}{.33\textwidth}
\centering
\includegraphics[width=1.\textwidth]{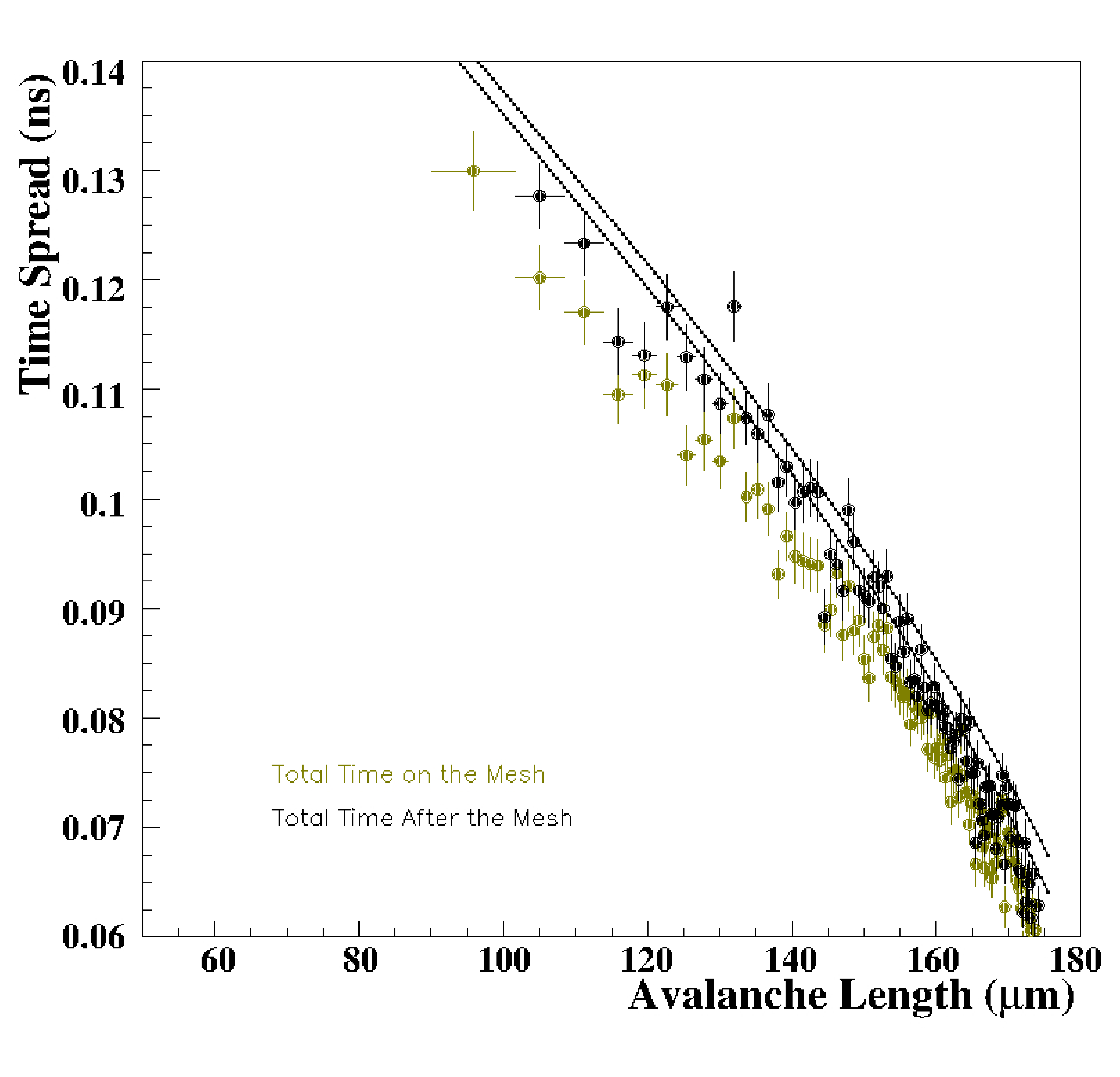}
\end{minipage}
\caption{The points represent GARFIELD++ simulation results concerning the spread of the total time on the mesh (golden points) and the spread of the total time  after the mesh (black points) versus the avalanche length. The solid lines represent predictions based on Eq. \ref{eq:eq68}. The double lines indicate the systematic uncertainty due  to the value of the $w$ parameter, discussed in Section \ref{modelaval3}. The voltage settings considered in these comparisons are: 450 V at the anode and drift voltage of 325 V (left plot), 350 V (center plot), and 400 V (right plot).}
\label{fig:fig16}
\end{figure}

\begin{figure}
\centering
\begin{minipage}{.45\textwidth}
\centering
\includegraphics[width=1.\textwidth]{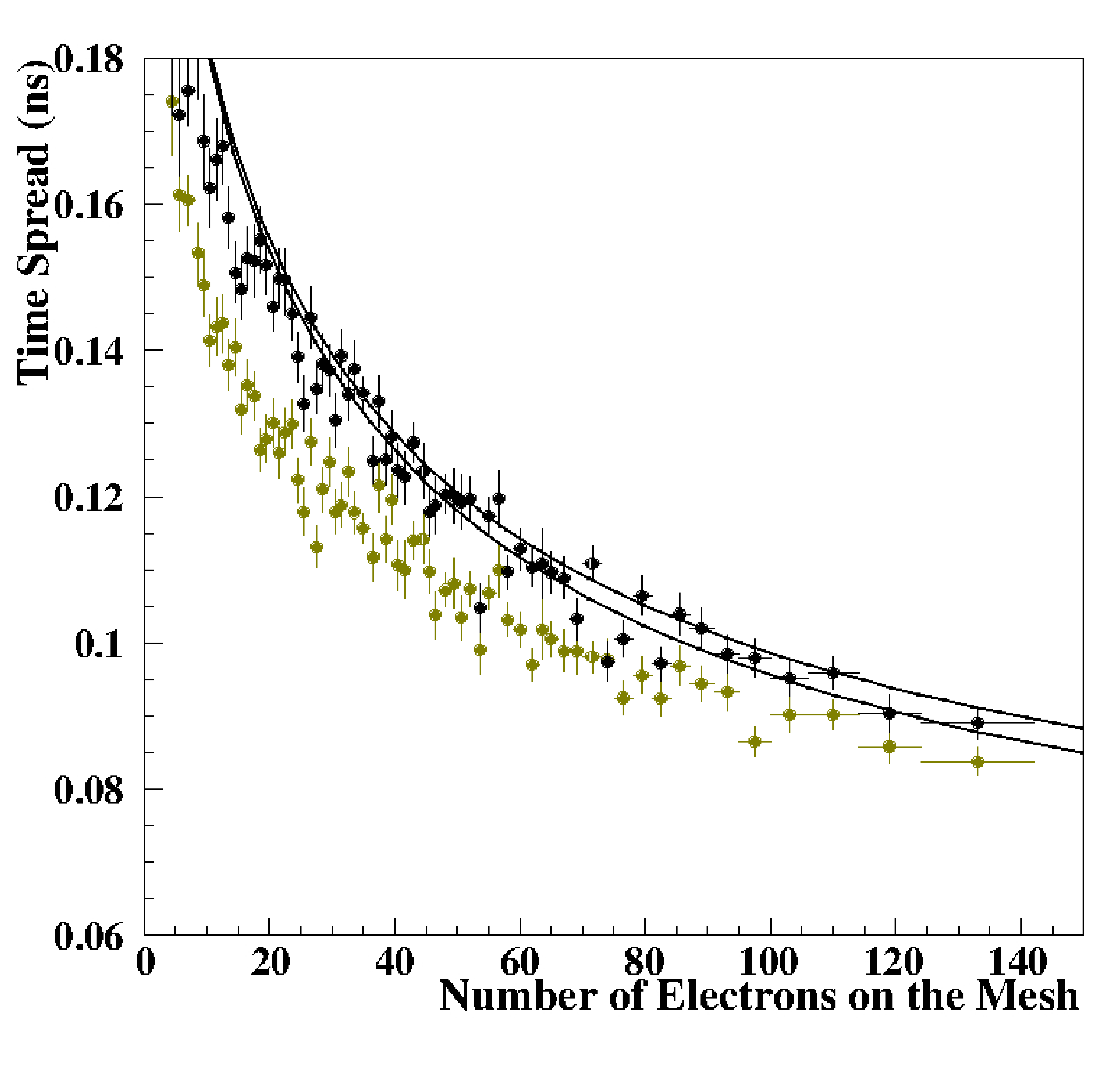}
\end{minipage}
\begin{minipage}{.45\textwidth}
\centering
\includegraphics[width=1.\textwidth]{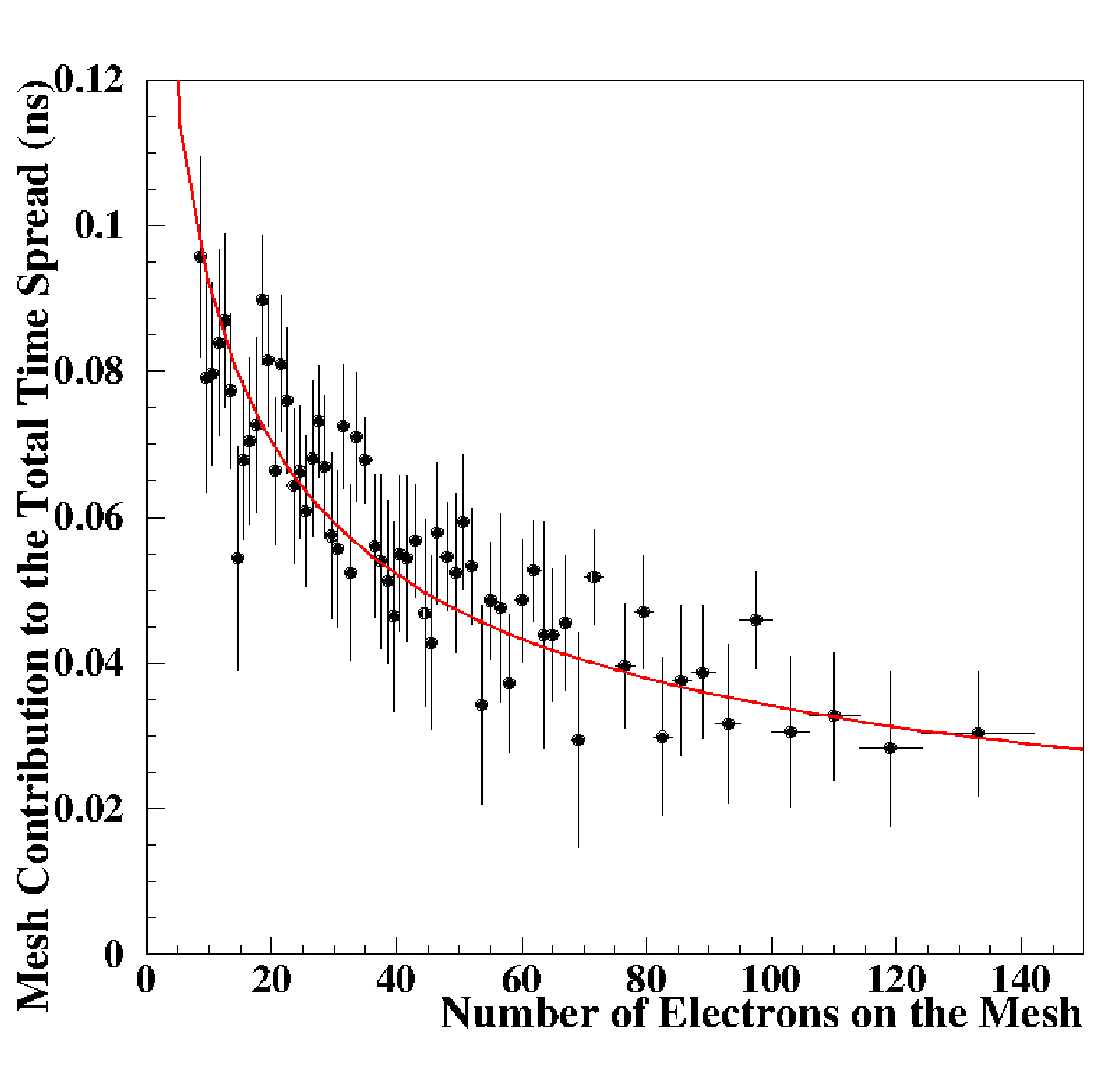}
\end{minipage}\\
\begin{minipage}{.45\textwidth}
\centering
\includegraphics[width=1.\textwidth]{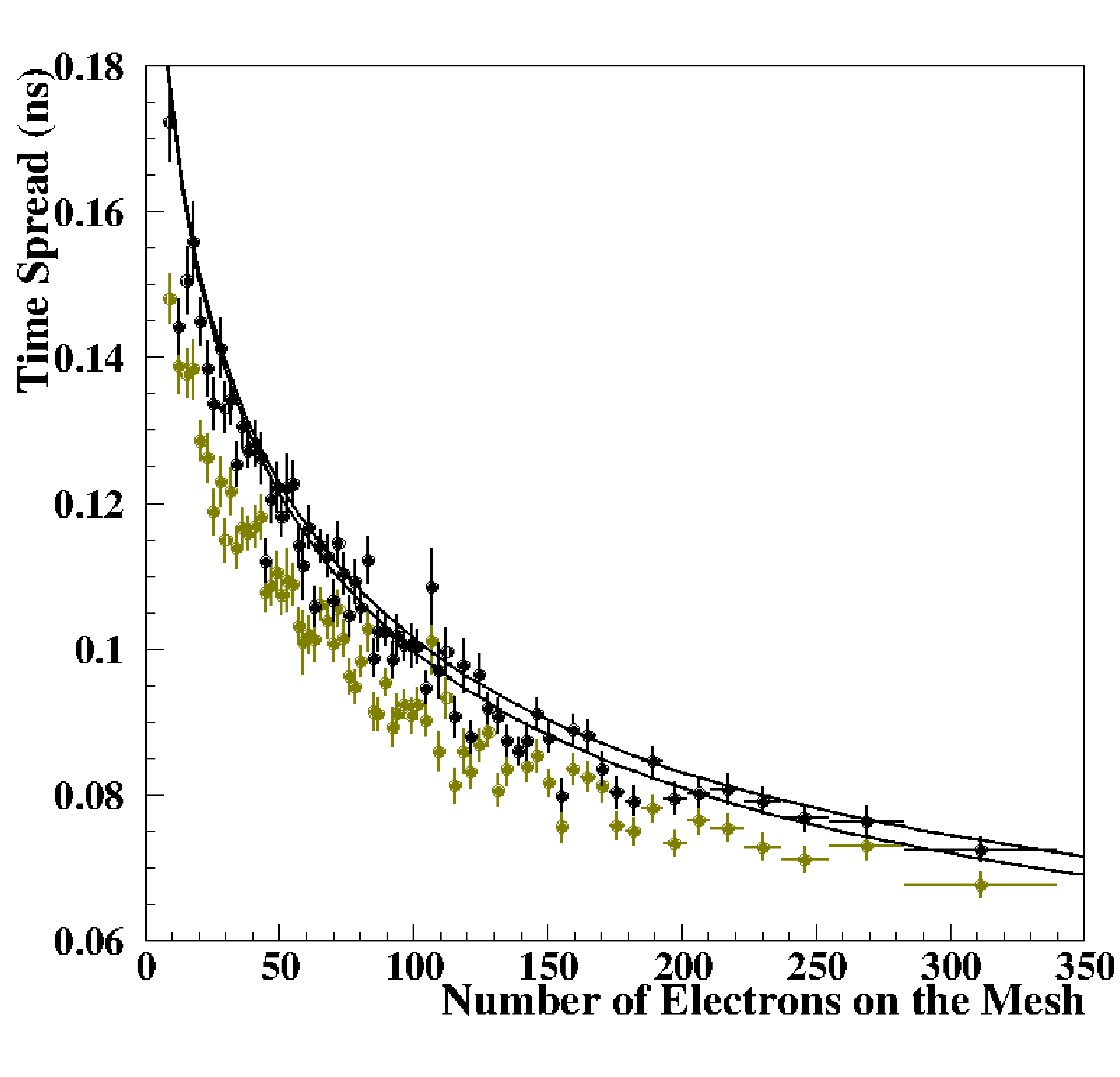}
\end{minipage}
\begin{minipage}{.45\textwidth}
\centering
\includegraphics[width=1.\textwidth]{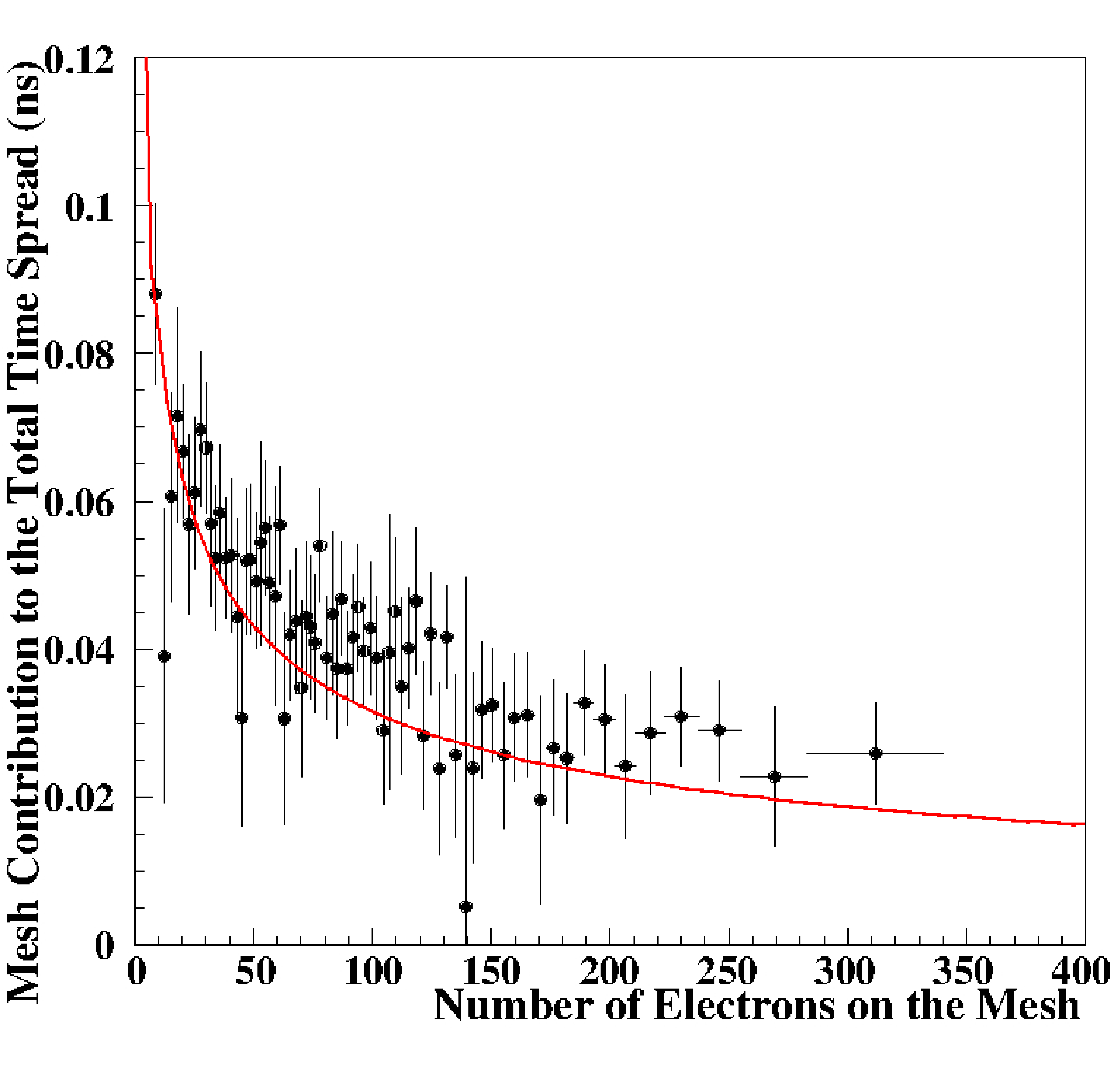}
\end{minipage}\\
\begin{minipage}{.45\textwidth}
\centering
\includegraphics[width=1.\textwidth]{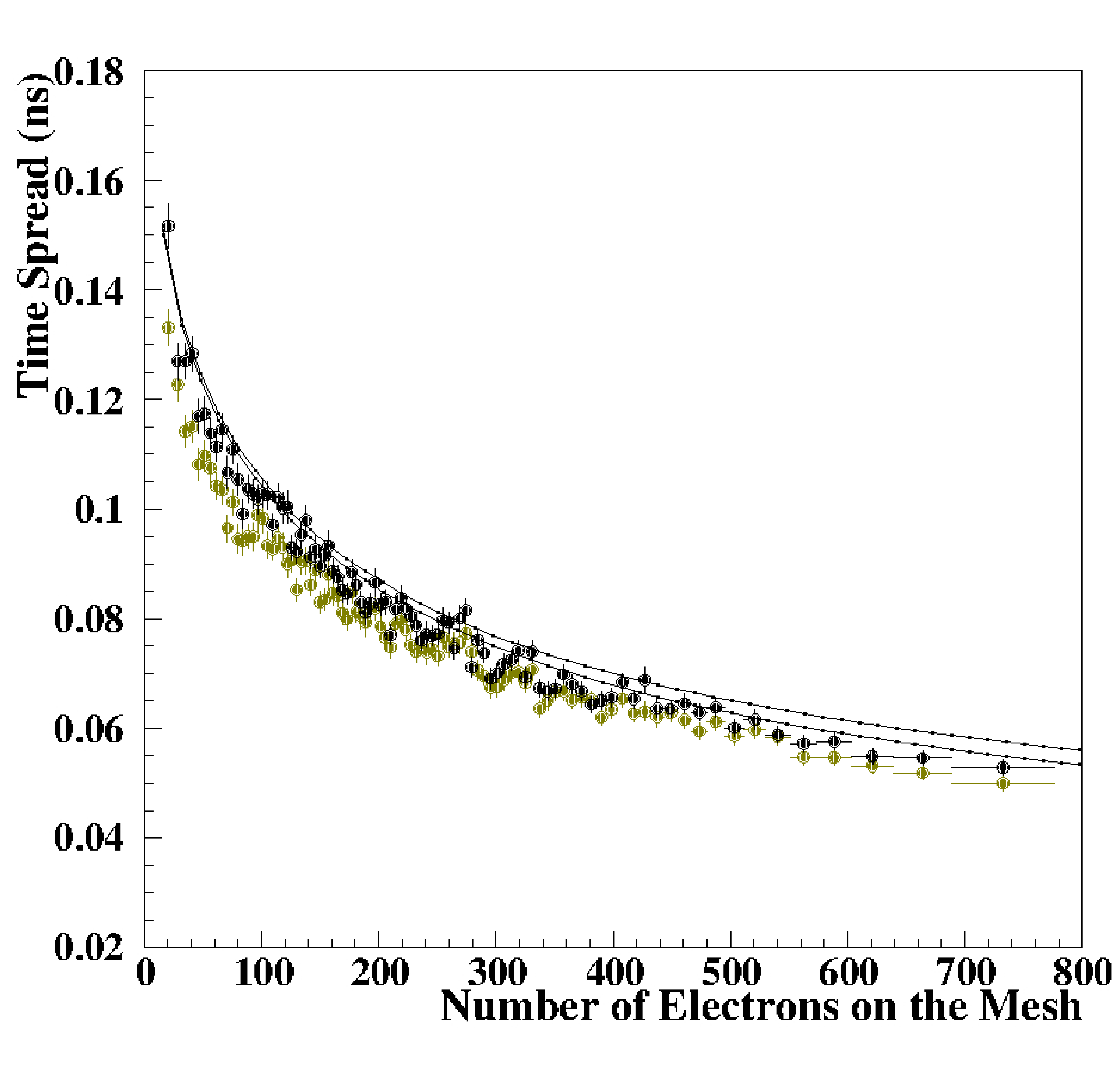}
\end{minipage}
\begin{minipage}{.45\textwidth}
\centering
\includegraphics[width=1.\textwidth]{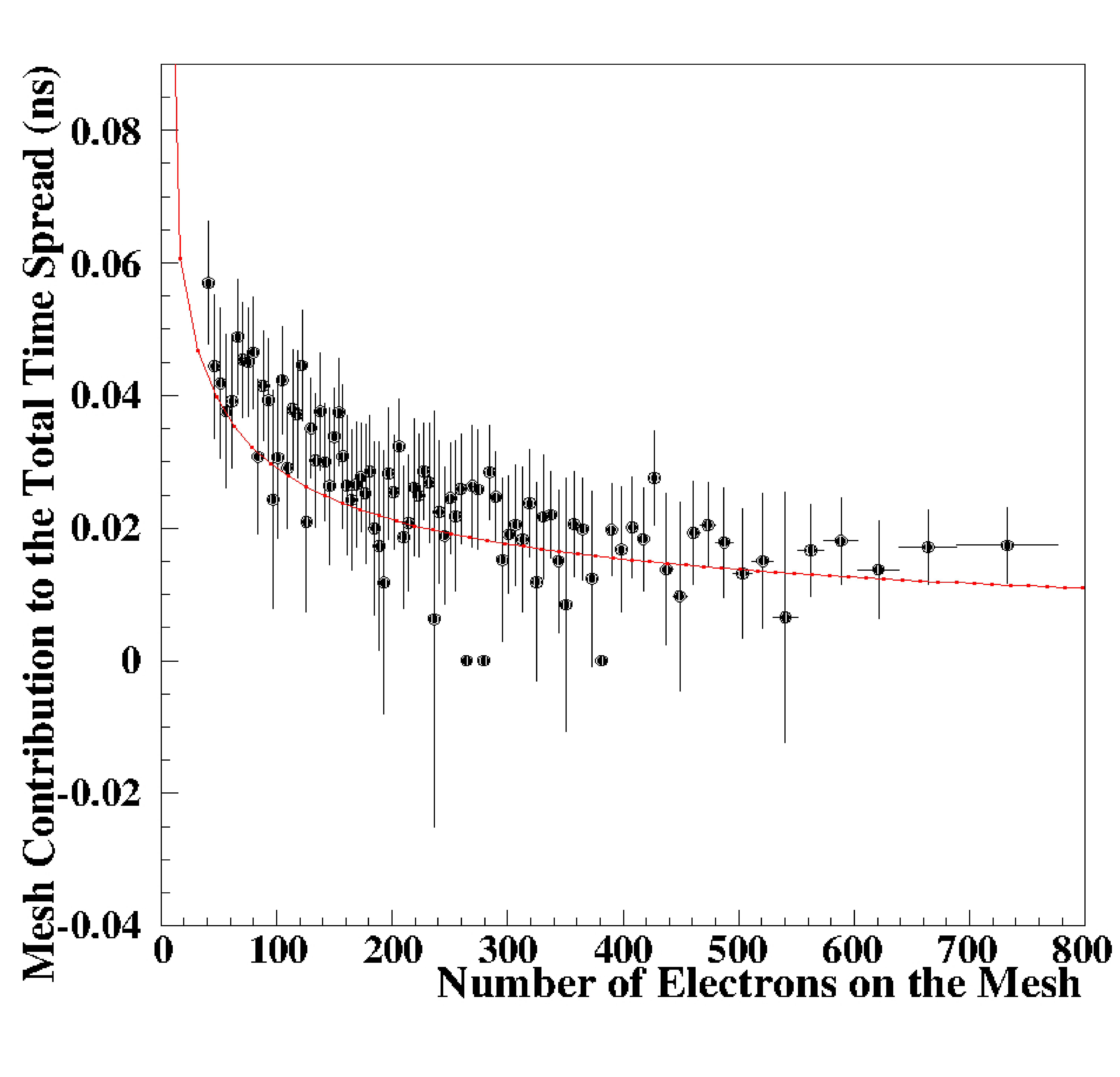}
\end{minipage}
\caption{The points represent GARFIELD++ simulation results. The left column plots show the spread of the total time on the mesh (golden points) and after the mesh (black points) versus the electron multiplicity on the mesh. The right column plots display the mesh contribution (i.e. the square root of the difference between the variance of the total time after and on the mesh) versus the electron multiplicity on the mesh. The solid lines represent  predictions of Eq. \ref{eq:eq70} and Eq. \ref{eq:eq71}. The double lines represent the systematic uncertainty due to the unknown value of the w model-parameter.  The voltage settings considered in these comparisons are 450 V at the anode and drift voltages of 325 V (top row), 350 V (middle row), and 400 V (bottom row).  }
\label{fig:fig17}
\end{figure}

	GARFIELD++ simulations have shown that, for all PICOSEC operational conditions considered in this study, the transport of the pre-amplification electrons through the mesh reduces their multiplicity by  a factor of four that is  independent of the avalanche length and of the electron multiplicity on the mesh (see Fig. \ref{fig:fig8} and related comments in Section \ref{modelaval1}).\\
	 As expected the passage of the electrons through the mesh adds a delay to the signal arrival time. Simulations show that the added delay depends only on the applied drift voltage, being independent of the pre-amplification avalanche length and the electron multiplicity on the mesh, as it is shown in Fig. \ref{fig:fig15}. However, the spread of the total time after the mesh is found to increase relative to the spread of the total time on the mesh, i.e. the process of electrons traversing the mesh deteriorates the PICOSEC timing resolution. This effect depends on the applied drift field, as well as on the avalanche characteristics, as it is shown in Fig. \ref{fig:fig16} and Fig. \ref{fig:fig17}. 
	 Although the mesh transparency ($\approx 25\%$) is found to be insensitive to the considered drift voltages,  this reduction of the number of electrons influences the timing resolution in a drift-voltage dependent way. This fact signifies the importance of the correlation between the individual arrival times of the pre-amplification electrons (on and after the mesh) in determining the mesh effect on the timing resolution.\\

	Consider a pre-amplification avalanche of length L with N electrons arriving on the mesh, and let $T_{tot}$ be the total time on the mesh and $V[T_{tot}]$ be its variance.
	Then,
\begin{equation}\label{eq:eq56}
T_{tot}(L,N)=T(L,N)+T_{p}(L)=\frac{1}{N}\sum\limits_{k=1}^{N}t_{k}+T_{p}(L)
\end{equation}
where $T_{p}$ is the photoelectron transmission time, depending only on its drift length (D-L) as in Eq. \ref{eq:eq18}, and $t_{k} (k=1,2,...,N)$ are the pre-amplification electron arrival times on the mesh, starting from the instant of the first ionization. 
According to Eq. \ref{eq:eq6}, the avalanche arrival time (and consequently the total time on the mesh) is a function of both L and N. 

Since $T_{p}$ is uncorrelated with every one of the $t_k$’s, the variance, $V[T_{tot}]$, is expressed as:
\begin{equation}\label{eq:eq57}
\begin{array}{l}
V[T_{tot}(L,N)]= 
V\Bigg[\frac{1}{N}\sum\limits_{k=1}^{N}t_{k}\Bigg]+V[T_{p}(L)]=
V\Bigg[\frac{1}{N}\sum\limits_{k=1}^{N}t_{k}\Bigg]+\underbrace{\sigma_{p}^{2}\cdot (D-L)+\Phi}_{V[T_{p}(L)]}
\end{array}
\end{equation}
where $V[T_{p}(L)]=\sigma_{p}^{2}\cdot (D-L)+\Phi$ is, according to Eq. \ref{eq:eq21},  the time variance of the photoelectron at the point of the first ionization. 
As discussed in Section 5, the arrival times of the pre-amplification electrons are heavily inter-correlated. 
The first term in Eq. \ref{eq:eq57} is expressed analytically as:
\begin{equation}\label{eq:eq58}
\begin{array}{l}
V\Bigg[\frac{1}{N}\sum\limits_{k=1}^{N}t_{k}\Bigg]=
%\dfrac{1}{N^2}\Bigg(\sum\limits_{k=1}^{N}V_{k}^{2}+\sum\limits_{i=1}^{N}\sum\limits_{j=1,j\neq i}^{N}\big(\big<t_{i}t_{j}\big>-\big<t_i\big>\big<t_j\big>\big)\Bigg)\\
%=\dfrac{1}{N^2}\Bigg(N\sigma_{0}^{2}\cdot L+\sum\limits_{i=1}^{N}\sum\limits_{j=1,j\neq i}^{N}C_{ij}\Bigg)=
\dfrac{\sigma_{0}^{2}\cdot L}{N}+\dfrac{1}{N^2}\sum\limits_{i=1}^{N}\sum\limits_{j=1,j\neq i}^{N}C_{ij}
\end{array}
\end{equation}
where $\sigma_{0}^{2}$  is defined in Section 5 as the variance per unit length of a single electron in the avalanche, and $C_{ij}$ symbolizes the covariance between the arrival times of the $i^{th}$ and $j^{th}$  avalanche electrons. 

	Ignoring any new electron production while  traversing the mesh, let M be the number of electrons entering the amplification region, $T_{m}$ be the total arrival time  after passing the mesh (i.e. the average of the M arrival times on a plane just after the mesh) and $V[T_m]$  be the corresponding variance. 
	Then,
\begin{equation}\label{eq:eq59}
\begin{array}{l}
T_{m}(L,N)=
%\frac{1}{M}\sum\limits_{k=1}^{M}(t_{k}+\Delta t_{k})+T_{p}(L)\\
\frac{1}{M}\sum\limits_{k=1}^{M}t_{k}+\frac{1}{M}\sum\limits_{k=1}^{M}\Delta t_{k}+T_{p}(L)
\end{array}
\end{equation}
where $\Delta t_k$  is the extra time needed by the $k^{th}$ electron to arrive at the plane just after the mesh. 
Assuming that each of the N electrons arriving on the mesh has the same probability, M/N, to pass through the mesh\footnote{Indeed, the passage of an electron through the mesh is determined by the position of its impact point on the mesh; consequently, if the same avalanche is shifted parallel to its longitudinal axis, a different subset of the N arriving electrons will pass through the mesh. This is equivalent to giving the same  probability, M/N, to each of the N arriving electrons to pass through the mesh.}, Eq. \ref{eq:eq59} is written as:
\begin{equation}\label{eq:eq60}
\begin{array}{l}
T_{m}(L,N)= \frac{1}{M}\frac{M}{N} \sum\limits_{k=1}^{N}t_{k} + \frac{1}{M}\frac{M}{N}\sum\limits_{k=1}^{N}\Delta t_{k}+T_{p}(L)
%=\frac{1}{N}\sum\limits_{k=1}^{N}t_{k}+\frac{1}{N}\sum\limits_{k=1}^{N}\Delta t_{k}+T_{p}(L)
=T_{tot}(L,N)+<\Delta t>
\end{array}
\end{equation}
where $<\Delta t>$ is the mean time needed by an electron to pass through the mesh. Eq. \ref{eq:eq60} predicts that the total arrival time after the mesh is the total arrival time on the mesh delayed by a constant time, which is independent of the avalanche characteristics, as observed in the detailed GARFIELD++ simulation. Naturally  $<\Delta t>$, being the drift time of an electron traversing the mesh,   depends on the electric field around the mesh.\\

	Due to the fact that the terms, $\frac{1}{M}\sum\limits_{k=1}^{M}t_k$, $\frac{1}{M}\sum\limits_{k=1}^{M}\Delta t_k$  and $T_p$, in Eq. \ref{eq:eq59}, are mutually uncorrelated, the variance of the total time after the mesh is expressed as:
\begin{equation}\label{eq:eq61}
V[T_{m}(L,N)]=V\Bigg[\frac{1}{M}\sum\limits_{k=1}^{M}t_k\Bigg]+V\Bigg[\frac{1}{M}\sum\limits_{k=1}^{M}\Delta t_{k}\Bigg]+V[T_p\big[(L\big)]
\end{equation}
The first term in Eq. \ref{eq:eq61} is written, in analogy to Eq. \ref{eq:eq58}, as:
\begin{equation}\label{eq:eq62}
V\Bigg[\frac{1}{M}\sum\limits_{k=1}^{M}t_{k}\Bigg]=\dfrac{\sigma_{0}^{2}\cdot L}{M}+ \dfrac{1}{M^2}\sum\limits_{i=1}^{M}\sum\limits_{j=1,j\neq i}^{M}C_{ij}
\end{equation}
where $C_{ij}$ are defined in Eq. \ref{eq:eq58}.\\ 
 Eq. \ref{eq:eq62} can be simplified  by exploring further the observation that any one of the  pre-amplification electrons has the same probability to traverse the mesh. Then, noticing that  the covariance term, $\sum\limits_{i=1}^{M}\sum\limits_{j=1,j\neq i}^{M}C_{ij}$, in Eq. \ref{eq:eq62} comprises $M(M-1)$ $C_{ij}$ terms while the corresponding term in Eq. \ref{eq:eq58} is the sum of $N(N-1)$  $C_{ij}$ terms,  Eq. \ref{eq:eq62} can be approximated as:
\begin{equation}\label{eq:eq63}
\begin{array}{l}
V\Bigg[\dfrac{1}{M}\sum\limits_{k=1}^{M}t_{k}\Bigg]=\dfrac{\sigma_{0}^{2}\cdot L}{M}+ \dfrac{1}{M^2}\dfrac{M(M-1)}{N(N-1)}\sum\limits_{i=1}^{N}\sum\limits_{j=1,j\neq i}^{N}C_{ij}\\
 \simeq \dfrac{\sigma_{0}^{2}\cdot L}{M}+ \dfrac{1}{N^2}\sum\limits_{i=1}^{N}\sum\limits_{j=1,j\neq i}^{N}C_{ij}
\end{array}
\end{equation}
Due to the fact that the times $\Delta t_k$ are mutually uncorrelated, the second term in Eq. \ref{eq:eq61}, is written as:
\begin{equation}\label{eq:eq64}
V\Bigg[\dfrac{1}{M}\sum\limits_{k=1}^{M}\Delta t_{k}\Bigg]=\dfrac{\delta^2}{M}
\end{equation}
where $\delta^2$ is the variance of the time taken by an electron to pass  through the mesh. Substituting Eq. \ref{eq:eq63} and \ref{eq:eq64} into Eq. \ref{eq:eq61}, the variance of the total time after the mesh is expressed as:
\begin{equation}\label{eq:eq65}
V[T_{m}(L,N)]=\dfrac{\sigma_{0}^{2}\cdot L}{M}+\dfrac{1}{N^2}\sum\limits_{i=1}^{N}\sum\limits_{j=1,j\neq i}^{N}C_{ij}+\dfrac{\delta^2}{M}+V[T_{p}(L)]
\end{equation}
Subsequently, Eq. \ref{eq:eq58} is used to eliminate the double sum of the covariance terms and the variance of the total time after the mesh is expressed by the following formula:
\begin{equation}\label{eq:eq66}
\begin{array}{l}
V[T_{m}(L,N)]
%=\dfrac{\sigma_{0}^{2}\cdot L}{M}+\Bigg(\dfrac{\sigma_{0}^{2}\cdot L}{N}-\dfrac{\sigma_{0}^{2}\cdot L}{N}\Bigg)+\dfrac{1}{N^2}\sum\limits_{i=1}^{N}\sum\limits_{j=1,j\neq i}^{N}C_{ij}+\dfrac{\delta^2}{M}+V[T_{p}(L)]\\
%=\dfrac{\sigma_{0}^{2}\cdot L}{M}-\dfrac{\sigma_{0}^{2}\cdot L}{N}+\dfrac{\delta^2}{M}+\Bigg(\dfrac{\sigma_{0}^{2}\cdot L}{N}+\dfrac{1}{N^2}\sum\limits_{i=1}^{N}\sum\limits_{j=1,j\neq i}^{N}C_{ij}+V[T_{p}(L)]\Bigg)\\
=\sigma_{0}^{2}\cdot L\big(\dfrac{1}{M}-\dfrac{1}{N}\big)+\dfrac{\delta^2}{M}+V[T_{tot}(L,N)]
\end{array}
\end{equation}
The average ratio M/N expresses the electron transparency, $tr$, of the mesh, which retains the same mean value at all the operational conditions considered in this work. 
Using the mesh transparency to eliminate M, Eq. \ref{eq:eq66} is simplified to:
\begin{equation}\label{eq:eq67}
V[T_m\big(L,N\big)]=\dfrac{1}{N}\Bigg[\sigma_{0}^{2}\cdot L\big(\dfrac{1}{tr}-1\big)+\dfrac{\delta^2}{tr}\Bigg]+V[T_{tot}\big(L,N\big)]
\end{equation} 
Eq. \ref{eq:eq67} predicts an increase of the total time variance,   $V[T_{m}(L,N)]-V[T_{tot}(L,N)]$, which  depends on the electron multiplicity, N, on the electron transparency of the mesh, $tr$, and on the avalanche length L. \\

By averaging properly Eq. \ref{eq:eq67} over all possible N, i.e. following the procedure described in Section \ref{modelaval3}, the variance of the total time after the mesh is expressed as a function of the avalanche length as:

\begin{equation}\label{eq:eq68}
V[T_{m}(L)]=\big<V[T_{m}(L,N)]\big>_{N}=\dfrac{\theta+1}{2\theta}\exp\big[-a_{eff}L\big]\cdot\Bigg[\sigma_{0}^{2}\cdot L\big(\dfrac{1}{tr}-1\big)+\dfrac{\delta^2}{tr}\Bigg]+V\big[T_{tot}(L)\big]
\end{equation}
where the Gamma distribution property  $\big<\frac{1}{N}\big>=\frac{\theta+1}{\theta<N>}=\frac{\theta+1}{2\theta}\exp[-a_{eff}L]$ is used, and the last term, $V[T_{tot}(L)]=<V[T_{tot}(L,N)]>_N$, is given by Eq. \ref{eq:eq44}. \\
Consequently, the mesh contribution to the total time variance, which determines the PICOSEC timing resolution, is given in terms of the avalanche length as:
\begin{equation}\label{eq:eq69}
\Delta V(L)=V[T_{m}(L)]-V[T_{tot}(L)]=\dfrac{\theta+1}{2\theta}\exp[-a_{eff}L]\cdot\Bigg[\sigma_{0}^{2}\cdot L\big(\dfrac{1}{tr}-1\big)+\dfrac{\delta^2}{tr}\Bigg]
\end{equation}

The variance of the total time after the mesh can be also expressed   as a function of the electron multiplicity on the mesh, having properly averaged Eq. \ref{eq:eq67} over all possible avalanche lengths, as:
\begin{equation}\label{eq:eq70}
V\big[T_{m}(N)\big] = \big<V[T_{m}(L,N)]\big>_{L} = \dfrac{1}{N}\Bigg[\sigma_{0}^{2}\cdot\big<L(N)\big>\big(\dfrac{1}{tr}-1\big)+\dfrac{\delta^2}{tr}\Bigg]+V[T_{tot}(N)]
\end{equation}
where the last term, $V[T_{tot}(N)] =\big<V[T_{tot}(L,N)]\big>_L$, is given by Eq. \ref{eq:eq55} and the averaged length $\big<L(N)\big>=(\int\limits_{x_{1}}^{x_{2}}L\cdot G(L\vert N) dL$)  is defined in Section 4.  
Then, the mesh contribution to the PICOSEC resolution is expressed as function of N, as:  
\begin{equation}\label{eq:eq71}
\Delta V(N)=V[T_{m}(N)]-V[T_{tot}(N)]=\dfrac{1}{N}\Bigg[\sigma_{0}^{2}\cdot\big<L(N)\big>\big(\dfrac{1}{tr}-1\big)+\dfrac{\delta^2}{tr}\Bigg]
\end{equation}
Eq. \ref{eq:eq70} and Eq. \ref{eq:eq71} can be easily reformulated as functions of the number, M, of the electrons that pass through the mesh, by using the transformation $M=tr\cdot N$; recall that the PICOSEC e-peak amplitude was found \citep{kostas}  proportional to M (see also Fig. \ref{fig:fig3}).\\
	
	In the above description of the electron transport through the mesh two sources contribute to the increase of the time variance: i) an extra time spread due to the electron drift in the inhomogeneous electric field around the mesh,  and ii) the statistical effect caused by the depletion of the number of mutually-correlated avalanche electrons.
	 The first contribution  is expressed by the term proportional to $\delta^2$  in Eq. \ref{eq:eq67} or equivalently in Eq. \ref{eq:eq69} and Eq. \ref{eq:eq71}.
	The time-spread $\delta$ depends on the PICOSEC operational conditions and it is treated as an input parameter in this model. 
	Values of $\delta$, which are  evaluated using GARFIELD++ simulations, assuming  several drift voltages,  are compiled in Table \ref{tab:tableA-8}, exhibiting  a decreasing functional dependence on the drift voltage.
	However, the terms  proportional to $\delta^2$ contribute  to the increase of the time variance (e.g. in Eqs. \ref{eq:eq67},  Eq. \ref{eq:eq69} and Eq. \ref{eq:eq71})  much weaker  than the other terms, which are related to statistical correlations. \\
	Due to  correlation terms, the variance of the total-time after the mesh (Eq. \ref{eq:eq66}) is not proportional to the variance of the total time on the mesh. 
	The mesh adds to the variance of the total time on the mesh  a term which is almost proportional to $L\cdot\exp[-a_{eff}L]$ when expressed as a function of L (see Eq. \ref{eq:eq69}), or almost proportional to $\frac{<L(N)>}{N}$ (see Eq. \ref{eq:eq71})  when expressed as a function of N. 
	As the drift voltage increases and the electron multiplication factor, $a_{eff}$, increases, both the above terms\footnote{In case that the electron multiplication factor increases, the average length of the avalanches that produce  N pre-amplification electrons, $<L(N)>$, decreases.} decrease for all L and N.  Thus,  the mesh influence on the timing resolution weakens as the drift field increases, as the GARFIELD++ simulations demonstrate. \\
	The model is also in  good agreement with the GARFIELD++ simulations in describing quantitatively the mesh effect on the timing resolution, for all the  PICOSEC operational conditions  considered in this work, as it is demonstrated in Figs. \ref{fig:fig16} and \ref{fig:fig17}.

%	Because the above terms are decreasing functions of L and N, while 
%	in connection to the fact that 
%	the average avalanche length and the average avalanche electron multiplicity are increasing functions of the drift field, it is expected that the observed influence of the  mesh to the resolution decreases at higher drift voltages. 
%	Indeed, as demonstrated in Fig. \ref{fig:fig16} and \ref{fig:fig17}, both the GARFIELD++ simulation and the model prediction agree that the contribution of the electron transport through the mesh to the PICOSEC  timing resolution is diminishing at high drift voltages.

\section{Discussion} \label{discl}
%This work employs the comparison of experimental data with detailed simulations, based on the  GARFIELD++ package and complemented with a statistical description of the electronic signal formation, to identify the microscopic quantities that determine the PICOSEC timing characteristics.
%	Subsequently, a stochastic model is developed that describes the  properties of the above  quantities, offering a phenomenological, microscopic interpretation of the observed timing properties of the detector. \\
%	The model is based on: i) the fact that an electron drifting in a gas under the influence of an homogeneous electric field achieves higher drift velocity when, in addition to  elastic scattering, undergoes inelastic interactions, and ii) the assumption that a newly produced electron through ionization acquires  a certain time-gain relative to its parent and subsequently drifts with the same velocity as the parent electron.  
%		The input parameters, compiled in Table \ref{tab:tableA-8}, are commonly used statistical variables\footnote{With the only exception of the time-gain parameter $\rho$, which has been introduced in this work.}, which have been evaluated by analysing GARFIELD++ simulation results. \\
		
\begin{figure}
\centering
\begin{minipage}{.34\textwidth}
\centering
\includegraphics[width=1.\textwidth]{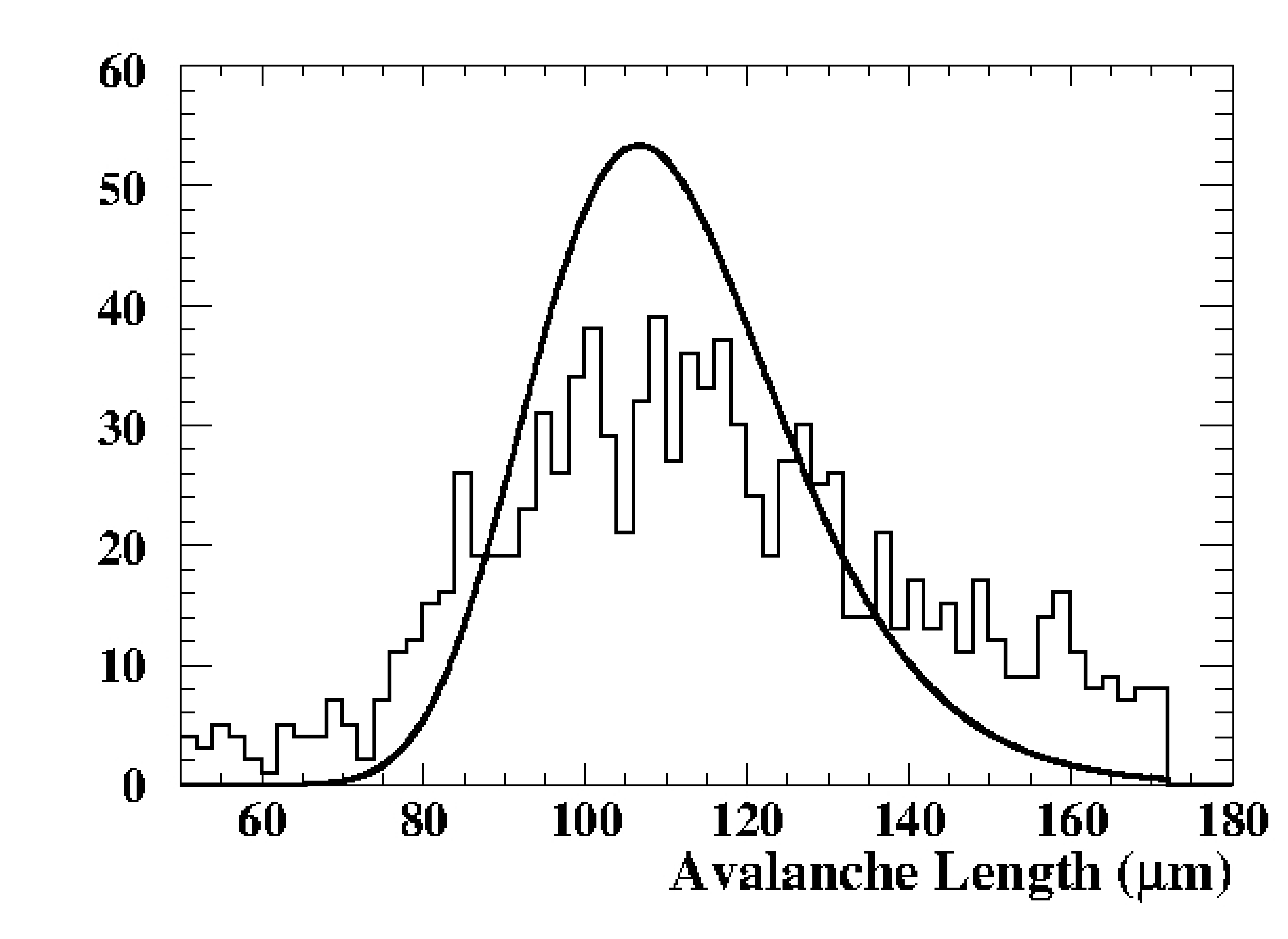}
\end{minipage}%
\begin{minipage}{.33\textwidth}
\centering
\includegraphics[width=1.\textwidth]{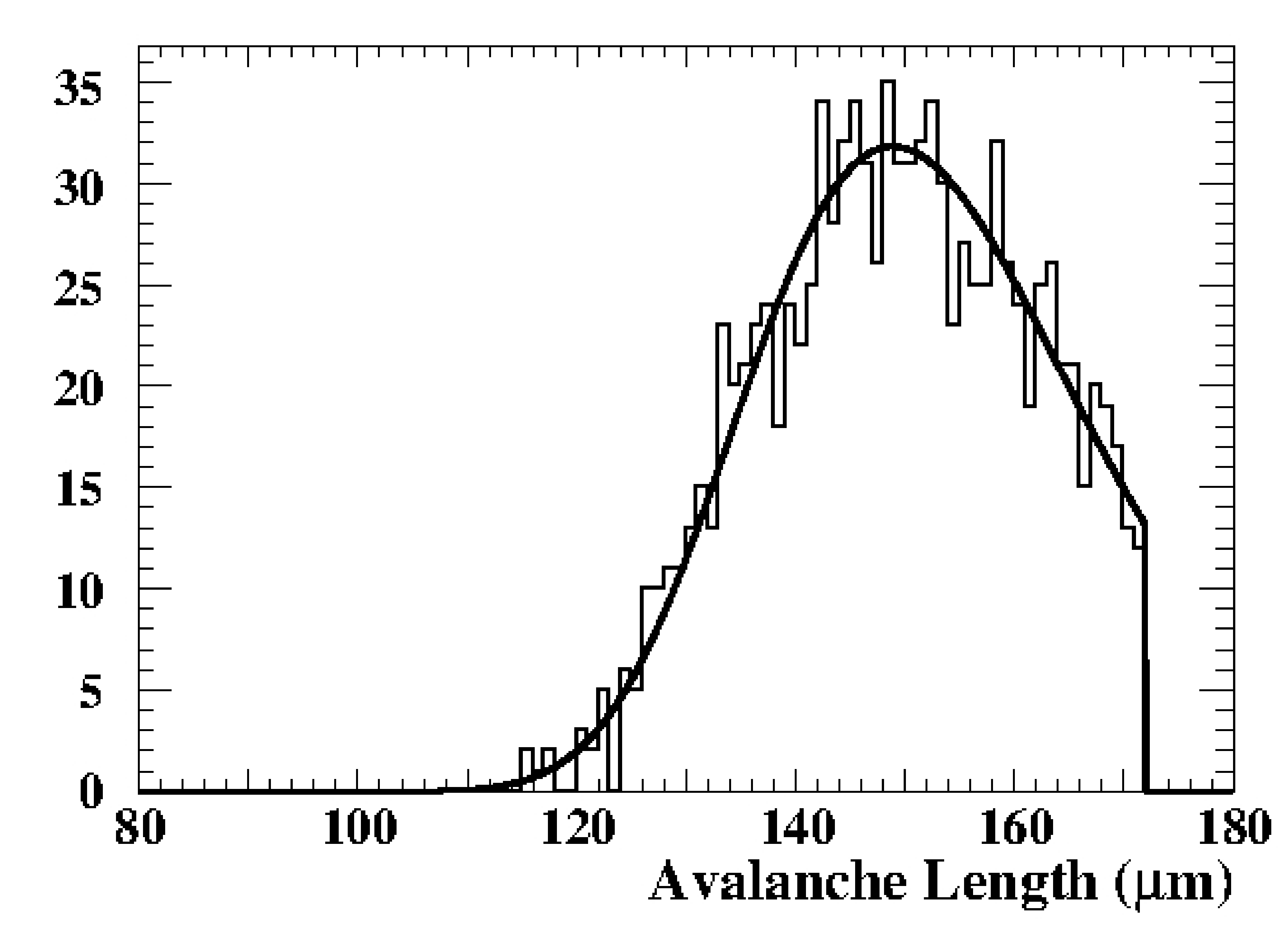}
\end{minipage}%
\begin{minipage}{.33\textwidth}
\centering
\includegraphics[width=1.\textwidth]{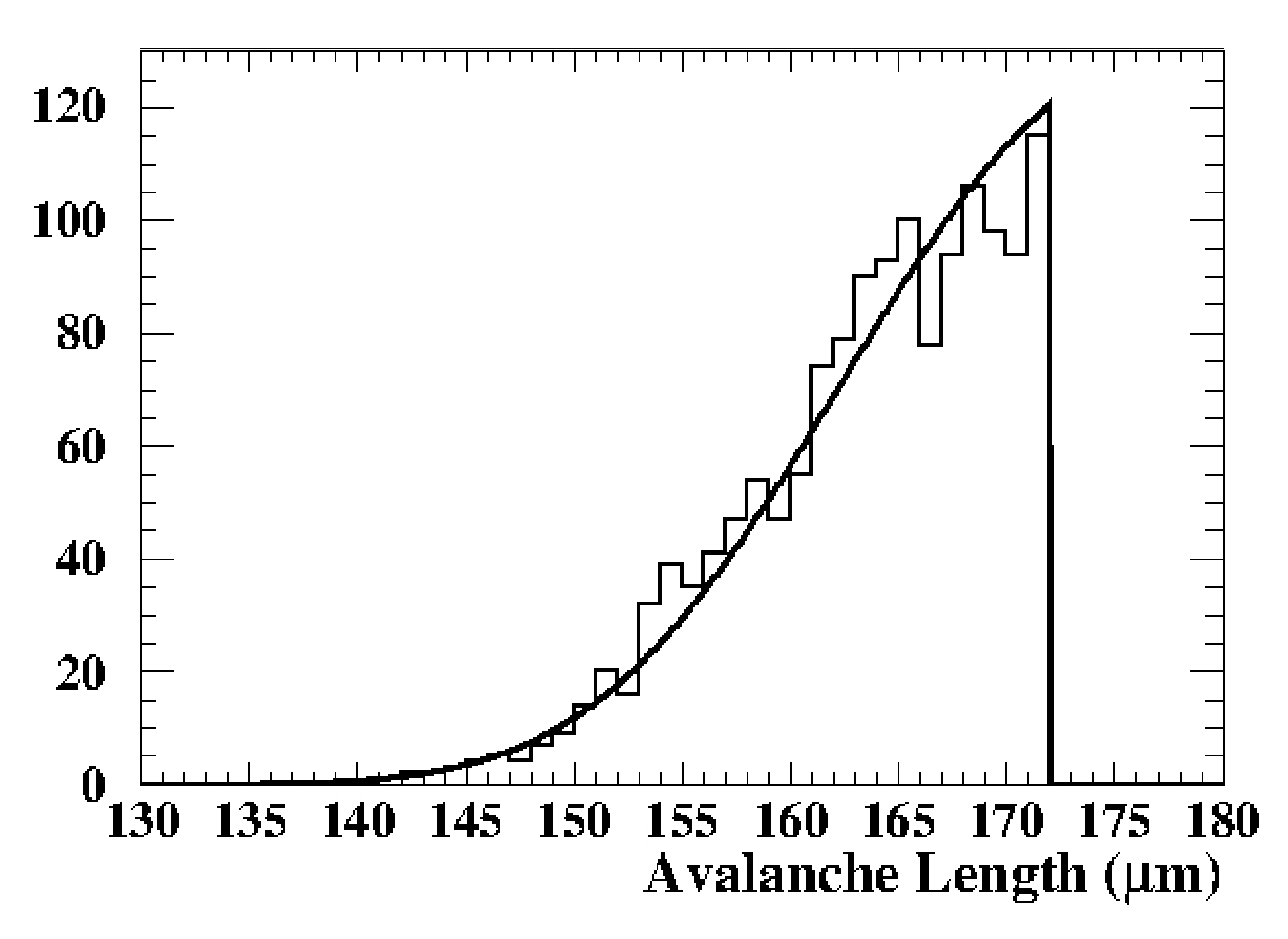}
\end{minipage}
\caption{Distributions of the avalanche length, produced by GARFIELD++ simulations (assuming 50\% Ptr, 425 V  and 450 V drift and anode voltage, respectively) in the case that the multiplicity of pre-amplification electrons is less than $120$ (left plot), between $400$ and $440$ (center plot) and $1230$ and $1300$ (right plot). 
%In the above distributions, only simulated events that the avalanche started further than $10\,\microm$ from the photocathode ($D-L>10\,\microm$) have been used. 
The solid lines represent the related predictions of the distribution function $G(L\vert N)$ defined by Eq. \ref{eq:eq16}. }
\label{fig:fig18}
\end{figure}

%The quantitative predictions of the model have been compared extensively with the related  GARFIELD++ simulation results  and found in a very good agreement at all operating PICOSEC conditions considered in this study, demonstrating the success of this stochastic interpretation.  
	A weak but systematic deviation of the model predictions from the GARFIE\-LD++ results has been observed at low electron multiplicities on the mesh.  
	Indeed, as shown in Fig. \ref{fig:fig10} and \ref{fig:fig14}, the model predictions of the mean value and the spread of avalanche time deviate from the GARFIELD++ points at avalanche electron multiplicities less than 300, for 50\% Ptr, 425 V drift and 450 V anode voltages. 
	As already stated, such deviations result from the inadequacy of the employed p.d.f.'s to approximate accurately the avalanche statistical properties at its very beginning (small avalanche length, low electron multiplicity). 
	As an example, the model predictions of both the mean value and the variance of the avalanche time, i.e. Eq. \ref{eq:eq17} and Eq. \ref{eq:eq53}, utilize the function $G(L\vert N)$. 
	Recall that this conditional p.d.f., defined in Section 4 by Eq. \ref{eq:eq16}, expresses the distribution of the length of an avalanche given that the avalanche electron multiplicity is N.  
	Predictions of Eq. \ref{eq:eq16} are compared to the respective distributions produced by GARFIELD++, in Fig. \ref{fig:fig18}. 
	Apparently, Eq. \ref{eq:eq16} approximates poorly the GARFIELD++ distributions at low $N$ but successfully describes the detailed-simulation results for higher values of electron multiplicity. 
	Therefore, the predictions of Eq. \ref{eq:eq17} and \ref{eq:eq53} suffer from the poor success of $G(L\vert N)$ to describe the GARFIELD++ results at low electron multiplicities.

	However, for practical reasons, PICOSEC data are collected with non-zero experimental, amplitude thresholds. 
	The data points shown in Fig \ref{fig:fig2}, in comparison with results based on simulated PICOSEC pulses,  were collected \citep{pico24} with  thresholds corresponding to e-peak charge greater than 3-4 pC, which translate (for 425 V drift and 450 V anode voltages, and 50\% Ptr) to $400-500$ electron multiplicity on the mesh. 
	At this  region of pre-amplification electron multiplicities, the model predictions are in an excellent agreement with the results of GARFIELD++ simulations, as shown in Fig. \ref{fig:fig10} and \ref{fig:fig14}. \\

\begin{figure}[h]
\centering\includegraphics[width=0.9\linewidth]{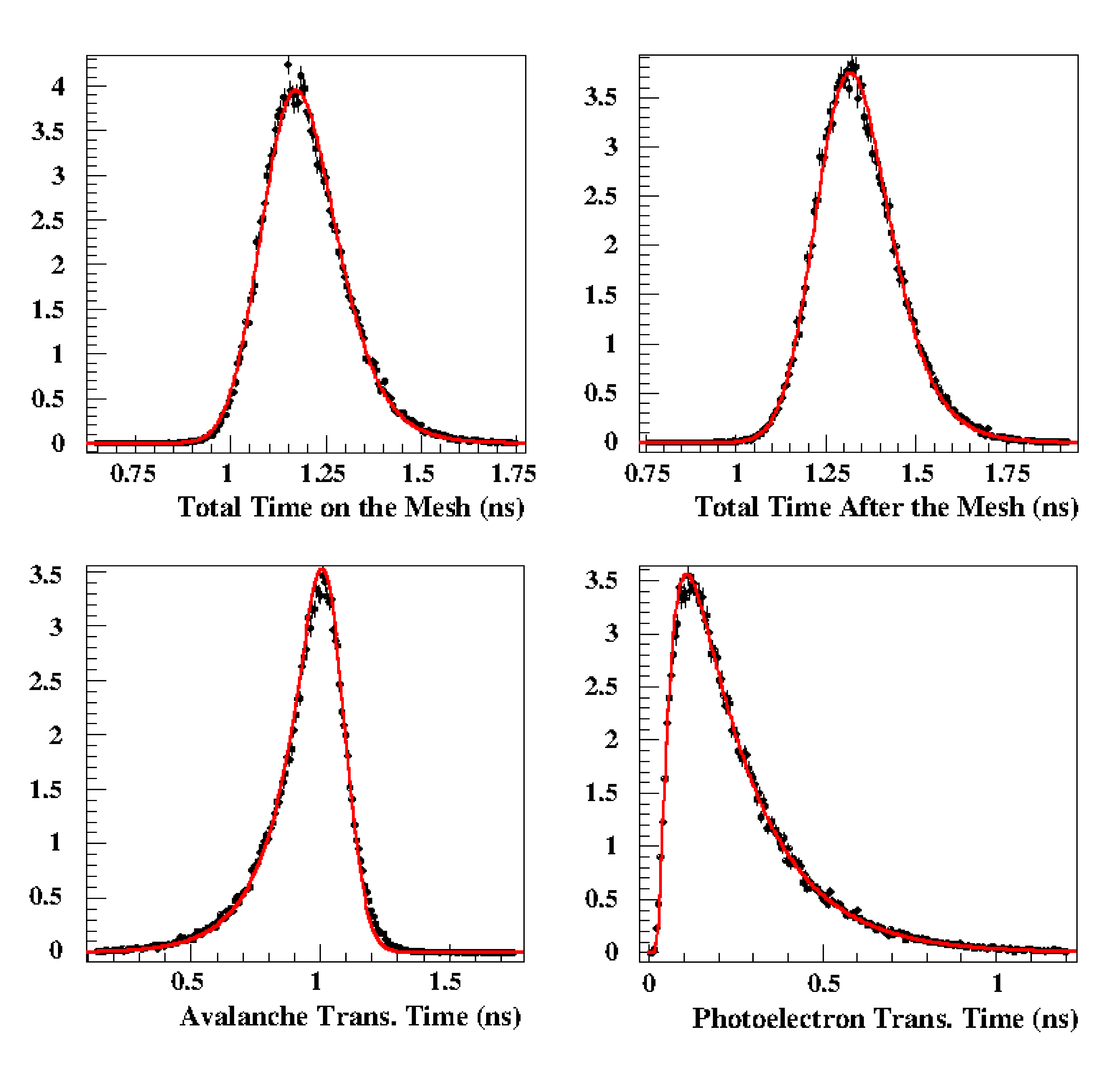}
\caption{Transmission time distributions for all events at 350 V and 450 V drift and anode voltage respectively and 50\% Ptr: (top-left) Total time on the mesh, (top-right) total-time after the mesh, (bottom-left) avalanche transmission time and (bottom-right) photoelectron transmission time. The points are results of GARFIELD++ simulations whilst the red lines represent the respective model predictions, as it is described in the text.}
\label{fig:fig19}
\end{figure}

	Up to this point, the model has been used to provide information on the mean value and the variance (i.e. to evaluate the first and second moments) of transmission time distributions. 
	However, it can also be used for more general statistical predictions, e.g. the complete probability density functions of the above time variables. 
	As an example, Fig. \ref{fig:fig19} shows the distributions, produced by GARFIELD++ simulations (black points), of the photoelectron, the avalanche and the total time (on and after the mesh), without any restriction on the avalanche length or on the electron multiplicity on the mesh. 
	The apparent left-right asymmetry and the long tails in these distributions are partially caused by the dependence of the mean transmission times on the length of the avalanche (or equivalently, on the length of the photoelectron drift-path, before the first ionization). 
	Nevertheless, the dependence of the respective variances on the length of the avalanche also contributes to the asymmetry and the tails. 
	In order to predict the functional form of the above asymmetric distributions, the model is complemented with the extra assumption that the related transmission times, corresponding to a certain avalanche length, follow an Inverse Gaussian distribution (Wald) function, which is expressed as:
\begin{equation}\label{eq:eq72}
f(x;\mu,\lambda)=\Bigg(\dfrac{\lambda}{2\pi x^3}\Bigg)^{1/2}\exp\Bigg[\dfrac{-\lambda(x-\mu)^2}{2\mu^{2}x}\Bigg]
\end{equation}
with the parameter $\mu$ to be the mean value and the shape parameter $\lambda$ to be related with the variance of the distribution as $V[x]=\mu^3/\lambda$. 
In general, the convolution of two Wald distributions is not a Wald distribution. Consequently, even if the photoelectron and avalanche transmission times are described by Wald distributions, it is not necessarily true that the total-times are distributed according to the same functional form. 
	However, GARFIELD++ simulation results indicate, see also Fig. \ref{fig:fig4}, that the distributions of the total-times, on and after the mesh, are very well approximated by Wald functions.

	Hereafter, the model assumes that the statistical properties of the photoelectron transmission time, $T_p$, and the avalanche transmission time, T,  are described by  Wald distributions as follows: 
\begin{equation}\label{eq:eq73}
\begin{array}{l}
f_p\big(T_{p};\mu_{p}(L),\lambda_{p}(L)\big)=\Bigg(\dfrac{\lambda_{p}(L)}{2\pi T^3}\Bigg)^{1/2}\cdot \exp\Bigg[\dfrac{-\lambda_{p}(L)\big(T_{p}-\mu_{p}(L)\big)^2}{2\mu_{p}^{2}(L)\cdot T_{p}}\Bigg]\\
f\big(T;\mu(L),\lambda(L)\big)=\Bigg(\dfrac{\lambda(L)}{2\pi \cdot T^3}\Bigg)^{1/2}\cdot \exp\Bigg[\dfrac{-\lambda(L)\cdot \big(T-\mu(L)\big)^2}{2\mu^{2}(L)\cdot T}\Bigg]
\end{array}
\end{equation}
where\\
$\mu_{p}(L)=\dfrac{D-L}{V_p}+d_{off}$, according to Eq.\ref{eq:eq18}\\
$\lambda_{p}(L)=\dfrac{\mu_{p}^{3}(L)}{(D-L)\cdot \sigma_{p}^{2}+\Phi}$. according to Eq. \ref{eq:eq21}\\
Similarly $\mu (L)=<T(L)>$ where $<T(L)>$ is given by Eq. \ref{eq:eq13}\\
$\lambda(L)=\dfrac{\mu^3(L)}{V\big[T(L)\big]}$ and $V\big[T(L)\big]$ is given by Eq. \ref{eq:eq43}.

Using the probability density $R(L;a)$, i.e. the p.d.f. to observe an avalanche of length L (defined by Eq. \ref{eq:eq15}) the distributions of $T_p$ and T for any possible value of L are given  by:
\begin{equation}\label{eq:eq74}
\begin{array}{l}
F_{p}(T_p)=\int\limits_{x_1}^{x_2}f_{p}\big(T_{p};\mu_{p}(L),\lambda_{p}(L)\big)\cdot R(L;a)dL\\
F(T)=\int\limits_{x_1}^{x_2}f\big(T;\mu(L),\lambda(L)\big)\cdot R(L;a)dL
\end{array}
\end{equation}

	The solid lines in the bottom-row plots of Fig. \ref{fig:fig19} represent graphically the model predictions expressed  by the respective p.d.f.s of Eq. \ref{eq:eq74}. 
	The model predictions are in excellent agreement with the GARFIELD++ simulation results.
	
Similarly, it is assumed that the total-time distributions, on and after the mesh ($T_{tot}$ and $T_m$, respectively) for a certain avalanche length, L, can be well  approximated by Wald functions, as: 
\begin{equation}\label{eq:eq75}
\begin{array}{l}
f_{tot}\big(T_{tot};\mu_{tot}(L),\lambda_{tot}(L)\big)=\Bigg(\dfrac{\lambda_{tot}(L)}{2\pi T_{tot}^3}\Bigg)^{1/2}\cdot \exp\Bigg[\dfrac{-\lambda_{tot}(L)\big(T_{tot}-\mu_{tot}(L)\big)^2}{2\mu_{tot}^{2}(L)\cdot T_{tot}}\Bigg]\\
f_{m}\big(T_{m};\mu_{m}(L),\lambda_{m}(L)\big)=\Bigg(\dfrac{\lambda_{m}(L)}{2\pi T_{m}^3}\Bigg)^{1/2}\cdot \exp\Bigg[\dfrac{-\lambda_{m}(L)\big(T_{m}-\mu_{m}(L)\big)^2}{2\mu_{m}^{2}(L)\cdot T_{m}}\Bigg]
\end{array}
\end{equation}
where\\
$\mu_{tot}(L)=\dfrac{D-L}{V_p}+d_{off}+\big<T(L)\big>$, according to Eq. \ref{eq:eq13} and \ref{eq:eq18}\\
$\lambda_{tot}(L)=\dfrac{\mu_{tot}^{3}(L)}{V\big[T_{tot}(L)]}$. according to Eq. \ref{eq:eq44}\\
Also $\mu_{m} (L)=\mu_{tot}(L)+<\Delta t>$ according to Eq. \ref{eq:eq60}, and \\
$\lambda_{m}(L)=\dfrac{\mu_{m}^{3}(L)}{V[T_{m}(L)]}$ where $V[T_{m}(L)]$ is given by Eq. \ref{eq:eq68}.

The predictions of Eq. \ref{eq:eq75} are shown in the top-row plots of Fig. \ref{fig:fig19} to be also in excellent  agreement with the GARFIELD++ simulation results. 
It addition it has been verified that the model predicts successfully the transmission time distributions at all drift voltage settings considered in this study.\\

%	As demonstrated through this work, the developed model is very successful in providing insights for the major microscopic mechanisms, which determine the timing characteristics of the detector, and in explaining coherently the unexpected behaviour of microscopic quantities, predicted by GARFIELD++ simulations.  
%	Due to the very good agreement of the model predictions with the detailed GARFIELD++ simulation results, the formulae developed in this work can be used easily as a tool for fast predictions, provided that the values of the model input-parameters, i.e. the parameters shown in Table \ref{tab:tableA-8}, are known for the considered operating conditions.  
%	This necessity, obviously limits the application of the developed model as a stand-alone tool. 
%	However, having available sets of input parameter values for certain operational settings, it is possible to derive an empirical parametrization of the input parameters, which can be used to provide input to the model for the whole region of operational settings covered by the above parameterization.

\section{Summary} \label{concl}
This work employs the comparison of experimental data with detailed simulations, based on the  GARFIELD++ package, complemented with a statistical description of the electronic signal formation, to identify the microscopic quantities that determine the PICOSEC timing characteristics.
	Subsequently, a stochastic model is developed that describes the  properties of the above  quantities, offering a phenomenological, microscopic interpretation of the observed timing properties of the detector. \\
	The model is based on: i) the fact that an electron drifting in a gas under the influence of an homogeneous electric field achieves higher drift velocity when, in addition to  elastic scattering, undergoes inelastic interactions, and ii) the assumption that a newly produced electron through ionization acquires  a certain time-gain relative to its parent and subsequently drifts with the same velocity as the parent electron.  
		The input parameters, compiled in Table \ref{tab:tableA-8}, are commonly used statistical variables\footnote{With the only exception of the time-gain parameter $\rho$, which has been introduced in this work.}, which have been evaluated by analyzing GARFIELD++ simulation results. \\
		
The quantitative predictions of the model have been compared extensively with the related  GARFIELD++ simulation results  and found in  very good agreement at all operating PICOSEC conditions considered in this study, demonstrating the success of this stochastic interpretation.  

	As demonstrated through this work, the developed model is very successful in providing insights for the major microscopic mechanisms, which determine the timing characteristics of the detector, and in explaining coherently the unexpected behavior of microscopic quantities, predicted by GARFIELD++ simulations.  
	Due to the very good agreement of the model predictions with GARFIELD++, the formulae developed in this work can be used easily as a tool for fast predictions, provided that the values of the model input-parameters, i.e. the parameters shown in Table \ref{tab:tableA-8}, are known for the considered operating conditions.  
	This necessity, obviously limits the application of the developed model as a stand-alone tool. 
	However, having available sets of input parameter values for certain operational settings, it is possible to derive empirical parametrizations of the input parameters, which can be used to provide input to the model for a broader region of operational settings covered by the above parameterization. \\

\bibliography{Model_MathDraft_v12}

\newpage
\appendix

\section{}\label{Appendix A}

\begin{table}[!h]
\begin{center}
\resizebox{1.0\textwidth}{!}{
\begin{tabular}{ c| c c c}
 & Ptr 0\% & Ptr 50\% & Ptr 100\%   \\ 
\hline 
Photoelectron Drift Velocity ($\microm/$ns) & $156.8\pm 0.4$ & $150.5\pm 0.8$ & $142.2\pm 1.0$ \\  [0.35cm]
Avalanche Drift Velocity ($\microm/$ns) & $181.4\pm 0.5$ & $184.8\pm 0.8$ & $188.2\pm 0.9$ \\  [0.35cm]
Avalanche-Electron Drift Velocity ($\microm/$ns) & $169.9\pm 0.2$ & $170.4\pm 0.2$ & $170.0\pm 0.2$ \\  [0.35cm]
\end{tabular}
}
\caption{The values of: the photoelectron drift velocity $V_{p}$, the avalanche drift velocity $V_{a}$ and the drift velocity $V_{ea}$, of an avalanche-electron, for three different values of Ptr and default high voltage settings.}
\label{tab:tableA-1}
\end{center}
\end{table}

\begin{table}[!h]
\begin{center}
\resizebox{1.0\textwidth}{!}{
\begin{tabular}{ c| c c c}
 & Ptr 0\% & Ptr 50\% & Ptr 100\%   \\ 
\hline 
First Townsend Coeff. ($\microm^{-1}$) & $0.0520 \pm 0.0003$ & $0.0695\pm 0.0005$ & $0.0893\pm 0.0008$ \\  [0.35cm]
\end{tabular}
}
\caption{The first Townsend coefficient,  estimated from GARFIELD++ simulations, for different Ptr values and the default drift voltage settings.}
\label{tab:tableA-2}
\end{center}
\end{table}

\begin{table}[!h]
\begin{center}
\resizebox{1.0\textwidth}{!}{
\begin{tabular}{ c| c c c}
 & Ptr 0\% & Ptr 50\% & Ptr 100\%   \\ 
\hline 
Mean time-gain, $\rho$ (ns) & $17.40\cdot10^{-3}\pm 0.3 \cdot 10^{-3}$ & $17.25\cdot 10^{-3}\pm 0.42\cdot 10^{-3}$ & $17.72\cdot 10^{-3}\pm 0.48\cdot 10^{-3}$ \\  [0.35cm]
Time Constant, C (ns) & $53.50\cdot 10^{-3}\pm 3.0\cdot 10^{-3}$ & $60.0\cdot 10^{-3}\pm 4.0\cdot 10^{-3}$ & $68.0\cdot 10^{-3}\pm 5\cdot 10^{-3}$ 
\end{tabular}
}
\caption{Mean values of the time-gain $\rho$ and values of the constant term C (see Eq. \ref{eq:eq7}), estimated for three Ptr values and the default drift voltage settings.}
\label{tab:tableA-3}
\end{center}
\end{table}

\begin{table}[!h]
\begin{center}
\resizebox{1.0\textwidth}{!}{
\begin{tabular}{ c| c c c}
\multicolumn{4}{c}{Number of Electrons on the Mesh}\\
 & Ptr 0\% & Ptr 50\% & Ptr 100\%   \\ 
\hline 
Constant Term &2 (fixed) & 2 (fixed) & 2 (fixed) \\  [0.35cm]
Multiplication Coeff., $a_{eff}$ ($\microm^{-1}$) & $32.47\cdot 10^{-3} \pm 0.01\cdot 10^{-3}$ & $39.12\cdot 10^{-3} \pm 0.01\cdot 10^{-3}$ & $45.30\cdot 10^{-3} \pm 0.02\cdot 10^{-3}$\\[0.35cm] \\
 \multicolumn{4}{c}{Number of Electrons after the Mesh}\\
 & Ptr 0\% & Ptr 50\% & Ptr 100\%   \\ \hline
Constant Term &$0.53\pm 0.01$ & $0.50\pm 0.02$ & $0.57\pm 0.02$ \\  [0.35cm]
Exponential Slope & $32.80\cdot 10^{-3} \pm 0.3\cdot 10^{-3}$&$39.40\cdot 10^{-3}\pm 0.2\cdot 10^{-3}$&$45.00\cdot 10^{-3}\pm 0.2\cdot 10^{-3}$
\end{tabular}
}
\caption{The exponential slopes and the constant terms that determine the number of electrons on and after the mesh, as estimated by GARFIELD++ simulations. (top) The exponential slope $a_{eff}$ is the avalanche multiplication coefficient. The mean number of electrons on the mesh (q), is given as a function of the avalanche length (L) by the expression $q\left( L;a_{eff}\right) = q_0 \cdot e^{a_{eff}L}$, where the constant term ($q_0$) is set to $q_0$=2, because the avalanche starts with two electrons. (bottom) The number of electrons passing through the mesh, is also expressed exponentially as a function of L. The passage through the mesh does not affect the exponential slope. However the constant term is found to be $\simeq 0.5$, which translates to $\sim 25\%$ mesh transparency.}
\label{tab:tableA-4}
\end{center}
\end{table}

\begin{table}[!h]
\begin{center}
\resizebox{1.0\textwidth}{!}{
\begin{tabular}{ c| c c c}
 & Ptr 0\% & Ptr 50\% & Ptr 100\%   \\ 
\hline 
On the Mesh& $0.510 \pm 0.005$ & $0.464 \pm 0.005$ & $0.422 \pm 0.005$ \\  [0.35cm]
After the Mesh& $0.530 \pm 0.01$ & $0.475 \pm 0.005$ & $0.430 \pm 0.005$ 
\end{tabular}
}
\caption{Ratio of the RMS over the mean value of the number of electrons in any given avalanche length. Notice that this ratio equals to $\big(1/(1+\theta)^{1/2}\big)$, where $\theta$ is the shape parameter of the Gamma distribution function.}
\label{tab:tableA-5}
\end{center}
\end{table}

\begin{table}[!h]
\begin{center}
\resizebox{1.0\textwidth}{!}{
\begin{tabular}{ c| c c c}
 & Ptr 0\% & Ptr 50\% & Ptr 100\%   \\ 
\hline 
Time Variance per unit length (ns$^{2}/\microm$)& $11.65\cdot 10^{-5}\pm 0.05\cdot 10^{-5}$ & $11.75\cdot 10^{-5}\pm 0.05\cdot 10^{-5}$ & $11.67\cdot 10^{-5}\pm 0.05\cdot 10^{-5}$ \\  [0.35cm]
Constant Term (ns$^{2}$) & $16.55\cdot 10^{-5}\pm 1.50\cdot 10^{-5}$ & $16.78\cdot 10^{-5}\pm 1.62\cdot 10^{-5}$ & $17.03\cdot 10^{-5}\pm 0.80\cdot 10^{-5}$ 
\end{tabular}
}
\caption{Diffusion properties of the avalanche electron.}
\label{tab:tableA-6}
\end{center}
\end{table}

\begin{table}[!h]
\begin{center}
\resizebox{1.0\textwidth}{!}{
\begin{tabular}{ c| c c c}
 & Ptr 0\% & Ptr 50\% & Ptr 100\%   \\ 
\hline 
Time Variance per unit length (ns$^{2}/\microm$)& $13.27\cdot 10^{-5}\pm 0.3\cdot 10^{-5}$ & $13.80\cdot 10^{-5}\pm 0.3\cdot 10^{-5}$ & $13.30\cdot 10^{-5}\pm 0.6\cdot 10^{-5}$ \\  [0.35cm]
Constant Term (ns$^{2}$) & $-47.27\cdot 10^{-5}\pm 6.80\cdot 10^{-5}$ & $-56.22\cdot 10^{-5}\pm 6.8\cdot 10^{-5}$ & $-67.64\cdot 10^{-5}\pm 13.4\cdot 10^{-5}$ 
\end{tabular}
}
\caption{Diffusion properties of a photoelectron before it initiates an avalanche.}
\label{tab:tableA-7}
\end{center}
\end{table} 

\begin{table}[!th]
\begin{center}
\resizebox{1.0\textwidth}{!}{
\begin{tabular}{ c| c c c c c}
\textbf{Penning Transfer Rate}&\multicolumn{5}{c}{\textbf{50\%}}\\
\textbf{Anode Voltage} &\multicolumn{5}{c}{\textbf{450 V}} \\ 
\hline 
\textbf{Drift Voltage} &\textbf{325 V} & \textbf{350 V} & \textbf{375 V} & \textbf{400 V}& \textbf{425 V}\\  \hline
$a$ ($10^{-2}\microm^{-1}$) &$3.607 \pm 0.018$ & $4.400 \pm 0.020$ &$5.208 \pm 0.027$  &$6.069 \pm 0.027$ &$6.950 \pm 0.032$ \\  [0.35cm]
$a_{eff}$ ($10^{-2}\microm^{-1}$) &$2.215\pm 0.001$ & $2.629\pm 0.001$ &$3.055 \pm 0.001$  &$3.484 \pm 0.001$ &$3.912 \pm 0.001$ \\  [0.35cm]
$\theta$ &$2.698\pm 0.142$ & $2.906\pm 0.154$ &$3.037 \pm 0.162$  &$3.313 \pm 0.179$ &$3.645 \pm 0.191$ \\  [0.35cm]
$V_{ea}^{-1}$ ($10^{-3} \textrm{ns}/\microm$) &$7.311\pm 0.003$ & $6.877\pm 0.003$ &$6.509 \pm 0.002$  &$6.173 \pm 0.002$ &$5.866 \pm 0.004$ \\  [0.35cm]
$V_{p}^{-1}$ ($10^{-3} \textrm{ns}/\microm$ &$8.065\pm 0.026$ & $7.678\pm 0.026$ &$7.266 \pm 0.028$  &$6.923 \pm 0.028$ &$6.643 \pm 0.031$ \\  [0.35cm]
$d_{off}$ ($10^{-2} \textrm{ns}$) &$-3.831 \pm 0.084$ & $-3.437 \pm 0.082$ &$-2.883 \pm 0.075$  &$-2.678 \pm 0.068$ &$-2.364 \pm 0.079$ \\  [0.35cm]
$\rho$ ($10^{-2} \textrm{ns}$) &$3.570\pm 0.054$ & $2.919\pm 0.027$ &$2.489 \pm 0.030$  &$2.185 \pm 0.028$ &$1.725 \pm 0.045$ \\  [0.35cm]
C ($10^{-2} \textrm{ns}$) &$7.555\pm 0.218$ & $7.511\pm 0.117$ &$7.668 \pm 0.166$  &$7.778 \pm 0.196$ &$7.001 \pm 0.516$ \\  [0.35cm]
$\sigma_{p}^{2}$ ($10^{-4} \textrm{ns}^{2}/\microm$) &$2.137\pm 0.054$ & $1.908\pm 0.046$ &$1.662\pm 0.073$  &$1.554\pm 0.050$ &$1.380\pm 0.063$ \\  [0.35cm]
$\Phi (10^{-4} \textrm{ns}^2)$ &$-9.967 \pm 2.417$ & $-7.936 \pm 1.395$ &$-6.40 \pm 1.650$  &$-7.525 \pm 1.343$ &$-5.622 \pm 1.284$ \\  [0.35cm]
$\sigma_{0}^{2}$ ($10^{-4} \textrm{ns}^{2}/\microm$) &$2.094 \pm 0.005$ & $1.778\pm 0.003$ &$1.543\pm 0.004$  &$1.341\pm 0.003$ &$1.175\pm 0.004$ \\  [0.35cm]
tr &$0.244\pm 0.009$ & $0.248\pm 0.044$ &$0.238\pm 0.011$  &$0.251\pm 0.009$ &$0.247\pm 0.009$ \\  [0.35cm]
$\delta (10^{-2} \textrm{ns})$ &$7.217 \pm 0.034$ & $6.871 \pm 0.032$ &$6.607 \pm 0.031$  &$6.305 \pm 0.030$ &$5.938 \pm 0.040$ \\  [0.35cm]
$\Delta t_{mesh} (10^{-1} \textrm{ns})$ &$1.521 \pm 0.005$ & $1.455 \pm 0.005$ &$1.400 \pm 0.004$  &$1.344 \pm 0.003$ &$1.303 \pm 0.004$\\ \hline
\multicolumn{6}{c}{\textbf{Control Parameters}}\\ \hline
\textbf{$x_{1}$} {($\microm$)} &0 & 0 &0  &0 &0 \\  [0.35cm]
\textbf{$x_{2}$} {($\microm$)} &164 & 167 &174  &174 &172 \\  [0.35cm]
\textbf{$w/\rho$} &1 & 1 & 1 & 1&1 \\  [0.35cm]
\textbf{$D$} {($\microm$)} &182 & 182 &182  &182 &182 \\  [0.35cm]
\textbf{$N_{max}$} &350 & 500 &1250  &1750 &3500 \\  [0.35cm]
\end{tabular}
}
\caption{Parameter values used in the model.}
\label{tab:tableA-8}
\end{center}
\end{table} 

\section{}
\label{Appendix B}
Let $y(L)$ be a measurement (random variable) of a physical variable Y, which depends on another physical variable, L, as $Y=f(L)$. Let also the statistical properties of $y$ depend on L, such that:
\begin{equation}\label{eq:eqB1}
\begin{array}{l}
\big<y(L)\big>=\int\limits_{\Omega_{y}}y\cdot H(y,L)dy=f(L)\\
\big<y^{2}(L)\big>-\big<y(L)\big>^{2}=\int\limits_{\Omega_{y}}\left[ y-\big<y(L)\big>\right] ^2\cdot H(y,L)dy=u(L)
\end{array}
\end{equation}
where $\Omega_y$ is the set of all possible values of y and H(y,L) is the p.d.f. describing the measurement process, which explicitly depends on the physical variable L, resulting to mean values and variances dependent on L as shown in Eq. \ref{eq:eqB1}. Furthermore, the physical variable L is distributed, for physics reasons, according to the p.d.f. g(L). Suppose an experiment in which several measurements y of the physical variable Y are performed but there is not any experimental way to know the corresponding value of L. In the following the expected variance of the measurements, y, for any possible L, is expressed in terms of f(L), u(L) and g(L). A possible outcome of a measurement in the above experiment will follow the p.d.f. h(y) given as
\begin{equation}\label{eq:eqB2}
h(y)=\int\limits_{\Omega_{L}}H(y,L)\cdot g(L)dL
\end{equation}
with $\Omega_L$ standing for the set of all possible values of L. The mean value of the measurements y, for any possible value of L, will be
\begin{equation}\label{eq:eqB3}
<y>=\int\limits_{\Omega_{y}}\int\limits_{\Omega_{L}}y\cdot H(y,L)\cdot g(L) dL dy=\int\limits_{\Omega_{L}}f(L)\cdot g(L) dL
\end{equation}
The second moment of y is expressed in the same way as:
\begin{equation}\label{eq:eqB4}
<y^{2}>=\int\limits_{\Omega_{y}}\int\limits_{\Omega_{L}}y^{2}\cdot H(y,L)\cdot g(L) dL dy=\int\limits_{\Omega_{L}}\big[u(L)+f^{2}(L)\big]\cdot g(L) dL
\end{equation}
where the definition of u(L) from Eq. \ref{eq:eqB1} has been used. Combining Eq. \ref{eq:eqB3} with Eq. \ref{eq:eqB4} the variance of y for any possible L is given by:
\begin{equation}\label{eq:eqB5}
\begin{array}{l}
V[y]=<y^{2}>-<y>^2\\
=\int\limits_{\Omega_L}\left[ u(L)+f^{2}(L)\right] \cdot g(L)dL-\left[ \int\limits_{\Omega_L}f(L)\cdot g(L)dL\right]^2\\
=\int\limits_{\Omega_L} u(L)\cdot g(L) dL+\left\lbrace \int\limits_{\Omega_L}f^{2}(L) \cdot g(L)dL-\left[ \int\limits_{\Omega_L}f(L)\cdot g(L)dL\right]^2\right\rbrace 
\end{array}
\end{equation}
where the first term expresses the proper averaging of the y variances each defined at specific L. However, the fact that the mean value of y depends on L results to an additional term. This second term expresses the variance of f(L) while L is distributed according to g(L).
%\end{linenumbers}
\end{document}